\documentclass[reprint,twocolumn]{revtex4}

\usepackage{graphicx}%
\usepackage{dcolumn}%
\usepackage{bm}%
\usepackage{epsfig,colordvi}
\usepackage[section]{placeins}
\usepackage{color} 
\usepackage{here}

 \def\pv{\mathbf p}

 \def\be{\begin{equation}}
 \def\ee{\end{equation}}
 \def\bea{\begin{eqnarray}}
 \def\eea{\end{eqnarray}}
 \def\bean{\begin{eqnarray*}}
 \def\eean{\end{eqnarray*}}
 \def\gsim{\mathrel{\rlap{\lower0.2em\hbox{$\sim$}}\raise0.2em\hbox{$>$}}}
 \def\ksim{\mathrel{\rlap{\lower0.2em\hbox{$\sim$}}\raise0.2em\hbox{$<$}}}
 \def\kg{\mathrel{\rlap{\lower0.25em\hbox{$>$}}\raise0.25em\hbox{$<$}}}

 \newcommand{\eq}[1]{(\ref{#1})}

\begin{document}

\title{Cluster and hyper-cluster production in relativistic heavy-ion collisions within the Parton-Hadron-Quantum-Molecular-Dynamics approach}

\author{Susanne Gl\"a{\ss}el$^{1}$, Viktar Kireyeu$^{2,3}$, Vadim Voronyuk$^{2,3}$, 
J\"org Aichelin$^{4,5}$, Christoph Blume$^{1}$,  Elena~Bratkovskaya$^{6,7,3}$, 
Gabriele Coci$^6$,  Vadim Kolesnikov$^2$ and  Michael Winn$^4$}

\affiliation{$^1$ Institut f\"ur Kernphysik, Johann Wolfgang Goethe-Universit\"at,
Max-von-Laue-Str. 1, 60438 Frankfurt am Main, Germany}
\affiliation{$^{2}$ Joint Institute for Nuclear Research, Joliot-Curie 6, 141980 Dubna, Moscow region, Russia}
\affiliation{$^3$ Helmholtz Research Academy Hessen for FAIR (HFHF), GSI Helmholtz Center for Heavy Ion Physics, Campus Frankfurt, 60438 Frankfurt, Germany}
\affiliation{$^4$ SUBATECH, Universit\'e de Nantes, IMT Atlantique, IN2P3/CNRS
4 rue Alfred Kastler, 44307 Nantes cedex 3, France}
\affiliation{$^5$ Frankfurt Institute for Advanced Studies, Ruth Moufang Str. 1, 60438 Frankfurt, Germany}
\affiliation{$^6$ GSI Helmholtzzentrum f\"ur Schwerionenforschung GmbH,
  Planckstr. 1, 64291 Darmstadt, Germany} 
\affiliation{$^7$ Institut f\"ur Theoretische Physik, Johann Wolfgang Goethe-Universit\"at,
Max-von-Laue-Str. 1, 60438 Frankfurt am Main, Germany}

\date{\today}

\begin{abstract} \noindent
We study cluster and hypernuclei production in heavy-ion collisions at relativistic energies employing
the Parton-Hadron-Quantum-Molecular-Dynamics (PHQMD) approach, a microscopic n-body transport model based on the QMD propagation of the baryonic degrees of freedom with density dependent 2-body potential interactions.
All other ingredients of PHQMD, including the collision integral and the treatment of the quark-gluon plasma (QGP) phase, are adopted from the Parton-Hadron-String Dynamics (PHSD) approach. In PHQMD the cluster formation occurs dynamically, caused by the interactions. The clusters are recognized by the Minimum Spanning Tree (MST) algorithm. 
We present the PHQMD results for cluster and hypernuclei formation  in comparison with the available experimental data 
at AGS, SPS, RHIC-BES and RHIC fixed target energies. We also provide predictions on cluster production for the upcoming FAIR and NICA experiments.
PHQMD allows to study the time evolution of formed clusters and the origin of their production, which helps to understand how such weakly bound objects are formed and survive in the rather dense and hot environment created in heavy-ion collisions. It offers therefore an explanation of the 'ice in the fire' puzzle.
\end{abstract}


\maketitle

 
\section{Introduction}
Cluster production is one of the challenging topics of heavy-ion physics.
Since light clusters with a mass $A \le 4$ have been discovered at midrapity at the Bevalac accelerator in Berkeley \cite{Gutbrod:1988gt,Nagamiya:1981sd} and later been extensively studied at the SIS accelerator at GSI \cite{FOPI:2010xrt}, their production has been a mystery which was even deepened by the discovery that also at higher beam energies, at AGS \cite{Armstrong:2000gz}, at SPS \cite{Anticic:2016ckv}, at RHIC \cite{Adam:2019wnb,Arsene:2010px} and lately as well at the LHC \cite{Acharya:2017bso}, light midrapidity clusters are created. This is due to the fact that the multiplicities and the transverse momentum distributions of hadrons, such as $\pi, K, \Lambda$  and $p$, observed at midrapidity, follow thermal model predictions, suggesting that the produced fireball is in thermal equilibrium.
Taking the temperature and the chemical potential as determined from the multiplicity of light hadrons, one can predict the multiplicity of these clusters. Surprisingly, these predictions agree quite well with the measured multiplicities \cite{Andronic:2010qu,Andronic:2017pug}. So it seems that these clusters and hadrons form a common thermal midrapidity source.
The temperature of this source is well above $T = 100$~MeV and increases with beam energy. 

On the other hand, light clusters, with a binding energy per nucleon  well below $10$~$MeV$, cannot live in a thermal heat bath of a temperature of more than $T = 100$~MeV. They would not survive a collision with one of the constituents of this thermal source. This "puzzle" has been  phrased as "ice in the fire" and so it is the challenge to understand how such weakly bound objects can be formed during a heavy-ion collision. 

If one assumes that the density in the thermalized system does not fluctuate, these observed clusters cannot be formed early, when collisions are frequent and the mean free path for collisions is small. They can also not be formed late because the baryon density decreases very quickly and therefore soon the distance between nucleons gets larger than the range of the nuclear potential. Simulations show that
the time interval between these two limits is very short.

There are two other conditions which hold for the creation of clusters and complicate their production process: 
a) to conserve energy and momentum, the formed cluster has to interact with a catalyst which is only available if the density is not too low and 
b) clusters like a deuteron, with a binding energy of 2.24~MeV  and a root-mean-squared (rms) radius of 1.95~fm, can hardly be formed, if the hadron density is that high that there are hadrons in between the cluster nucleons which screen the interaction between the proton and neutron of the deuteron. 
All these conditions point to a very limited interval in course of the expansion in which these clusters can be formed and it is not easy to understand how the clusters, after production, can still come to a thermal equilibrium with the expanding system, as their multiplicity seems to indicate.

Almost all single particle observables, like rapidity and transverse momentum distributions, as well as collective variables, like the radial and elliptic flow, can nowadays be reasonably well reproduced by transport approaches like PHSD \cite{Cassing:2008sv,Cassing:2009vt}, EPOS \cite{Pierog:2013ria}, AMPT \cite{Lin:2004en} or UrQMD  \cite{Bass:1998ca,Bleicher:1999xi}, which simulate the entire heavy-ion reaction on the computer. 
However, for a description of cluster production this is not the case. The reason is that the first three 
approaches are based on the time evolution of the single particle Wigner density, which conceptually does not 
allow for studying the dynamical production of clusters due to genuine many body correlations. 
In UrQMD (in the standard version) baryons do not interact by potential interactions. This excludes also  the dynamical formation of clusters in the course of the time evolution of the system.

To address cluster formation in such transport approaches, the coalescence model has been applied to extract clusters from the distribution of single baryons (cf. e.g. \cite{Botvina:2014lga,Botvina:2016wko, Abelev:2010rv,Zhu:2015voa,Zhao:2021dka}). The coalescence model itself has a long history. It has been advanced by Butler and Pearson \cite{Butler:1963pp} who calculated that in a static potential the momentum distribution of the deuteron $n_d(\pv)$ is given by $n_d(\pv) \propto \frac{1}{\pv^2}n_n^2(\frac{\pv}{2})$, where $n_n(\pv)$ is the nucleon distribution function if one takes into account that deuteron production is a three body process. Later-on, simpler versions have been advanced which also predicted $n_d(\pv) \propto K\  n_n^2(\frac{\pv}{2})$, however with very different constants of proportionality, $K$.
These models did not take into account anymore the three body nature of deuteron production.
In  Ref.~\cite{Schwarzschild:1963zz} $K\propto p_0^3$, where $p_0$ is a coalescence radius in momentum space,
in Ref.~\cite{Kapusta:1980zz} $K\propto V$, the fireball volume in which the deuteron is produced and in Ref.~\cite{Bond:1977fd} $K\propto \frac{1}{V}$, where $V $ is the volume at which a sudden transition from a strongly interacting system to a non-interacting system occurs.
In the meantime experiments have shown \cite{Adam:2019wnb} that the ratio between the deuteron momentum distribution and that of the nucleons $ \frac{n_d(\pv)}{n_n^2(\frac{\pv}{2})}$
is not given by a single constant but by a momentum dependent function.

More recently, a model has been advanced in Ref.~\cite{Scheibl:1998tk} which relates the single particle phase space distribution functions of protons $f_p(r,p)$ and neutrons $f_n(r,p)$ with the momentum distribution of the deuterons via the Wigner density of the relative motion of the deuteron $W(\mathbf{r,q})$:
\bea
\frac{dN}{d^3P_d}&=&\frac{3}{(2\pi)^3}\int d^3r_d\int \frac{d^3r d^3q}{(2\pi)^3} W({\mathbf{r,q}})\nonumber\\  &\cdot& f_p({\mathbf{r_+,q_+}})f_n({\mathbf{r_-,q_-}})
\eea
with ${\mathbf{r}_\pm}= {\mathbf{r}_d}\pm {\mathbf{r}/2}$ and ${\mathbf{q}_\pm}= {\mathbf{p_d}/2}\pm {\mathbf{q}}$. ${\mathbf{r}_d}$ and $\mathbf{p_d}$ are the center of mass coordinates, $\mathbf{r}$ and 
$\mathbf{q}$ the relative coordinates.
In actual calculations the deuteron wave function is replaced by a Gaussian wave function with the rms radius of the deuteron. For this wave function also the Wigner density is a Gaussian and easy to apply.    
This Wigner density technique has been frequently applied in transport approaches, in which baryons have no potential interaction \cite{Sombun:2018yqh,Ko:2010zza,Sun:2020uoj}, to predict the deuteron momentum distribution. 
It has also been applied to transport approaches, in which baryons interact via a mean field, the so-called Boltzmann-\"Uhling-Uhlenbeck (BUU) \cite{Aichelin:1984asp,Kruse:1985pg,Weil:2016zrk} approaches, and in which the nucleons are represented by a swarm of point-like test particles (cf. \cite{Mohs:2020awg}),
although it is not obvious how a transport theory, which follows the time evolution of the single particle phase space distribution, can be suited for the investigation of clusters. The necessity of the presence of a third body, when a deuteron is produced, is neglected in the Wigner density approach as well as the possibility that in between the nucleons of the deuteron other hadrons can be present which screen the interaction. 

Since the medium formed in heavy-ion collisions is rapidly expanding, the deuteron yield depends on 
the time when the coalescence model is applied. There are different possible choices for which we
refer to the original articles. If one assumes that the coalescence takes place
when the last of the cluster nucleons had its last hadronic collision \cite{Zhu:2015voa}, 
a quite reasonable agreement with the measured the yields is obtained.

In order to study the cluster production in a dynamical way, created by the potential interactions, which are present during the whole heavy-ion reaction, and hence without additional assumptions as inherent in the coalescence model, one has to start out from transport theories which propagate not the 1-body but the n-body phase space density. Such models, namely the family of QMD models \cite{Hartnack:1997ez}, have been developed in the eighties and can predict cluster formation at midrapidity as well as in the projectile/target domain. 
They have been successfully applied to describe heavy-ion collisions at lower energies of a few 100~$A$MeV \cite{Zbiri:2006ts} as well
as at the Bevalac/SIS energies of a few $A$GeV.   

Recently we advanced the Parton-Hadron-Quantum-Molecular-Dynamics (PHQMD) approach \cite{Aichelin:2019tnk}, a QMD type microscopic transport model which can  also be applied to relativistic energies. We have demonstrated that PHQMD reproduces single particle observables for energies between $E_{kin} = 500$~$A$MeV and $\sqrt{s_{NN}} = 200$~GeV and describes the cluster formation as well. There we compared the PHQMD results
to the available experimental data  at $E_{kin} \approx 600$ $A$MeV - 1.5~$A$GeV and 11~$A$GeV. 
In PHQMD clusters are formed dynamically due to the potential interactions between baryons. The clusters  are identified 
by the  MST (Minimum Spanning Tree) \cite{Aichelin:1991xy} or the SACA (‘Simulated Annealing Cluster Algorithm’) \cite{Puri:1996qv,Puri:1998te}
algorithm, which  finds the most bound configuration of nucleons and clusters. 

In this paper we present a detailed study of the cluster production at relativistic energies, with a special focus on 
the energies which are relevant for the upcoming NICA and FAIR experiments. We confront the PHQMD results with available experimental 
data on cluster production at midrapidity from AGS, SPS and RHIC BES (beam energy scan)  experiments.
After verifying that these data are well reproduced we advance to the study of the origin of clusters. In particular, we 
identify the time when they are formed and the way in which they are formed, either by disintegration of heavier clusters or 
by assembling protons and neutrons.    

This paper is organized as follows:
In Section II we briefly evoke the basic ideas of the PHQMD model. In Section III we provide the model description of the PHQMD and the MST cluster recognition procedure. In Sections IV and V we present the results for cluster and hypernuclei production in heavy-ion collisions for different energies, while in Section VI we investigate the time evolution of the cluster production.
Finally, in Section VII we summarize our findings.

\section{Model description: The Parton-Hadron-Quantum-Molecular-Dynamics (PHQMD) approach}

In this section we recall the basic ideas of the Parton-Hadron-Quantum-Molecular-Dynamics (PHQMD) approach.
Here we discuss only those aspects of PHQMD which are necessary to understand the results,
while further details can be found in Ref.~\cite{Aichelin:2019tnk}. 

The PHQMD is a n-body microscopic transport approach for the description of the heavy-ion collisions 
including cluster production. The dynamically formed clusters are identified by the SACA and MST algorithms.
The PHQMD unites the collision integrals of the  Parton-Hadron-String Dynamics 
(PHSD) approach (in version 4.0) with density dependent 2-body potential between baryons,
similar to the Quantum Molecular Dynamics (QMD) approach. Baryons are described by Gaussian wave functions which propagate under the influence of mutual density dependent 2-body forces (and not due to a mean field). This approach respects 'actio' is equal to 'reactio' and therefore energy and momentum are strictly conserved. In the numerical approach the violation of the total energy conservation is found to be less than 0.5\%.

\subsection{The Collision integral}

The collision integral of PHQMD is adopted from the PHSD approach.
We remind that the Parton-Hadron-String Dynamics is a nonequilibrium microscopic transport approach 
\cite{Cassing:2008sv,Cassing:2008nn,Cassing:2009vt,Bratkovskaya:2011wp,Linnyk:2015rco}
 which describes the strongly interacting partonic and hadronic
medium in- and out-of equilibrium in terms of off-shell massive quasiparticles (quarks and gluons) 
and off-shell hadrons.
It is based on a solution of the Cassing-Juchem generalised off-shell transport equations for test particles \cite{Cassing:1999wx,Cassing:1999mh}, based on the  Kadanoff-Baym equations \cite{KadanoffBaym} in first-order gradient expansion \cite{Juchem:2004cs,Cassing:2008nn}. 

The hadronic part is based on the early development of the HSD transport approach \cite{Ehehalt:1996uq,Cassing:1999es}, which includes the baryon octet and decouplet, the ${0}^{-}$ and ${1}^{-}$ meson nonets and higher resonances. 
In PHSD the description of multi-particle production in elementary baryon-baryon ($BB$),
meson-baryon ($mB$) and meson-meson ($mm$) reactions is realized 
within the Lund model \cite{NilssonAlmqvist:1986rx}, in terms of the
event generators FRITIOF~7.02 \cite{NilssonAlmqvist:1986rx,Andersson:1992iq} and PYTHIA~6.4 \cite{Sjostrand:2006za}. 
We note that in PHSD the Lund event generators (FRITIOF~7.02 and PYTHIA~6.4)
are "tuned", i.e. adjusted, to get a better agreement with experimental data 
on elementary $p+p$ collisions, especially at intermediate energies (cf. Ref.~\cite{Kireyeu:2020wou}). 
It contains also chiral symmetry restoration via the Schwinger 
mechanism for the string decay \cite{Cassing:2015owa,Palmese:2016rtq} in a dense medium, 
as well as in-medium effects such as a collisional broadening of the vector meson spectra functions \cite{Bratkovskaya:2007jk} and the modification of strange degrees of freedom in line with many-body G-matrix calculations
\cite{Cassing:2003vz,Song:2020clw}. 
We mention also that the implementation of detailed balance on the level of  $2\leftrightarrow 3$ 
reactions is realized for the main channels of strangeness production/absorption 
by baryons ($B=N, \Delta, Y$) and pions \cite{Song:2020clw}, as well as for the multi-meson fusion reactions for the formation of $B +\bar B$  pairs \cite{Cassing:2001ds}.

 In PHSD  the partonic, i.e. the QGP phase, is based on the Dynamical Quasi-Particle Model (DQPM) \cite{Cassing:2007yg,Cassing:2007nb} which describes the properties of QCD (in equilibrium) in terms of resummed single-particle Green's functions. The gluons and quarks in PHSD are massive, strongly-interacting quasi-particles, described by a spectral function. 
 The widths and pole positions of the spectral functions are defined by the real and imaginary parts of the parton  self-energies and the effective, temperature dependent coupling strength in the DQPM is fixed by fitting respective lQCD results from Refs. \cite{Aoki:2009sc,Cheng:2007jq,Borsanyi:2015waa}.
 The QGP phase is then evolved using the off-shell transport equations with self-energies and cross sections from DQPM. When the fireball expands the probability of the partons to hadronize increases close to the phase boundary (crossover at all RHIC energies), the hadronization takes place using covariant transition rates. The resulting hadronic system is further-on governed by the off-shell HSD dynamics incorporating (optionally) self-energies for the hadronic degrees-of-freedom \cite{Cassing:2003vz}.

We note also that in PHQMD the QMD dynamics is applied only for the baryonic degrees of freedom
(cf. following Section III.C), while the propagation of mesonic and partonic degrees of freedom
follows the PHSD dynamics.

\subsection{Initialization of the nuclei}

In PHQMD a baryon $i$ is represented by the single-particle Wigner density,
which is given by  
\begin{equation} \label{fdefinition}
 f ({\bf r}_i, {\bf p}_i,{\bf r}_{i0},{\bf p}_{i0},t)
 = \frac{1}{\pi^3 \hbar^3 }
{\rm e}^{-\frac{2}{L} ({\bf r}_i - {\bf r}_{i0} (t) )^2 }
 {\rm e}^{-\frac{L}{2\hbar^2} ({\bf p}_i - {\bf p}_{i0} (t) )^2}.
\end{equation}
The Gaussian width $L$ is taken as  $L=8.66$~fm$^2$.
We will use the $\hbar = c = 1$  convention for further consideration. The corresponding single particle density is obtained by integrating the single-particle Wigner density over the
momentum of the baryon. The total one-body density is the sum of the densities of all baryons. The n-body Wigner density is the direct product of the one body densities. At the beam energies considered here antisymmetrization can be neglected at midrapidity.

For the details of the initialization of the nucleons we refer to
\cite{Aichelin:2019tnk}. The centroids of the Gaussian wave function, which represent the nucleons, are distributed initially such that they reproduce the experimental rms radius of the nuclei,  
the momentum distribution of the $P_{i0}$ is given by a local Thomas-Fermi model and the average binding energy is found to be close to the binding energy given by the Bethe-Weizs\"acker mass formula.

\subsection{QMD Propagation}
For the time evolution of the wave function
we use the Dirac-Frenkel-McLachlan approach \cite{raab:2000,broeck:1988}
which is based on the variation
\be
\delta \int_{t_1}^{t_2} dt <\psi(t)|i\frac{d}{dt}-H|\psi(t)> = 0
\ee
and has been developed in chemical physics. It has also been applied in nuclear physics for QMD like models
\cite{Feldmeier:1989st,Aichelin:1991xy,Ono:1992uy,Hartnack:1997ez}. This approach
conserves the correlations in the system and does not suppress fluctuations as mean-field
calculations do.  Since clusters are n-body correlations it is well suited to address the creation and time evolution of clusters.
With our assumption that the wave functions have a Gaussian form and that the width of the wave function is time independent, one obtains for the time evolution of the centroids of the Gaussian single particle Wigner density two equations which resemble the equation-of-motion of a classical particle with the phase space coordinates ${\bf r_{i0},p_{i0}}$ \cite{Aichelin:1991xy}.
The difference is that here the expectation value of the quantal Hamiltonian is used and not a classical Hamiltonian: 
\begin{equation}
\dot{r_{i0}}=\frac{\partial\langle H \rangle}{\partial p_{i0}} \qquad
\dot{p_{i0}}=-\frac{\partial \langle H \rangle}{\partial r_{i0}} \quad .
\label{prop}
\end{equation}
The Hamiltonian of the nucleus is the sum of the Hamiltonians of the nucleons, composed of kinetic and two-body potential energy.
\be
H = \sum_i H_i  = \sum_i  (T_i+\sum_{j\neq i}  V_{i,j}).
\ee
The interaction between the nucleons has two parts, a local Skyrme type interaction and a Coulomb interaction
\begin{eqnarray}
&&\phantom{a}\hspace*{-5mm}
V_{i,j}= V({\bf r}_i, {\bf r}_j,{\bf r}_{i0},{\bf r}_{j0},t)  = V_{\rm Skyrme}+ V_{\rm Coul}  \label{EP} \\
&& \phantom{a}\hspace*{-5mm}
=\frac{1}{2} t_1 \delta ({\bf r}_i - {\bf r}_j)  +  \frac{1}{\gamma+1}t_2 \delta ({\bf r}_i - {\bf r}_j)  \,     
    \rho^{\gamma-1}({\bf r}_i,{\bf r}_j, {\bf r}_{i0},{\bf r}_{j0},t) \nonumber \\  
&&\phantom{a}\hspace*{-5mm}
 +\frac{1}{2}  \frac{Z_i Z_j e^2}{|{\bf r}_i-{\bf r}_j|} , \nonumber
\end{eqnarray}
with the density $ \rho({\bf r}_i,{\bf r}_j,{\bf r}_{i0}{\bf,r}_{j0},t)$ defined as
\bea
& &\rho({\bf r}_i,{\bf r}_j,{\bf r}_{i0},{\bf r}_{j0},t) =\nonumber \\
&=&C\frac{1}{2}\Big[ \sum_{j,i\neq j} \Big(\frac{1}{\pi L}\Big)^{3/2}e^{-\frac{1}{L}({\bf r} _i-{\bf r}_j-{\bf r}_{i0}(t)+{\bf r}_{j0}(t))^2} \nonumber \\
&+&\sum_{i,i\neq j} \Big(\frac{1}{\pi L}\Big)^{3/2}e^{-\frac{1}{L}({\bf r}_i-{\bf r}_j-{\bf r}_{i0}(t)+{\bf r}_{j0}(t))^2}\Big].
\label{dens1}
\eea
$C$ is a correction factor which is discussed in \cite{Aichelin:2019tnk}.
For the Skyrme  potential we  use the approximative form (for a discussion we refer to \cite{Aichelin:2019tnk}) 
\begin{equation} \label{eosinf} \langle V_{Skyrme}({\bf r_{i0}},t) \rangle \,=\, \alpha
\left(\frac{\rho_{int} ({\bf r}_{i0},t)}{\rho_0}\right) +
        \beta \left(\frac{\rho_{int} ({\bf r}_{i0},t)}{\rho_0}\right)^{\gamma}.
\label{Upot}
\end{equation}
$\rho_{int}({\bf r_{i0}})$ is the sum of the Gaussian single particle densities of all baryons $j\ne i$ at ${\bf r_{i0}}$, but with twice the width $(L\to 2L)$. The expectation value  of the Coulomb interaction can also be calculated analytically.

The expectation value of the  Hamiltonian which enters \eq{prop} is finally given by
\begin{eqnarray}
\langle H \rangle &=& \langle T \rangle + \langle V \rangle
\label{ham} \\
&=& \sum_i \big(\sqrt{p_{i0}^2+m^2}-m\big) \\ \nonumber &+&\sum_{i}  \langle V_{Skyrme}({\bf r}_{i0},t)+V_{coul}({\bf r}_{i0},t)\rangle .
\nonumber 
\end{eqnarray}
In infinite matter the parameters $\alpha(t_1,t_2),\beta (t_1,t_2) $ and $ \gamma$ can be related to the nuclear equation-of-state (EOS). 
Two of the three parameters are determined by the condition that the energy per nucleon has 
a minimum of $\frac{E}{A}=-16$~MeV at $\rho_0$, the ground state density of nuclear matter. The third one is usually expressed by the compression modulus $K$
of nuclear matter,  the inverse of the compressibility $\chi = \frac{1}{V}\frac{d V}{d P}$, where $P$ is the pressure in the system of volume $V$. Here we employ $K = 380$~MeV. Such an EOS is usually called hard.

With increasing bombarding energies relativistic dynamics becomes more important. The relativistic formulation of molecular dynamics has been developed in Ref.~\cite{Marty:2012vs}. However, the numerical realization of this method for realistic heavy-ion calculations is not feasible with present day computers.
Therefore, in order to extend our approach for relativistic energies, we introduce
the modified single-particle Wigner density $\tilde f$ of the nucleon $i$
\bea
&& \tilde f ({\bf r}_i, {\bf p}_i,{\bf r}_{i0},{\bf p}_{i0},t) =  \label{fGam} \\
&& =\frac{1}{\pi^3} {\rm e}^{-\frac{2}{L} ({\bf r}_{i}^T(t) - {\bf r}_{i0}^T (t) )^2} 
  {\rm e}^{-\frac{2\gamma_{cm}^2}{L} ({\bf r}_{i}^L(t) - {\bf r}_{i0}^L (t) )^2}.  \nonumber \\
&& \times {\rm e}^{-\frac{L}{2} ({\bf p}_{i}^T(t) - {\bf p}_{i0}^T} (t) )^2 
  {\rm e}^{-\frac{L}{2\gamma_{cm}^2} ({\bf p}_{i}^L(t) - {\bf p}_{i0}^L (t) )^2},  \nonumber 
\eea
which accounts for the Lorentz contraction of the nucleus in the beam $z$-direction
in coordinate and momentum space by the inclusion of $\gamma_{cm} =1/\sqrt{1-v_{cm}^2}$, where $v_{cm}$ is the velocity of the
bombarding nucleon in the initial $NN$ center-of-mass system.   
Accordingly, the interaction density modifies as 
\bea
 \tilde \rho_{int} ({\bf r}_{i0},t) 
 &\to & C  \sum_j \Big(\frac{1}{\pi L }\Big)^{3/2} \gamma_{cm} \
 {\rm e}^{-\frac{1}{L} ({\bf r}_{i0}^T(t) - {\bf r}_{j0}^T (t) )^2} \nonumber \\ 
 &&\times  {\rm e}^{-\frac{\gamma_{cm}^2}{L} ({\bf r}_{i0}^L(t) - {\bf r}_{j0}^L (t) )^2}.  
 \label{densGam}
\eea
With these modifications we obtain
\begin{equation}
\langle \tilde H \rangle =
\sum_i (\sqrt{p_{i0}^2+m^2}-m)  +\sum_{i}  \langle \tilde V_{Skyrme}({\bf r}_{i0},t)\rangle .
\end{equation}
with
\begin{equation} \label{eosinfb} \langle \tilde V_{Skyrme}({\bf r}_{i0},t) \rangle \,=\, \alpha
\left(\frac{\tilde \rho_{int} ({\bf r}_{i0},t)}{\rho_0}\right) +
        \beta \left(\frac{\tilde \rho_{int} ({\bf r}_{i0},t)}{\rho_0}\right)^{\gamma}.
\label{Upot1}
\end{equation}
This potential enters the time evolution equations (\ref{prop}). 

With increasing bombarding energy the dynamics of midrapidity particles starts to be dominated by collisions rather than by the potential interaction between two collisions and if the local energy density $\varepsilon$ is larger than the critical energy density of $\varepsilon_C \simeq 0.5$~GeV/fm$^3$ the transition from hadronic to partonic degrees-of-freedom occurs \cite{Moreau:2019vhw}.
Potential interactions among baryons are therefore relevant mainly for 'corona' particles (or spectators), as well as for the interacting baryons in the fireball after hadronization. Shortly after hadronization hadronic collisions are still frequent and the momentum transfer due to collisions is large in comparison to the momentum transfer due to the potential interactions between the collisions. After the collision have ceased, because the mean free path became too long, the nucleons can still interact by potential interactions.
Because collisions destroy weakly bound small clusters, this is also the moment when clusters can be formed. In this expanding region the inverse slope parameters of the transverse energy spectra of the
baryons are of the order of 100~MeV and therefore for all baryons we are in an
approximately non-relativistic regime.

\subsection{Cluster identification}

The formation of clusters in PHQMD is a consequence of nucleon-nucleon interactions acting during the time 
evolution of the heavy-ion collisions. The interplay between hadronic collisions and the potential interactions, which are attractive at baryonic densities around or below normal nuclear matter density, leads to the formation of clusters. We call this {\it dynamical} cluster formation because the formation of clusters does not happen at a given time, but is a continuous process due to the interactions among the baryons. In comparison to other approaches PHQMD does not require to employ an additional model, like a coalescence model or a statistical model, on top of the baryon dynamics to determine clusters. Coalescence models may use some quantal information about the clusters which are not contained in our approach 
like the average distance between nucleons, however they neglect that - due to energy conservation - the formation of a bound state needs another body in the exit channel and provide  no information about the formation time.

As explained in Ref.~\cite{Aichelin:2019tnk},
the cluster identification or recognition is realized in PHQMD by the  Minimum Spanning Tree (MST) procedure \cite{Aichelin:1991xy} or by the Simulated Annealing Clusterization Algorithm (SACA) \cite{Puri:1996qv,Puri:1998te}.
In this study we employ the MST procedure to identify clusters. 

In the MST algorithm only the coordinate space information is used to define clusters: Two nucleons are considered as part of a cluster if their distance is less than $r_0 = 4$~fm in the cluster rest system obtained by a Lorentz transformation from the computational frame. Nucleons which are more distant do not feel anymore an attractive interaction. 

Nucleons with a large relative momentum do not stay together for a long time. Consequently, additional cuts in momentum space change the cluster distribution only little \cite{Kireyeu:2021igi}. The MST procedure has been recently applied to study the cluster production in different transport approaches \cite{Kireyeu:2021igi}. This study showed that in QMD like approaches, where the few-body correlations are kept, more light clusters at midrapidity are produced than in mean-field or cascade models which give both qualitatively similar results. 
\begin{figure}
        \includegraphics[scale=0.4]{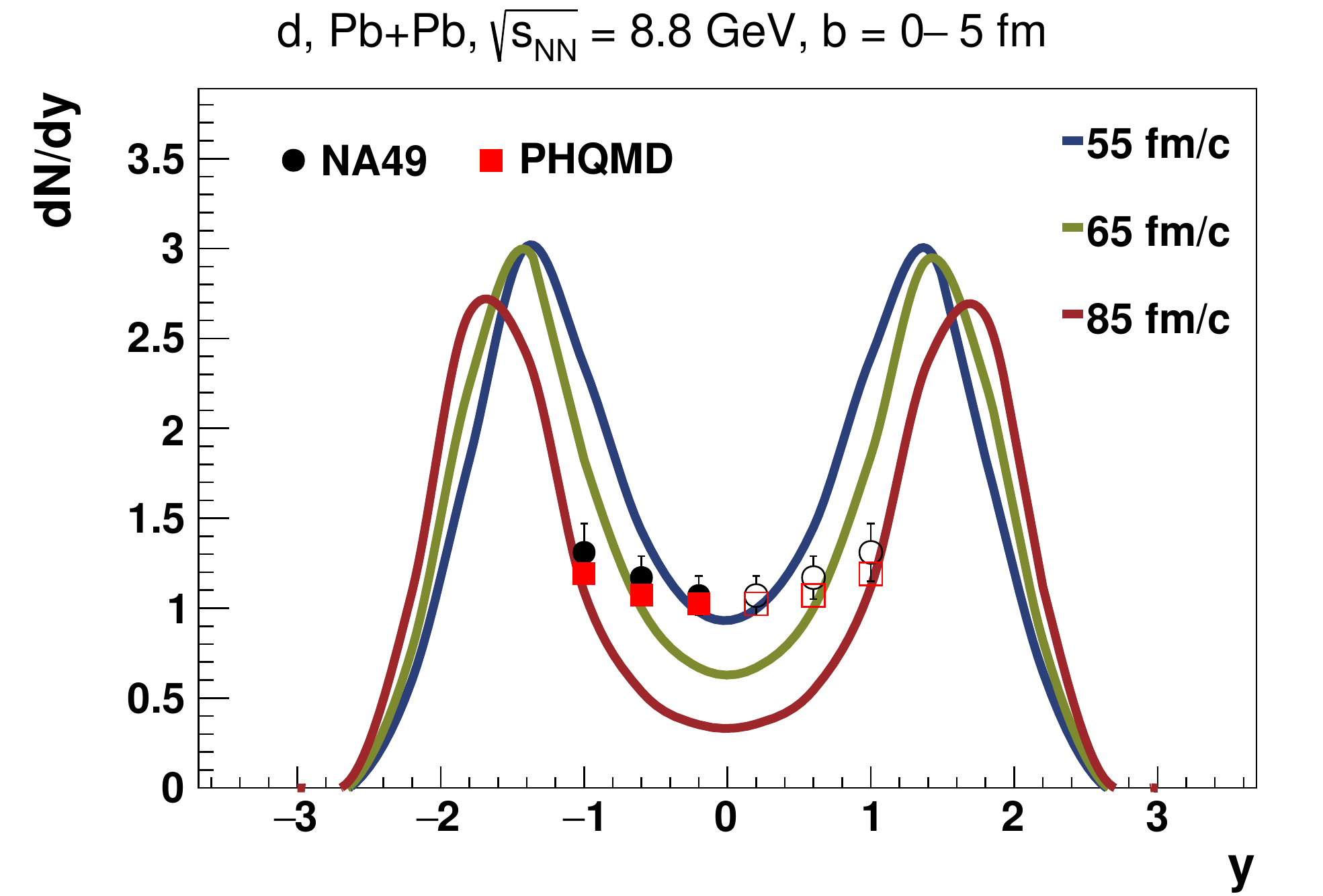}
\caption{\label{fig:8.8_dndy_cosh_time}
The rapidity distribution of deuterons for central Pb+Pb collisions at $E_{kin}= 40$~$A$GeV.  The full dots indicate the experimental data from the NA49 Collaboration \cite{Anticic:2016ckv}, the red squares show the PHQMD results taken at the physical time $t= t_0 \cosh(y)$ for $t_0 = 53$~fm/$c$. The solid lines represent PHQMD results at the rapidity independent times $t_0 = 55$~fm/$c$ (blue line), $t_0 = 65$~fm/$c$ (green line) and $t_0 = 85$~fm/$c$ (red line) measured in the center-of-mass of the heavy-ion reaction (computational frame).}
\end{figure}
The cluster recognition by MST does not influence the time evolution of the heavy-ion reaction as the underlying PHQMD propagates (in the version used here) only baryons, not clusters. 

Because each nucleon
has a unique identification number in the PHQMD code, the PHQMD approach allows for studying the time evolution of clusters 
by applying MST at different times during the simulation of a heavy-ion reaction. Starting at the end of the simulation we can go back in time through the recorded history of the clusters and their constituents. We can identify the point in time at which a cluster has been identified for the first time and investigate the environment at the creation point.

Semi-classical models (such as QMD) cannot project the n-body density onto the quantum ground state of a cluster. For very late times the differences between a fully quantal and our semi-classical approach may therefore influence the cluster distribution, because in the ground state of a cluster - being a quantum system of fermions - all states up to the Fermi energy
are occupied by nucleons and thus no
energy or momentum transfer between the nucleons is possible. This is not the case
in semi-classical approaches where one of the cluster nucleons may gain sufficient energy
to overcome the binding energy.
 
 Also, the necessity to neglect the time component of the Lorentz transformation when calculating the interaction density, \eq{densGam}, which enters the time evolution equation, may create energy uncertainties of the order of the binding energy (which is very small as compared to the typical particle energies).

For both reasons clusters which are formed may disintegrate again in this semi-classical approach and we have therefore to fix a cluster freeze-out time at which we consider the cluster production as terminated. This time is large as compared to the time of the last collision which the cluster nucleons suffer and slightly different for different clusters. The cluster freeze-out time is defined in the cluster rest system. Due to time dilatation it depends in the center-of-mass frame of the heavy-ion reaction, the frame in which the time evolution of the system is calculated, on the center-of mass rapidity of the cluster rest system: $t=t_0\cdot \cosh{y_{cm}}$, where $t_0 $ is the cluster freeze-out time at midrapidity. We call $t$ "physical time" because it marks identical times in the rest systems.
The time $t_0 $ is determined such that we reproduce the total experimental multiplicity of the clusters at  mid-rapidity. We have verified that only the multiplicity, but neither the form of the rapidity distribution nor that of the transverse momentum distribution or of the $B_2$ ratios, are affected by this choice. 

In order to illustrate the procedure discussed above, we present in Fig.~\ref{fig:8.8_dndy_cosh_time} 
the rapidity distribution $\frac{dN}{dy}$ of deuterons for central Pb+Pb collisions at $\sqrt{s_{NN}}=8.8$~GeV. The full lines show the rapidity distribution at three different (rapidity independent) computational times ($t_0 = 55$~fm/$c$ (blue), $t_0 = 65$~fm/$c$ (green) and $t_0 = 85$~fm/$c$ (red)). We display as well as red squares the rapidity distribution of deuterons at the physical time $t=t_0\, \cosh{y_{cm}}$ with $t_0 = 53$~fm/$c$, which provides a good description of the experimental data of the NA49 Collaboration \cite{Anticic:2016ckv}.

\section{The PHQMD results for cluster production in heavy-ion collisions}

\subsection{Rapidity and $p_T$ spectra of light clusters for Au+Pb at $E_{kin} = 10.6$~$A$GeV}

\begin{figure}
\centering
\includegraphics[width=7.0cm]{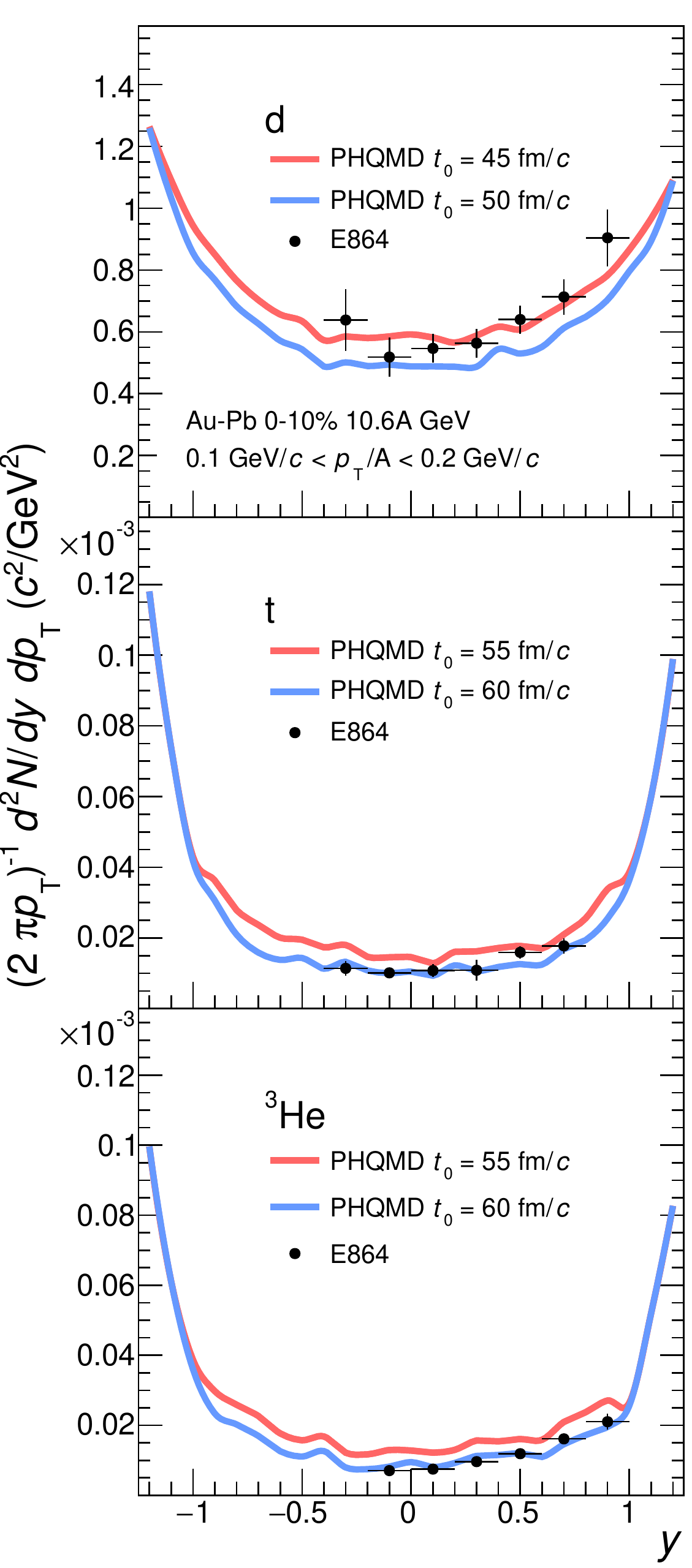}
\caption{\label{fig:1} The rapidity distribution of deuterons (top), tritons (middle) and ${}^3$He (bottom) in  central Au+Pb collisions at beam energy $ E_{kin} = 10.6$~$A$GeV. 
The dots indicate the experimental data from the E864 Collaboration \cite{Armstrong:2000gz}. The PHQMD results are taken at the physical time $t= t_0 \cosh(y)$ for $t_0 = 45$~(55)~fm/$c$ (red lines) and 50~(60)~fm/$c$ (blue lines) for d (${}^3$He, t). 
The PHQMD calculations are acceptance corrected. }  
\end{figure}

We start out our investigation with the cluster production in Au+Pb collisions at a beam energy of $E_{kin} = 10.6$~$A$GeV
which has been measured by the E864 Collaboration at the AGS accelerator in Brookhaven \cite{Armstrong:2000gz}. 
In Fig.~\ref{fig:1} we display the PHQMD results for the rapidity distributions of deuterons (top), tritons (middle) and $^3$He (bottom)  for the 10\% most central collisions. As discussed above, in our approach the number of clusters decreases as a function of time due to instabilities of our semi-classical clusters. 
Therefore, we have to determine a time at which we identify the clusters. Consequently, the rapidity and transverse momentum distribution is given in our approach up to an overall factor. The time at which we identify the clusters is given in the figures and is around 50~fm/$c$ for deuterons and around 60~fm/$c$ for tritium and ${}^3$He. 
Fig.~\ref{fig:1} shows that the rapidity distributions of light clusters are quite well described, especially the flatness of the distributions for the larger clusters and the increase with rapidity for the deuterons. 

\begin{figure}
\centering
\includegraphics[width=8.6cm]{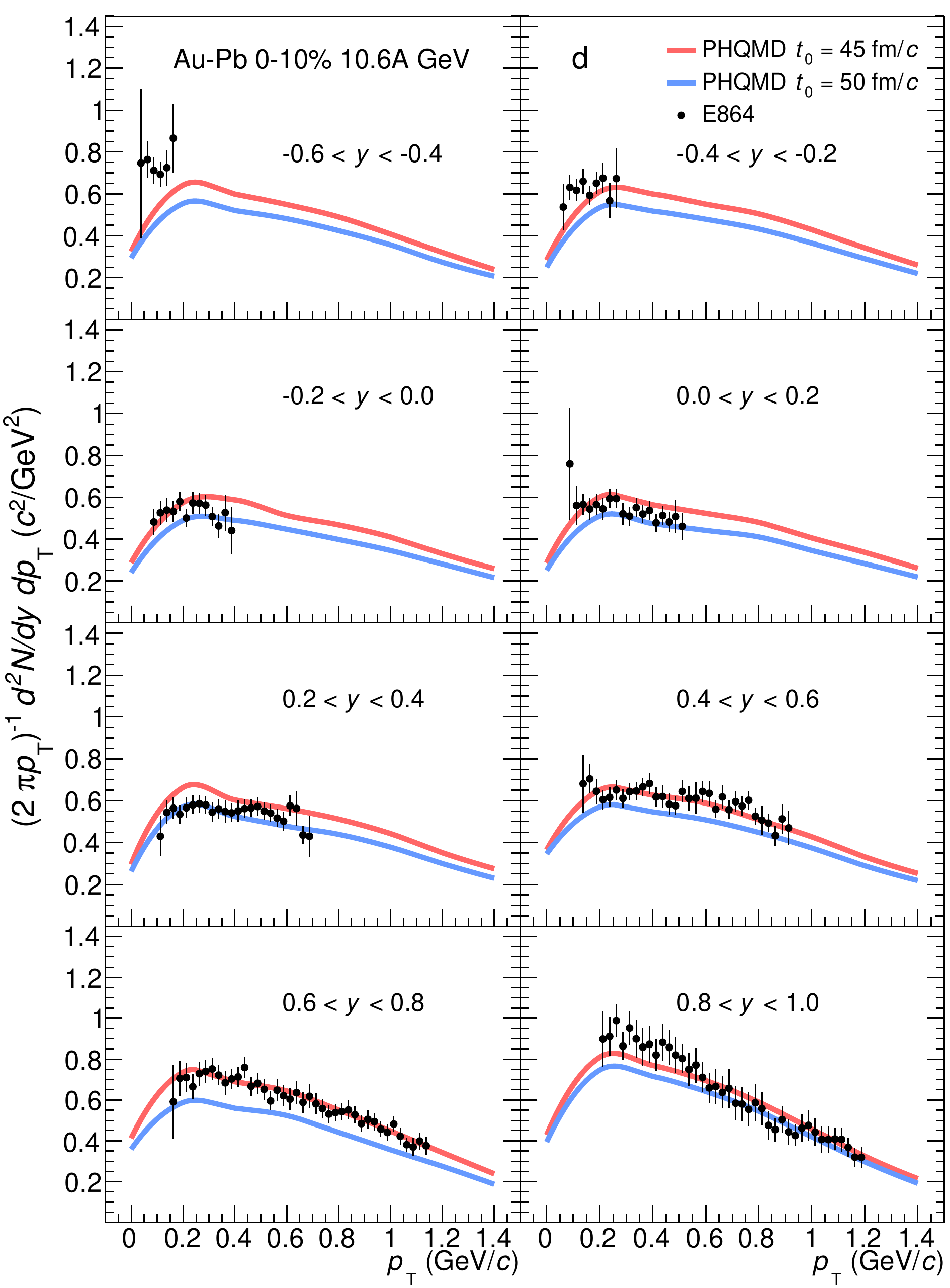}
\caption{\label{fig:2} The transverse momentum distribution of deuterons in central Au+Pb collisions at beam energy 
$ E_{kin} = 10.6$~$A$GeV for different rapidity intervals (as indicated in the legends).
The dots indicate the experimental data from the E864 Collaboration \cite{Armstrong:2000gz}. The PHQMD results are taken at the physical time $t= t_0 \cosh(y)$ for $t_0 = 45$~fm/$c$ (red lines) and 50~fm/$c$ (blue lines). }
\end{figure}

The transverse momentum distribution of the deuterons for the different rapidity intervals is displayed in Fig.~\ref{fig:2}. For most of the rapidity intervals the transverse momentum distribution is quite well reproduced. Only in the most negative and the most positive rapidity bin the calculations are outside of the error bars. 
%

\begin{figure}
\centering
\includegraphics[width=8.6cm]{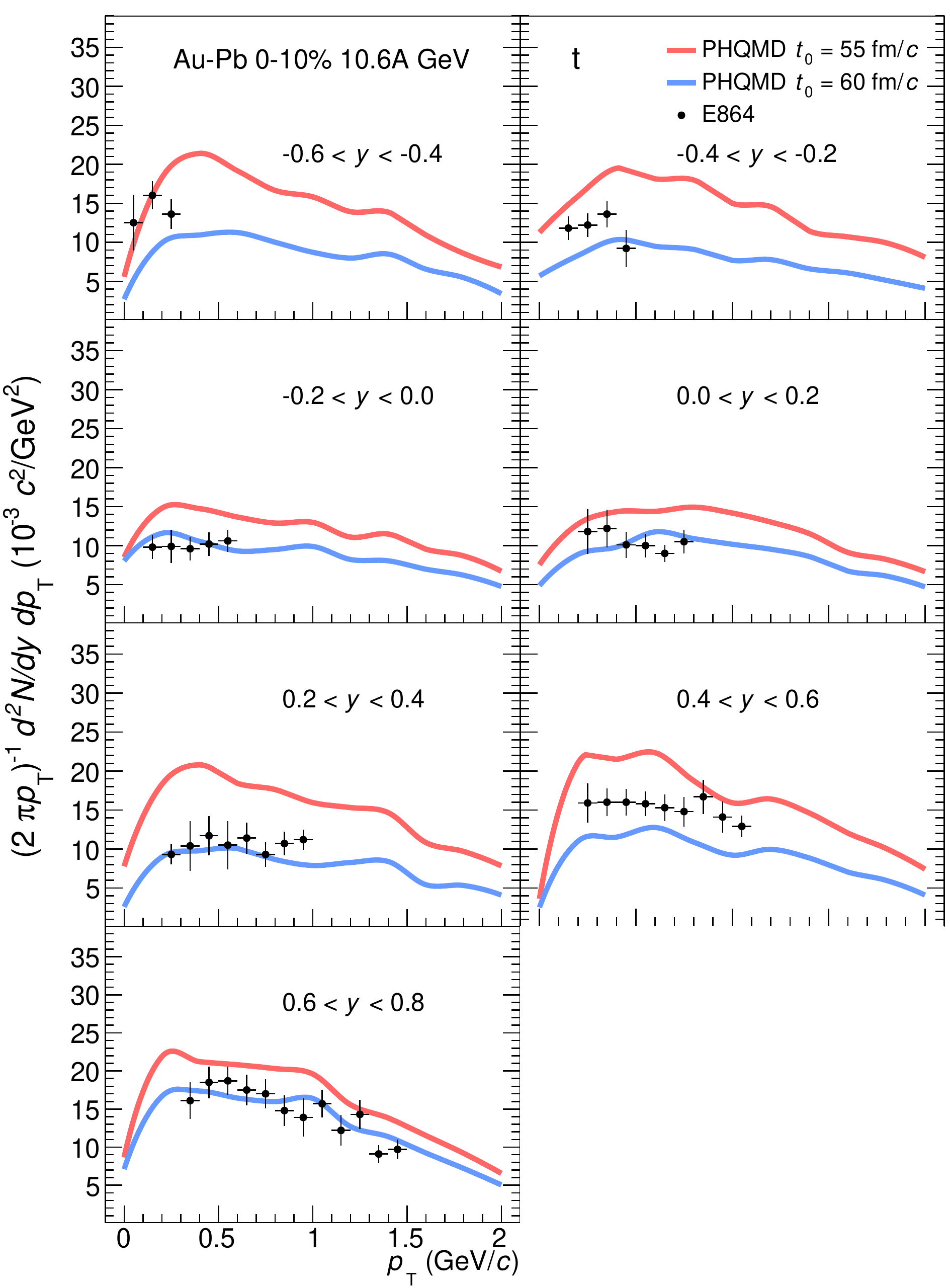}
\caption{\label{fig:3} 
The transverse momentum distribution of tritons in central Au+Pb collisions at beam energy $E_{kin} = 10.6$~$A$GeV for different rapidity intervals (as indicated in the legends).
The dots indicate the experimental data from the E864 Collaboration \cite{Armstrong:2000gz}. The PHQMD results are taken at the physical time $t= t_0 \cosh(y)$ for $t_0 = 55$~fm/$c$ (red lines) and 60~fm/$c$ (blue lines). }
\end{figure}

\begin{figure}
\centering
\includegraphics[width=8.6cm]{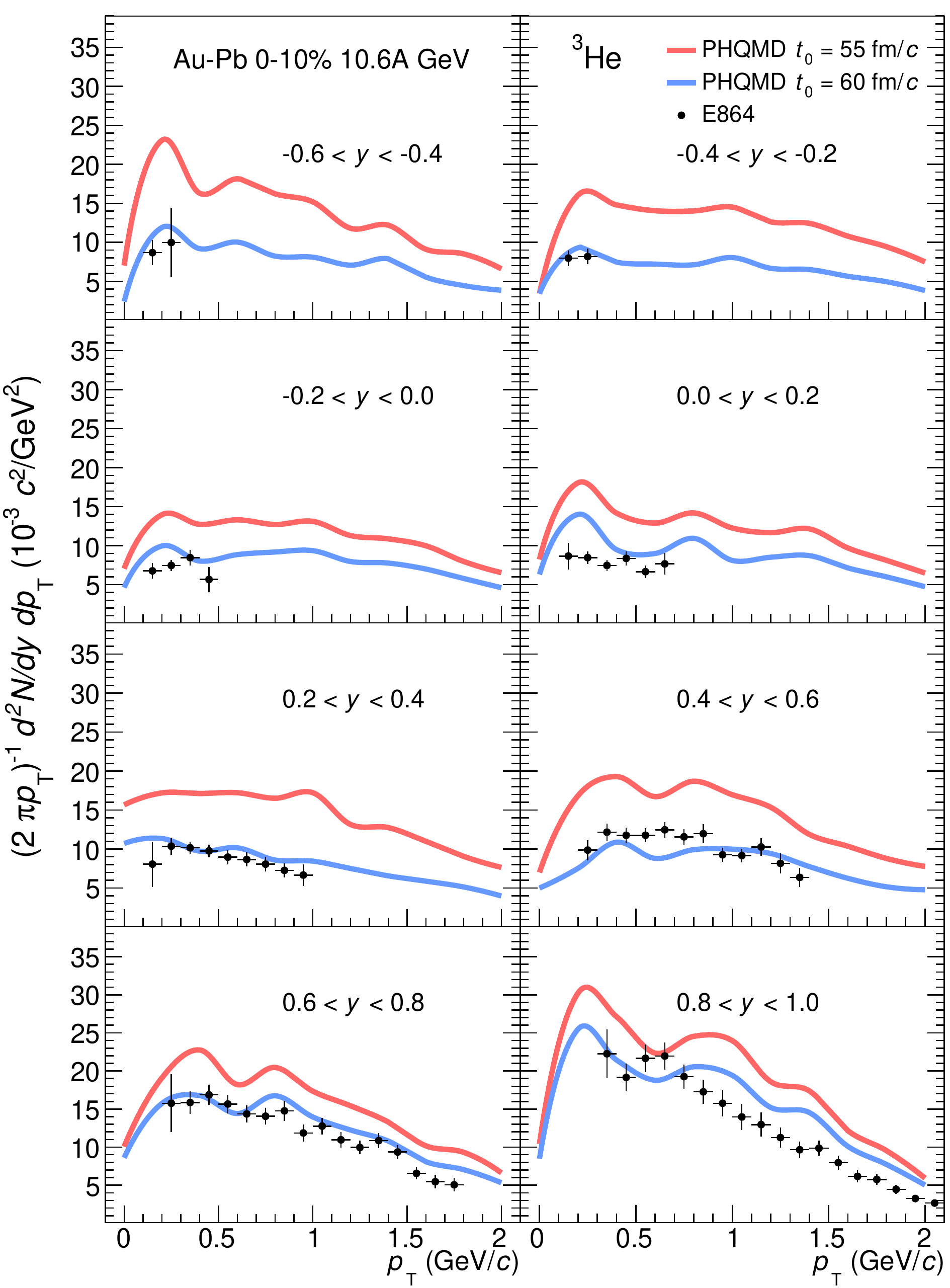}
\caption{\label{fig:4} 
The transverse momentum distribution of ${}^3$He in central Au+Pb collisions at beam energy $E_{kin} = 10.6$~$A$GeV for different rapidity intervals (as indicated in the legends).
The dots indicate the experimental data from the E864 Collaboration \cite{Armstrong:2000gz}. The PHQMD results are taken at the physical time $t= t_0 \cosh(y)$ for $t_0 = 55$~fm/$c$ (red lines) and 60~fm/$c$ (blue lines). }
\end{figure}

Figures~\ref{fig:3} and \ref{fig:4} present the $p_T$ distributions of tritons and ${}^3$He in the 10\% most central Au+Pb events. The slopes of the distributions are rather different compared to that of the deuterons.
Also for the $A=3$ cluster the $p_T$ spectra are quite well reproduced.

In order to compare the $p_T$ spectra of clusters with those of free nucleons one has introduced the covariant coalescence function, $B_A$:
\be
B_A = \frac{E\frac{d^3N_A}{dP^3}}{\Big(E\frac{d^3N_{neutrons}}{dp^3}\Big)^N \Big(E\frac{d^3N_{protons}}{dp^3}\Big)^Z}
\ee
when a cluster with baryon number $A$ and momentum $P = p A$ is formed out of $Z$ protons and $N$ neutrons.
Assuming that the proton and neutron momentum distributions are identical, $B_A$ can be reduced for the Au+Pb reaction to \cite{Armstrong:2000gz} 
\be
B_A = \frac{E\frac{d^3N_A}{dP^3}}{1.19^N \Big(E\frac{d^3N_{proton}}{dp^3}\Big)^A}.
\label{eq:BA}
\ee
If $B_2$ is independent of $p$, the result is compatible with the assumption that the probability that two nucleons, which carry half of the momentum of the deuteron, form a deuteron is independent of the momentum of the nucleons. 

\begin{figure}
\centering
\includegraphics[width=0.45\textwidth]{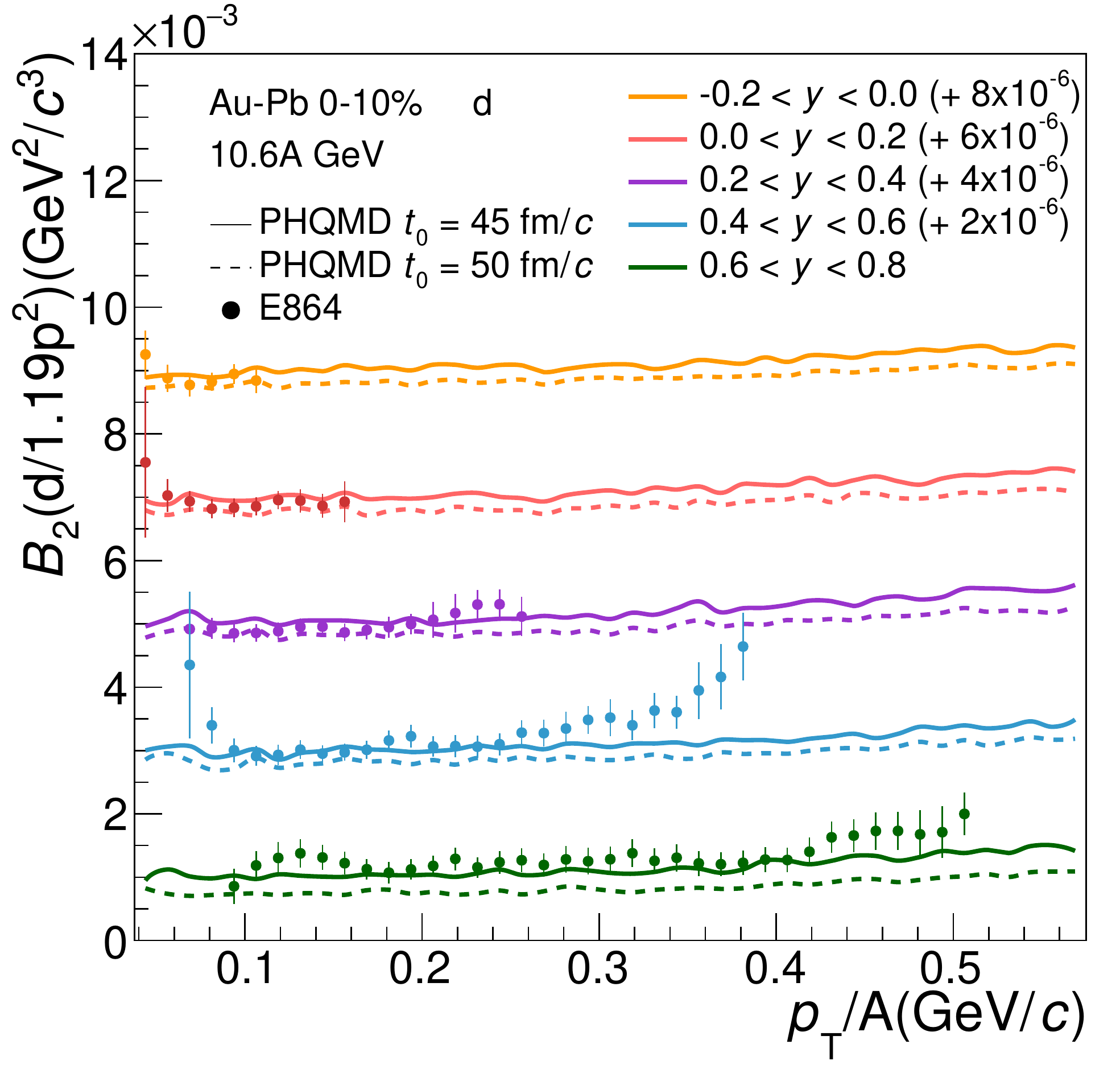}
\includegraphics[width=0.45\textwidth]{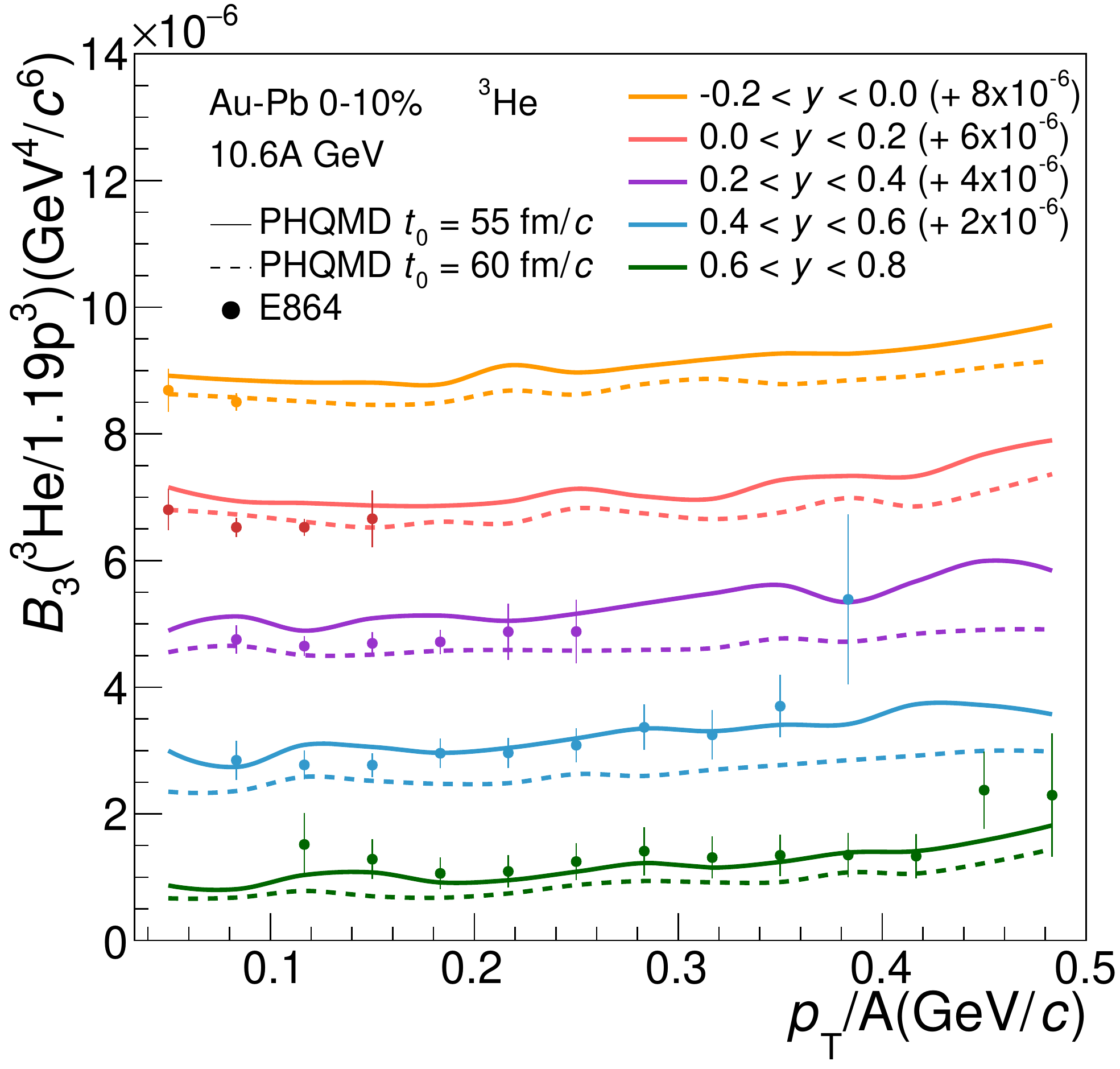}
\caption{\label{fig:10.6_BA} 
Coalescence function $B_2$ of deuterons (top) and ${}^3$He (bottom) in central Au+Pb collisions at beam energy $E_{kin} = 10.6$~$A$GeV, with an assumed neutron to proton ratio of 1.19, shown as function of transverse momentum in several rapidity intervals (as indicated in the legend).
The filled dots indicate the experimental data from the E864 Collaboration \cite{Armstrong:2000gz}. The PHQMD results are taken at the physical time $t= t_0 \cosh(y)$ for $t_0 = 45$~fm/$c$ (full lines) and 50~fm/$c$ (dashed lines) for deuterons and at $t_0 = 55$~fm/$c$ and 60~fm/$c$ for ${}^3$He. }
\end{figure}

In Fig.~\ref{fig:10.6_BA} we display $B_2$ (top) and $B_3$ (bottom). We compare for different rapidity intervals the PHQMD results for Au+Pb (dashed and full lines) with the $E_{kin} = 10.6$~$A$GeV Au+Pb data from the E864 Collaboration \cite{Armstrong:2000gz} (full points). The PHQMD results are taken at the physical time $t= t_0 \cosh(y)$ for $t_0 = 45$~fm/$c$ (solid lines) and 50~fm/$c$ (dashed lines) for deuterons and at $t_0 = 55$~fm/$c$ and 60~fm/$c$ for ${}^3$He.

We see that neither the experimental data nor the calculations show a remarkable structure for $B_2$ as a function of $p_T$. There is, however, a tendency that $B_2$ increases slightly with momentum. The calculations follow this trend with the exception of the experimentally observed strong increase at large $p_T$ in the interval $0.4\le y\le0.6$. 

This increase with $p_T$ is also visible for $B_3$. Also here the calculations follow this trend which means that high energetic nucleons have a higher chance to end up in a ${}^3$He than those with a lower momentum.

\begin{figure}
\centering
\includegraphics[width=0.45\textwidth]{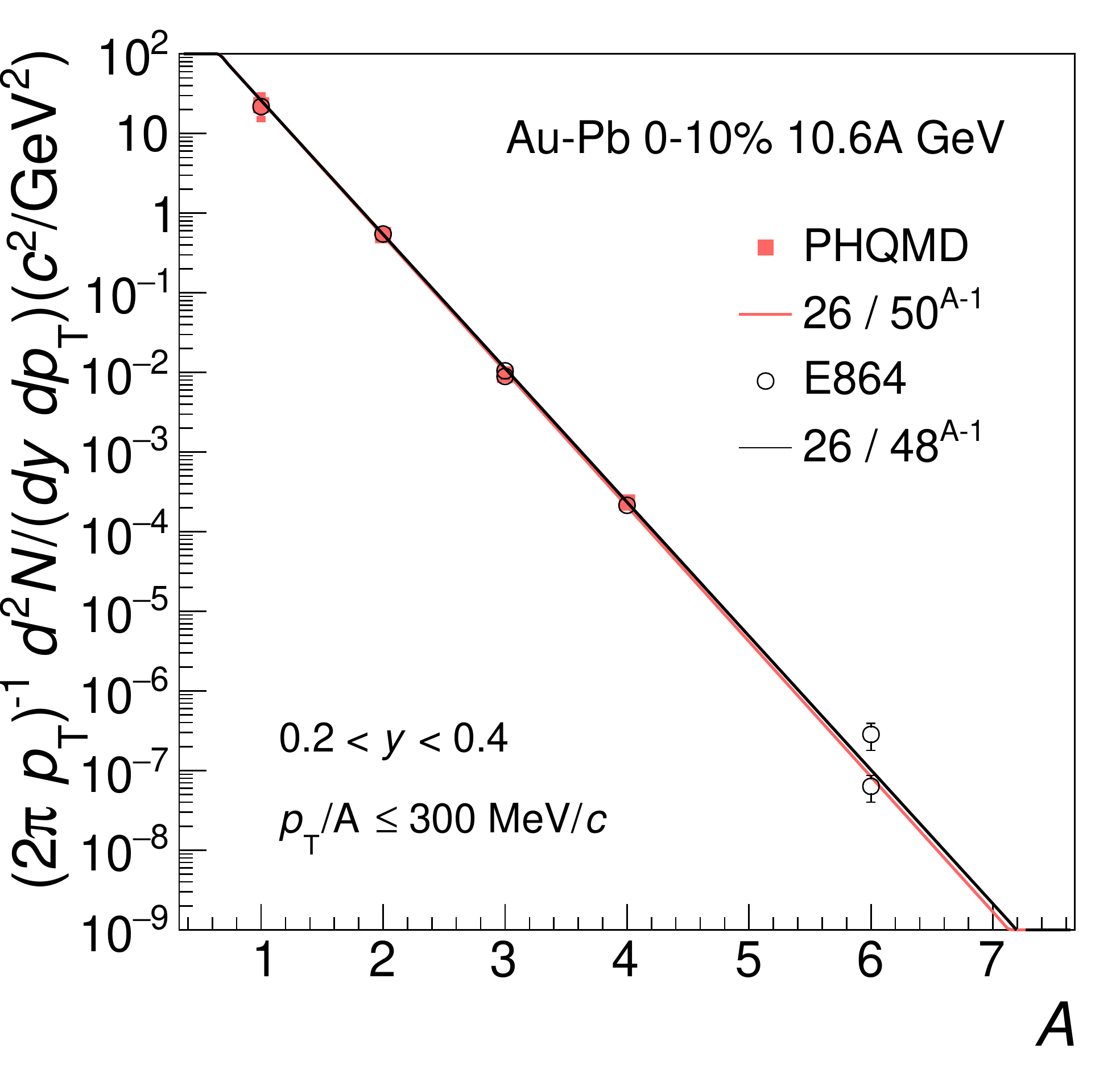}
\caption{\label{fig:10.6_penalty} 
Mass dependence of invariant yields of light nuclei in central Au+Pb collisions at beam energy $E_{kin} = 10.6$~$A$GeV in the small kinetic region of $0.2<y<0.4$ and $p_T / A < 0.3$~GeV$/c$. The open circles indicate the experimental data from the E864 Collaboration \cite{Armstrong:2000gz} (protons are corrected for hyperon feed-down). The PHQMD results (filled squares) are taken at the physical time $t= t_0 \cosh(y)$ with $t_0 = 50$~fm/$c$ for deuterons,  $t_0 = 60$~fm/$c$ for tritons as well as for ${}^3$He and $t_0 = 70$~fm/$c$ for ${}^4$He. Experimental and PHQMD data are fitted with an exponential function. For the fit of the PHQMD data (red line), the constant prefactor is fixed to 26 for a better comparison of the penalty factor with the one from experiment (black line). The penalty factor for each additional nucleon is approximately 48 for the experimental data and approximately 50 for the PHQMD data.}
\end{figure}

Another way to compare protons and cluster is to calculate the penalty factor for cluster production. This factor describes
the reduction of the absolute yield in a $(p_T, y)$ bin as a function of the cluster size.
This penalty factor is displayed in Fig.~\ref{fig:10.6_penalty} and compared with experimental data. This figure shows the mass dependence of invariant yields of light nuclei in central Au+Pb collisions at beam energy $E_{kin} = 10.6$~$A$GeV in the small kinetic region of $0.2<y<0.4$ and $p_T / A < 0.3 $~GeV$/c$. The open circles indicate the experimental data from the E864 Collaboration \cite{Armstrong:2000gz} (protons are corrected for hyperon feed-down). The PHQMD results (filled squares) are taken at the physical time $t= t_0 \cosh(y)$ with $t_0 = 50$~fm/$c$ for deuterons,  $t_0 = 60$~fm/$c$ for tritons as well as for ${}^3$He and $t_0 = 70$~fm/$c$ for ${}^4$He. Experimental and PHQMD data are fitted with an exponential function. For the fit of the PHQMD data (red line), the constant prefactor is fixed at 26 for a better comparison of the penalty factor with the one from experiment (black line). The penalty factor for each additional nucleon is approximately 48 for the experimental data and approximately 50 for the PHQMD data.

Fig.~\ref{fig:10.6_penalty} shows that our approach is able to quantitatively describe clusters up to a mass $A=4$, the limit we can reach with the presently available computers. To predict the yield of $A=6$ clusters one needs about 1000 times more CPU time than we have presently available.

Thus, we can conclude that PHQMD, once the cluster freeze-out time is fixed, describes quite well all observables which have been measured in central Au+Pb collisions at $E_{kin} = 10.6$~$A$GeV. These observables include
rapidity and $p_T$ distribution at different rapidities for clusters of different size, the ratio of the $p_T$ spectra of clusters and free nucleons as well as the penalty factor for larger clusters. Having established that the model works for the lower end of the beam energy interval of interest we proceed now to the upper end.

\subsection{Light clusters produced in Pb+Pb collisions at $\sqrt{s_{NN}} = 8.8$~GeV}

\begin{figure}
        \includegraphics[scale=0.4]{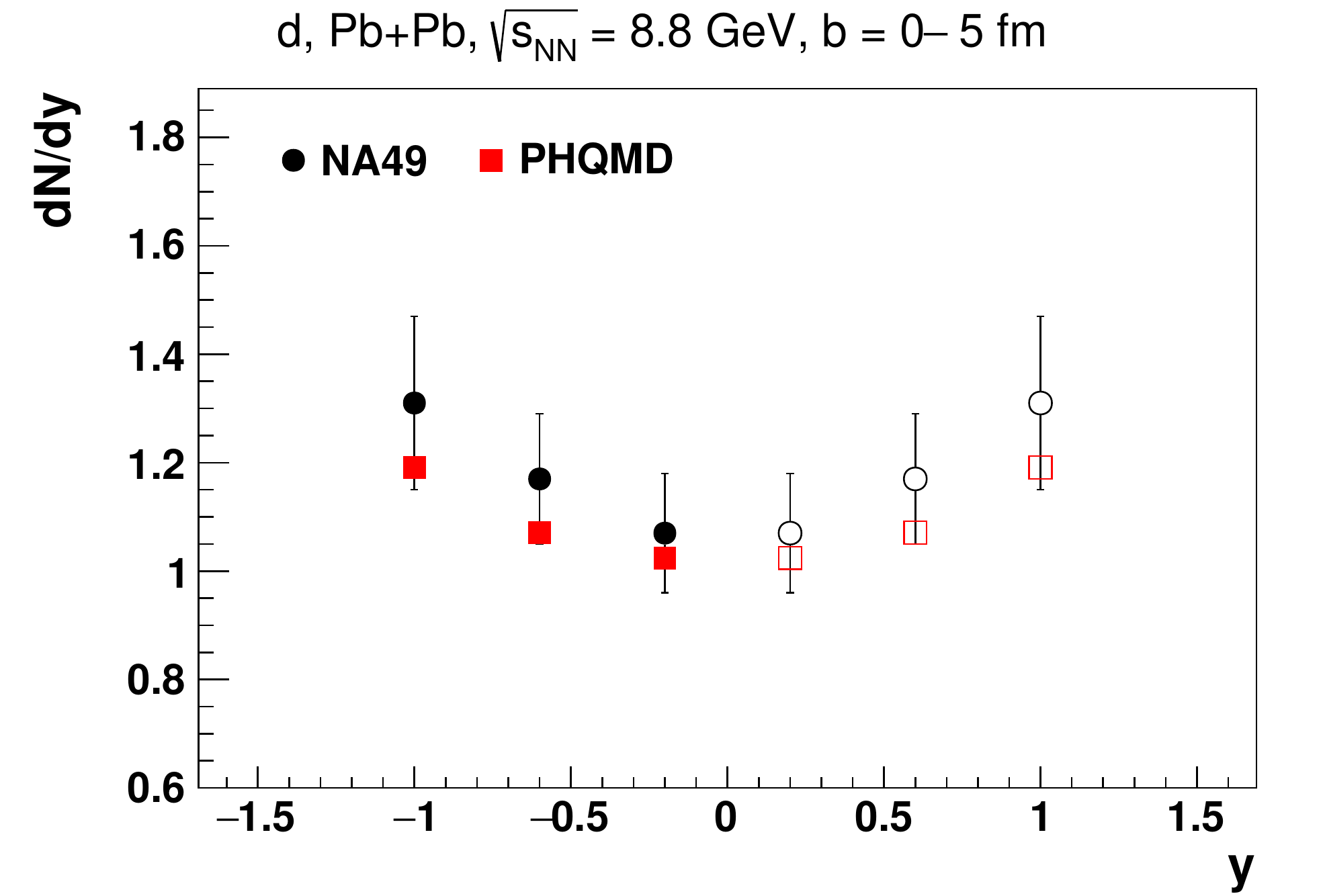} 
\caption{\label{fig:5} The rapidity distribution of deuterons for central Pb+Pb collisions at beam energy $E_{kin} = 40$~$A$GeV.  The dots indicate the experimental data from the NA49 Collaboration \cite{Anticic:2016ckv}, the red squares show the PHQMD results taken at the physical time $t= t_0 \cosh(y)$ for $t_0 = 53$~fm/$c$. }
\end{figure}

\begin{figure}
        \includegraphics[scale=0.3]{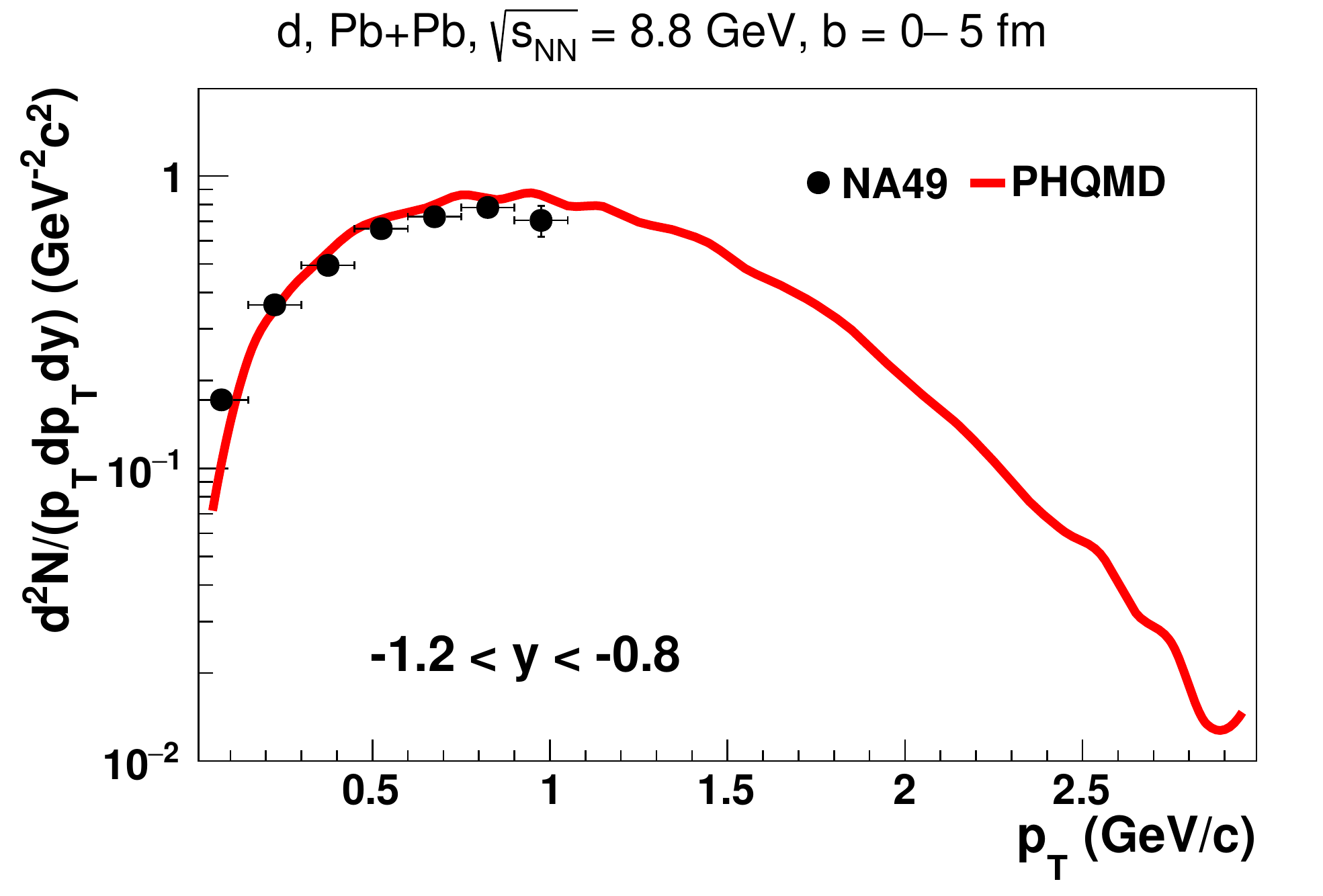} \\
        \includegraphics[scale=0.3]{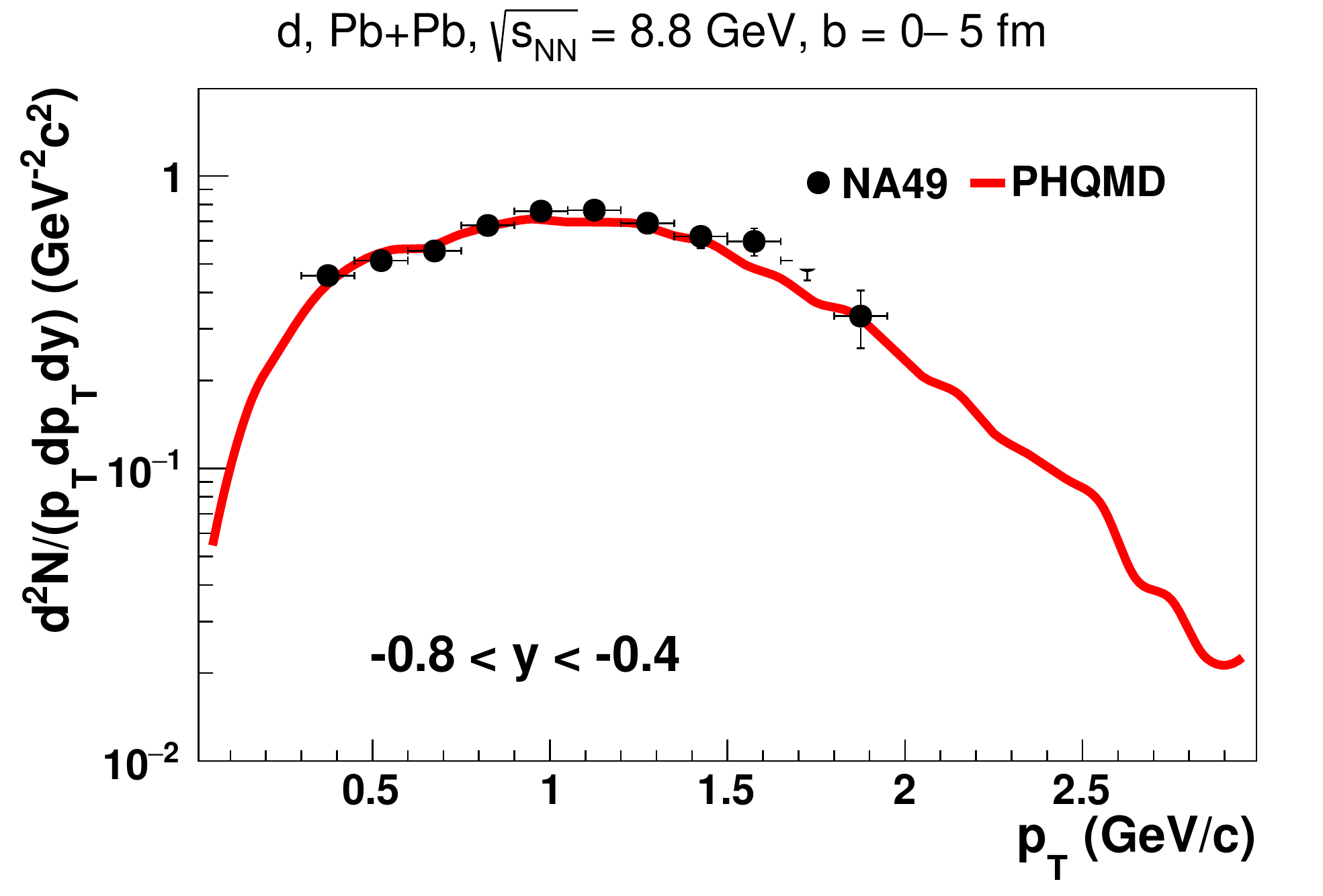} \\
        \includegraphics[scale=0.3]{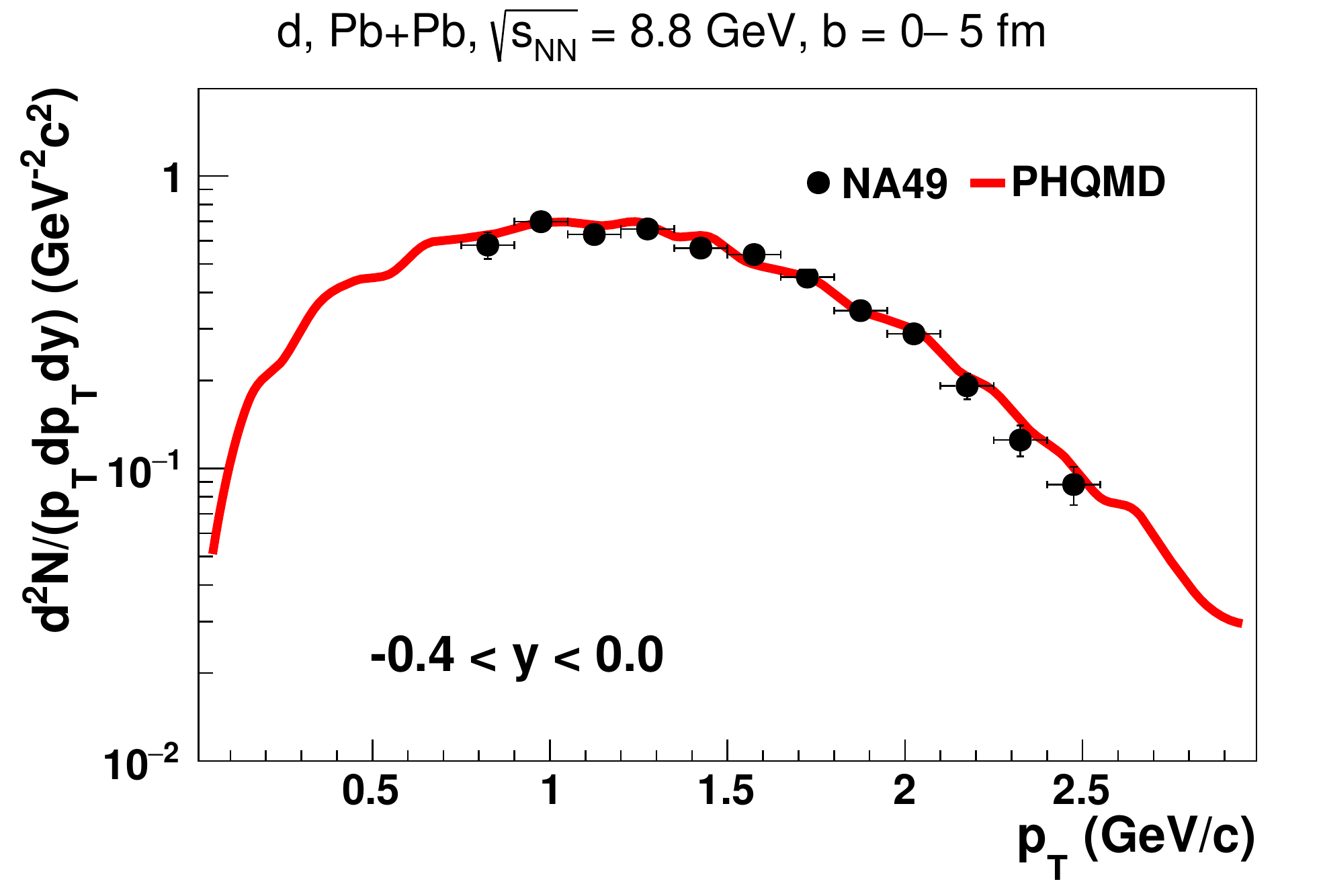} 
\caption{\label{fig:deut_pt_8.8} The transverse momentum spectra of deuterons for Pb+Pb central collisions at $\sqrt{s_{NN}}=8.8$~GeV for different rapidity intervals: $-1.2<y<0.8$ (upper),  $-0.8<y<0.4$ (middle),  $-0.4<y<0.0$ (bottom). The dots indicate the experimental data from the NA49 Collaboration \cite{Anticic:2016ckv}, the red lines show the PHQMD results.}  
\end{figure}

Midrapidity clusters have also been measured in heavy-ion collisions at the SPS at CERN. For the comparison with PHQMD results we select the Pb+Pb data at $E_{kin} = 40$~$A$GeV, an energy which is close to the lower limit of the RHIC beam energy scan ($\sqrt{s_{NN}}=7.7$~GeV), where in near future new data can be expected. At this energy the NA49 Collaboration has measured the rapidity and transverse momentum distribution of deuterons and ${}^3$He for the 7\% most central collisions. For deuterons the rapidity and transverse momentum distribution are displayed in Figs.~\ref{fig:5} and \ref{fig:deut_pt_8.8}. Here we identify the clusters at $t_0 = 53$~fm/$c$ and 67~fm/$c$, times which are very close to that employed at the lower energy. Also at this energy the experimental rapidity distribution is reproduced within the measurement uncertainties. In the whole $p_T$ range, where experimental data are available, the transverse momentum spectra for different rapidity bins are remarkable well reproduced by PHQMD. 

\begin{figure}
        \includegraphics[scale=0.4]{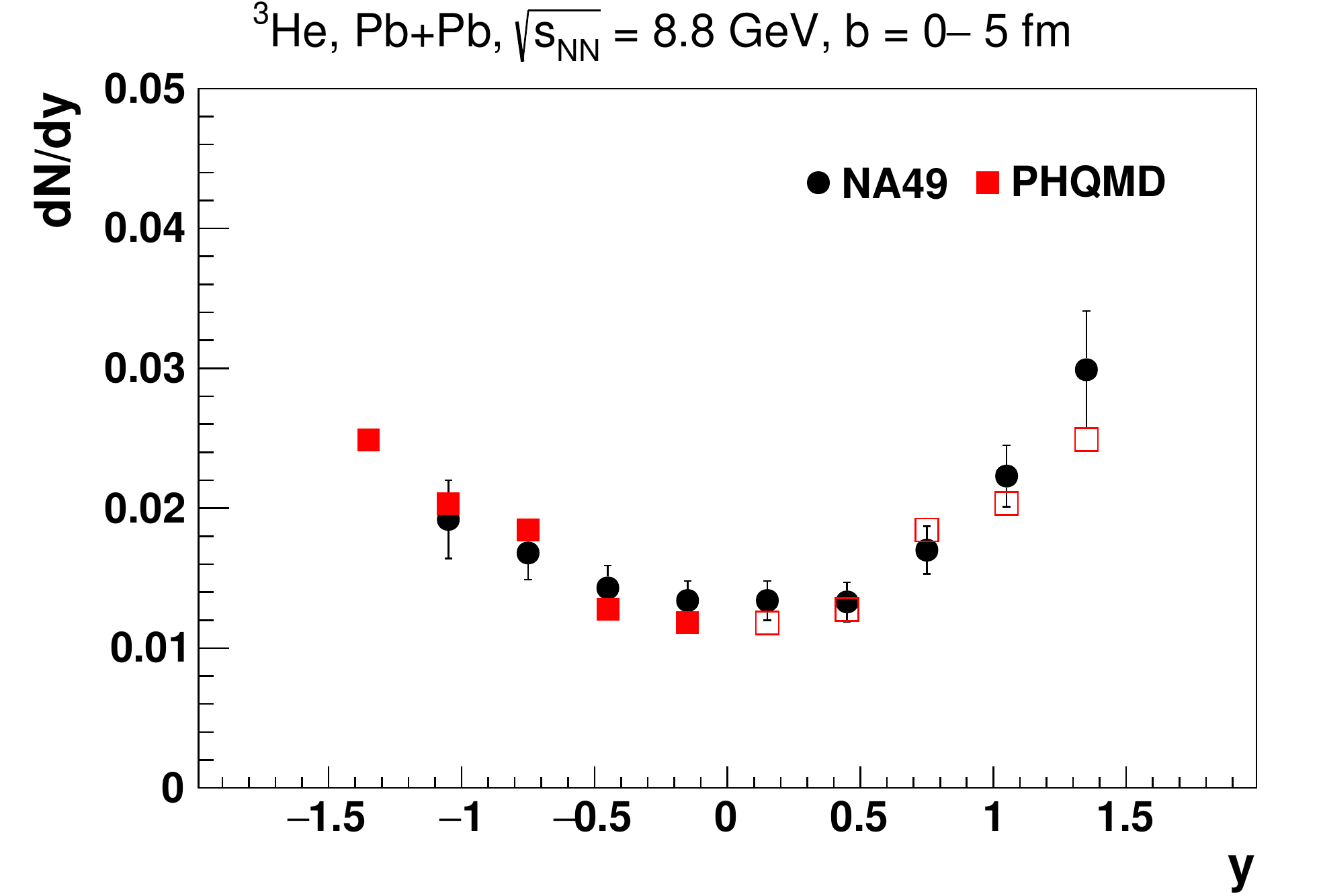} 
\caption{\label{fig:he3_dndy_8.8} The rapidity distribution of ${}^3$He from Pb+Pb central collisions at $\sqrt{s_{NN}}=8.8$~GeV.
The dots indicate the experimental data from the NA49 Collaboration \cite{Anticic:2016ckv}, the red lines show the PHQMD results
taken at the physical time $t= t_0 \cosh(y)$ for $t_0 = 67$~fm/$c$.}  
\end{figure}

\begin{figure}
        \includegraphics[scale=0.3]{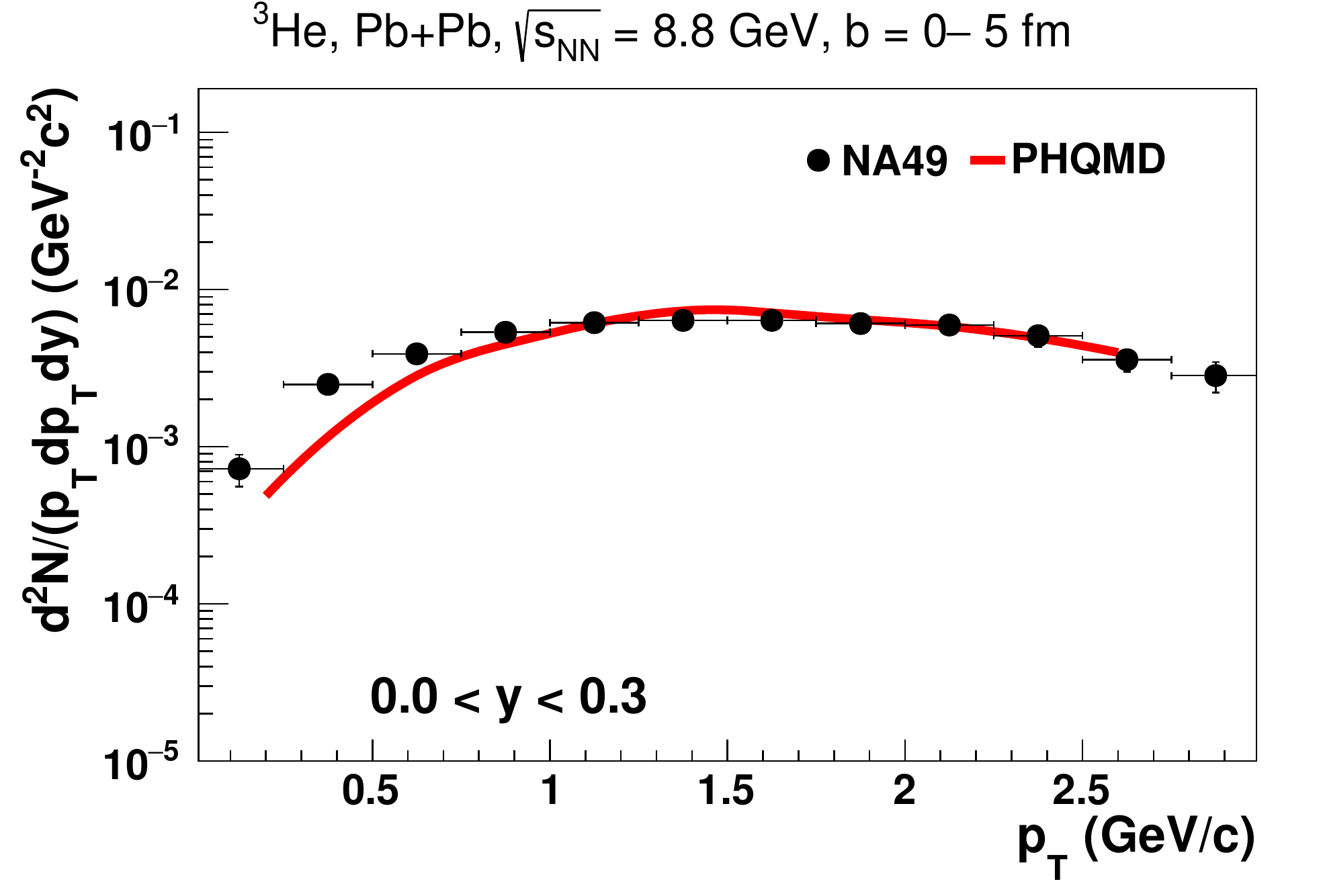} \\
        \includegraphics[scale=0.3]{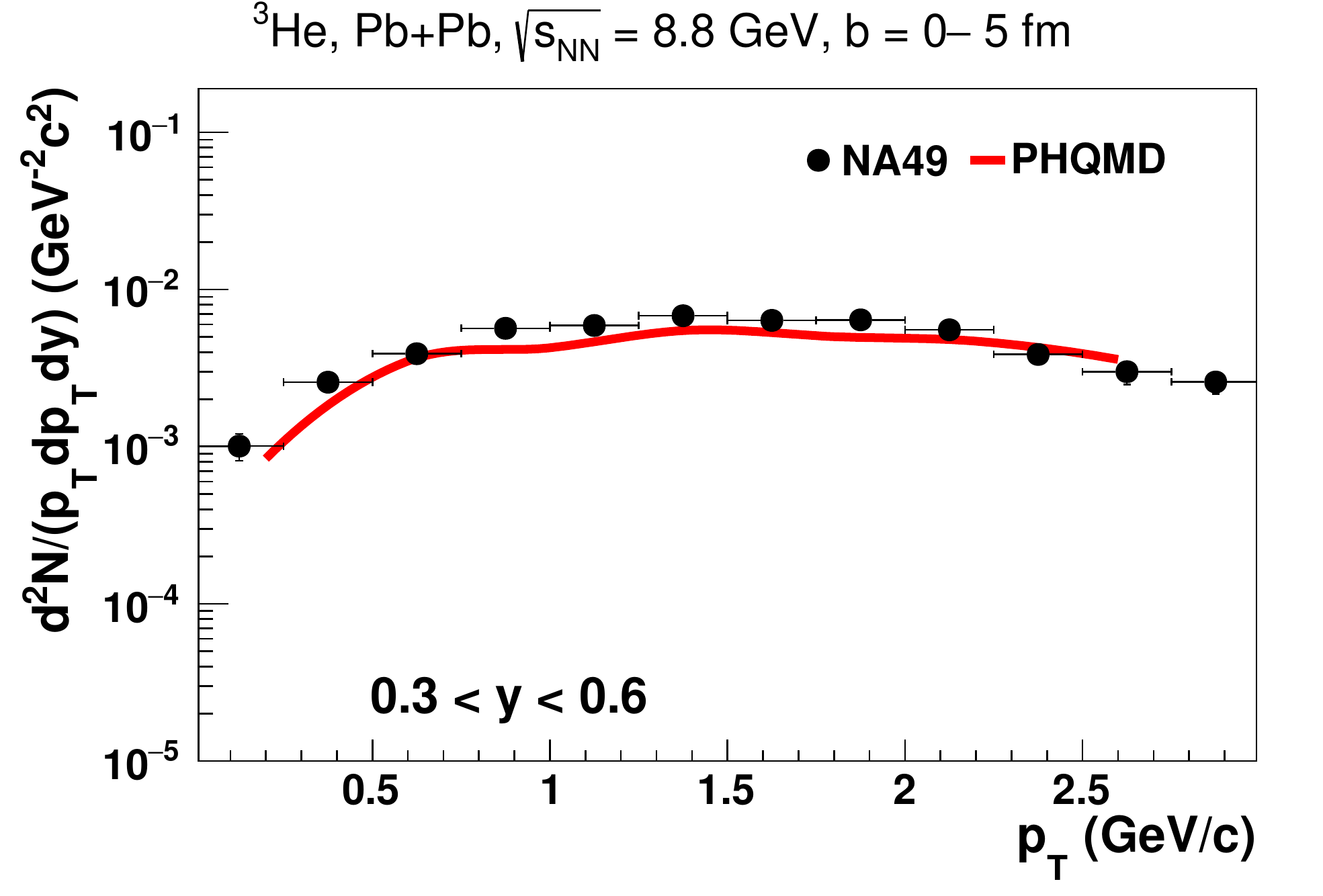} \\
        \includegraphics[scale=0.3]{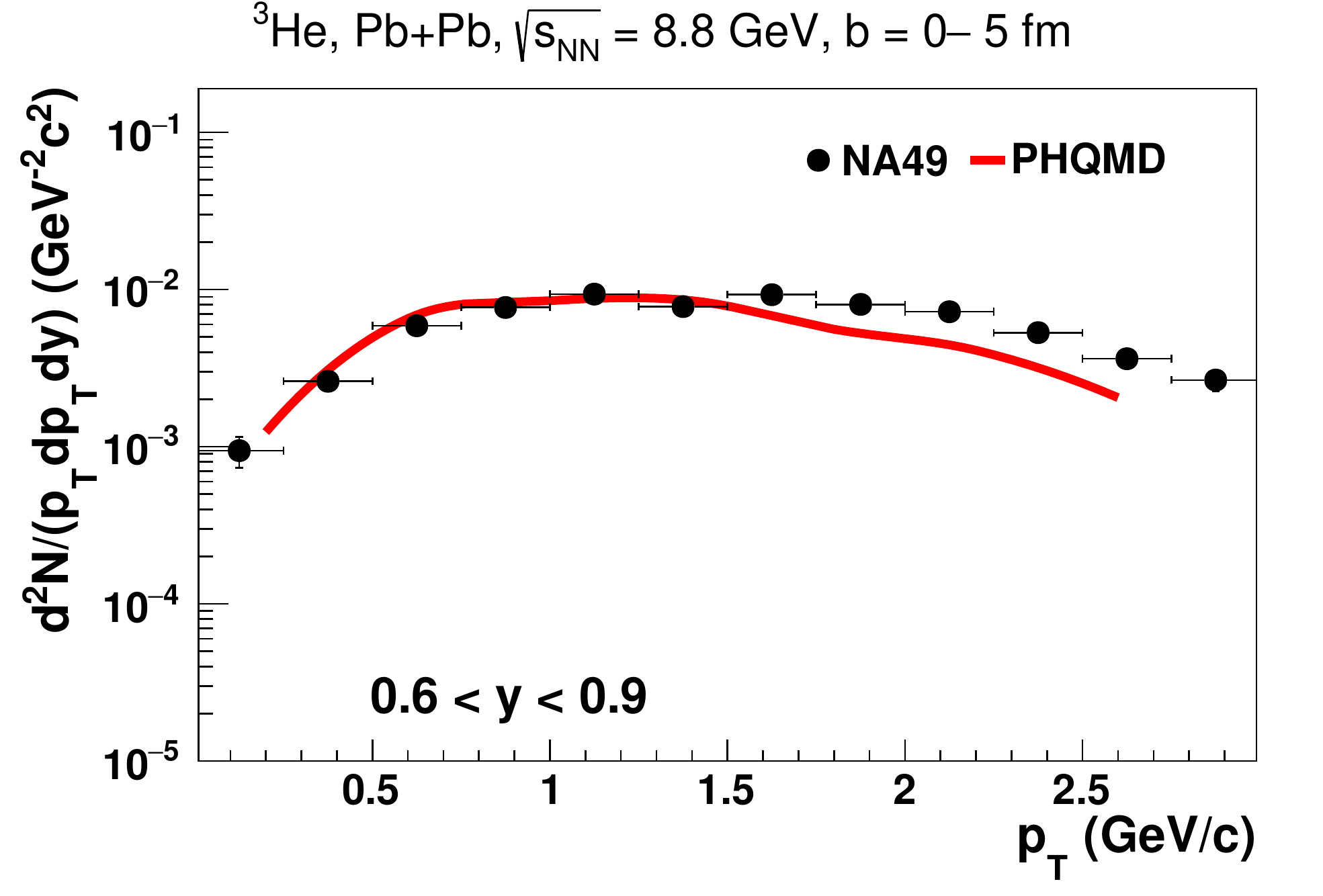} \\
        \includegraphics[scale=0.3]{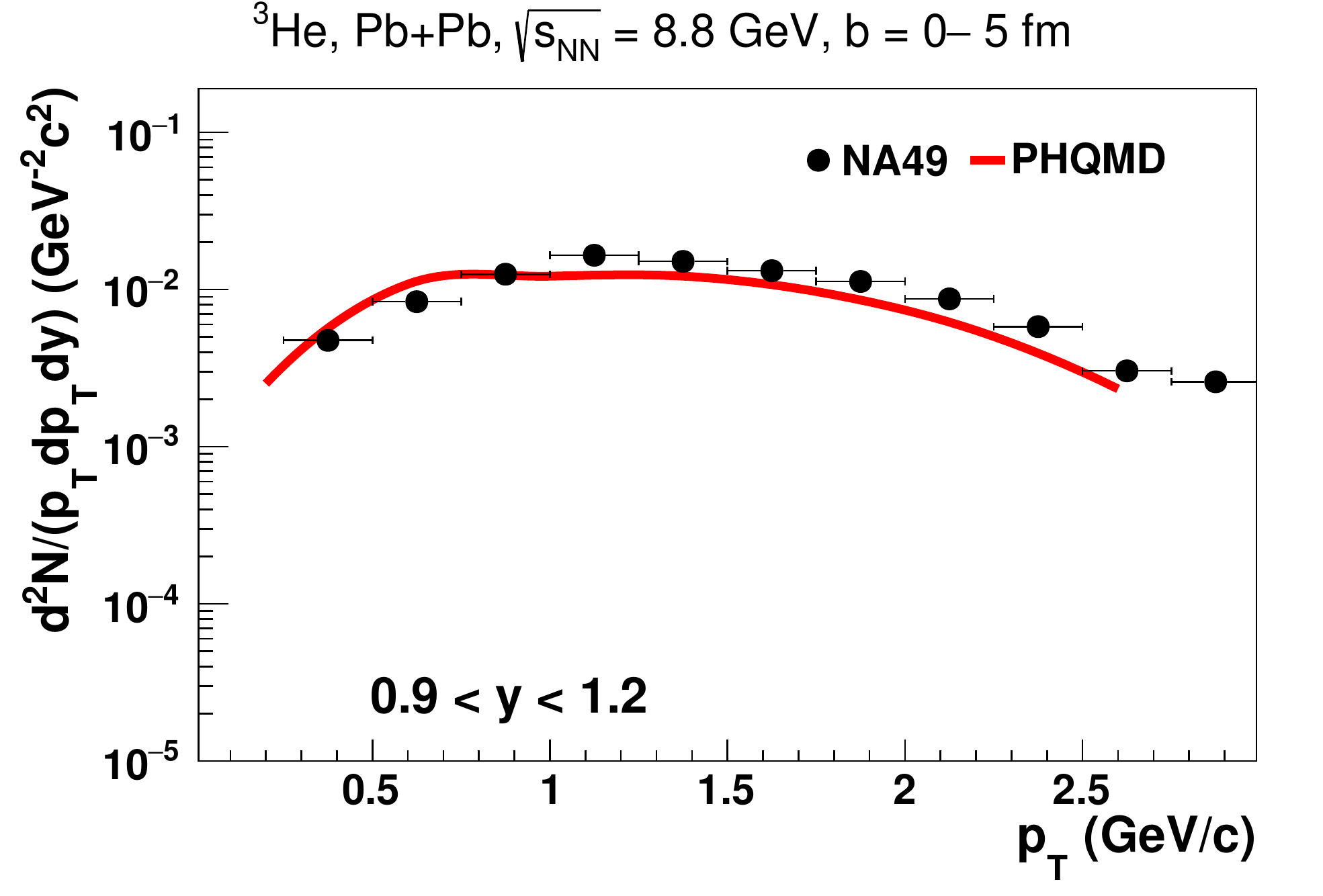} \\
        \includegraphics[scale=0.3]{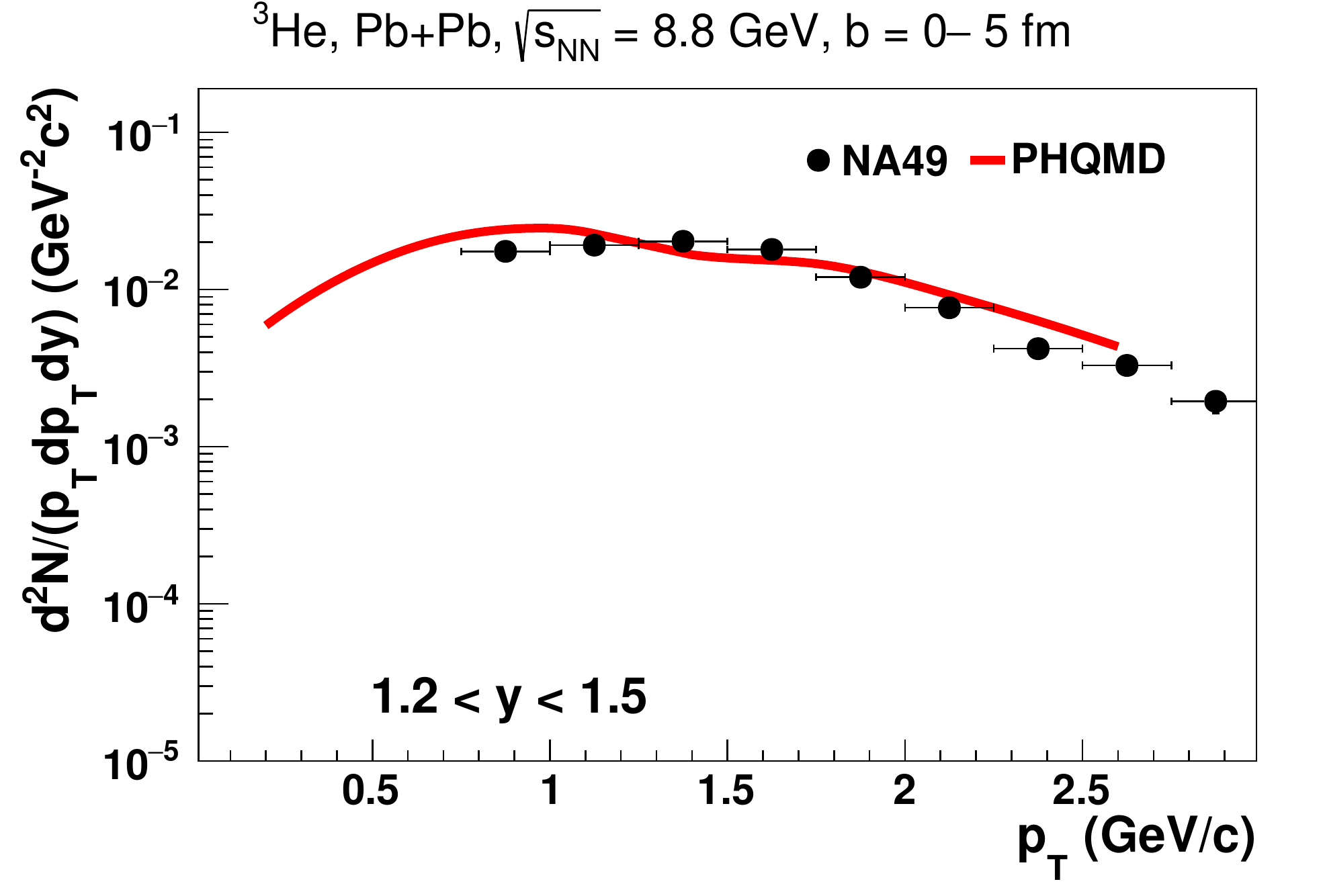} 
\caption{\label{fig:he3_pt_8.8} 
The transverse momentum spectra of ${}^3$He for Pb+Pb central collisions at $\sqrt{s_{NN}}=8.8$~GeV
for different rapidity intervals (from top to bottom): $0.0<y<0.3$,  $0.3<y<0.6$,  $0.6<y<0.9$,
 $0.9<y<1.2$,  $1.2<y<1.5$.
The dots indicate the experimental data from the NA49 Collaboration \cite{Anticic:2016ckv}, the red lines show the PHQMD results.}  
\end{figure}

\begin{figure}
    \includegraphics[scale=0.4]{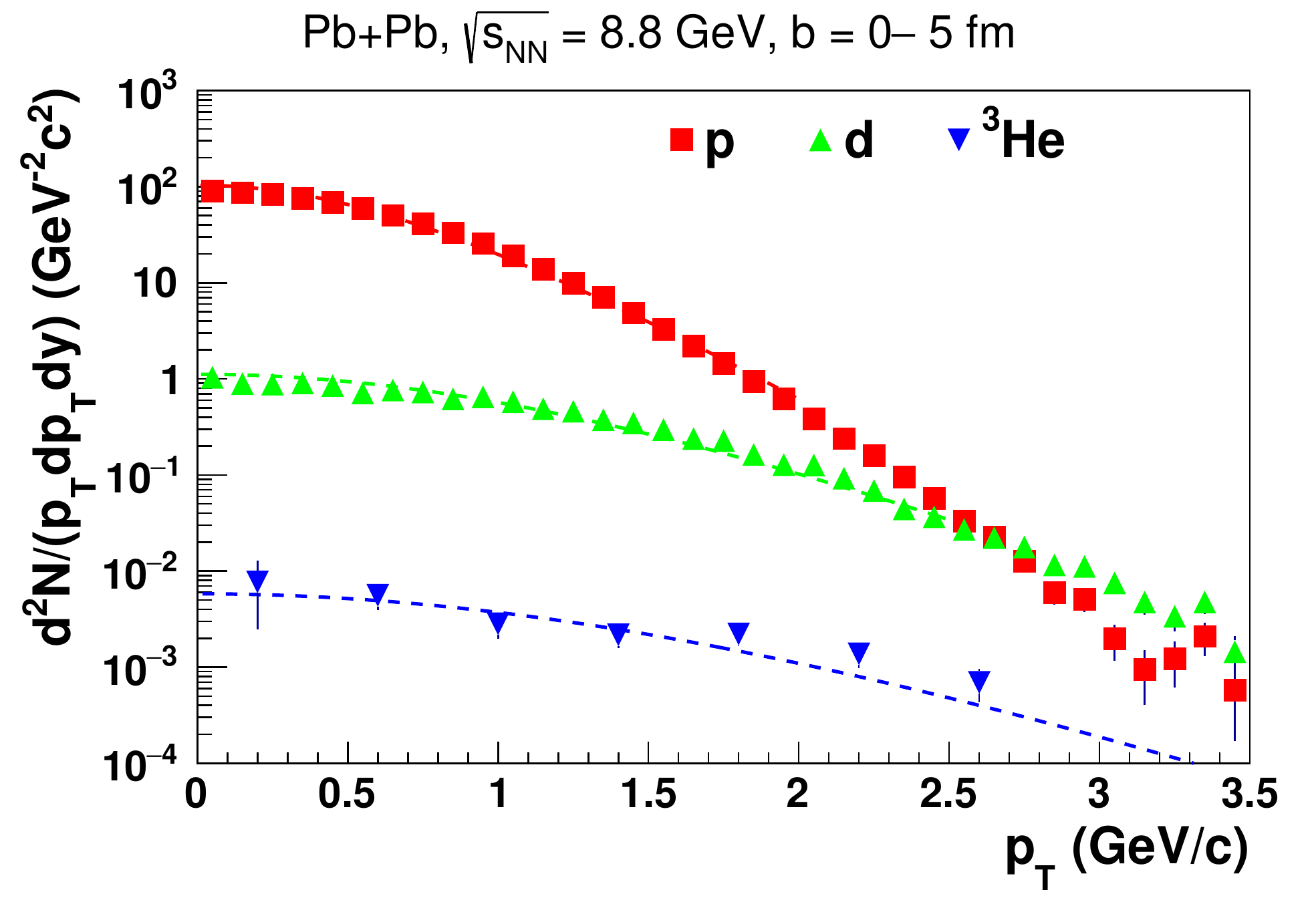}
\caption{\label{fig:8.8_d2nptdptdy}
Invariant mid-rapidity PHQMD $p_{T}$ spectra of $p$, $d$ and ${}^3$He from central Pb+Pb collisions at $\sqrt{s_{NN}} = 8.8$~GeV. Single exponential fits are plotted by the dashed curves.}
\end{figure}

\begin{figure}
    \includegraphics[scale=0.4]{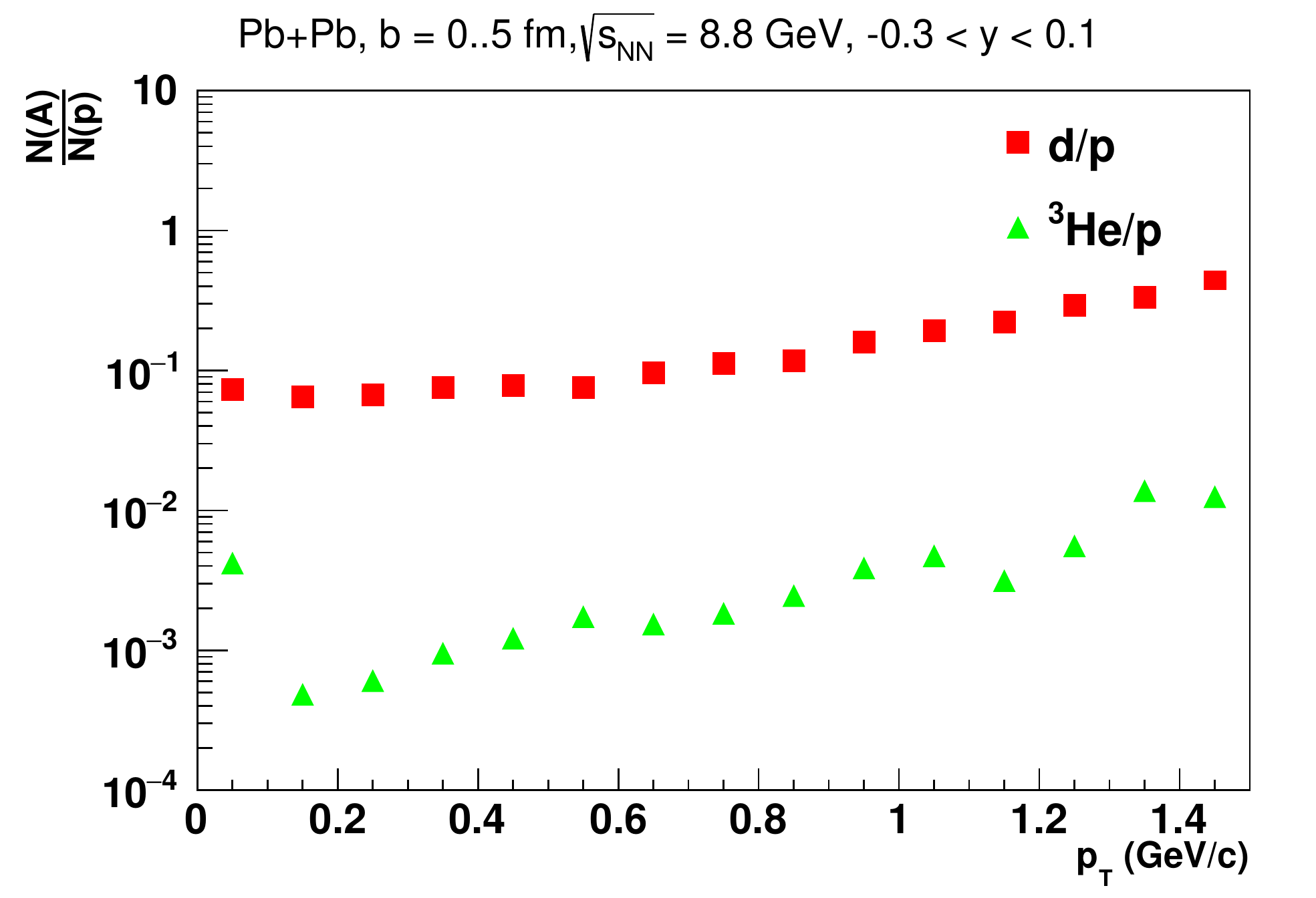}
\caption{\label{fig:8.8_pt_ratios} 
The ratios $N(d)/N(p)$ (red squares) and $N({}^{3}$He$)/N(p)$ (green triangles) from PHQMD as a function of $P_{T}/A$ at midrapidity, $-0.3 < y < 0.1$, in  Pb+Pb central collisions at $\sqrt{s_{NN}}=8.8$~GeV.}
\end{figure}

Also, the PHQMD rapidity distribution of ${}^3$He, displayed in Fig.~\ref{fig:he3_dndy_8.8}, shows a good agreement with the experimental data over the whole rapidity range and the same is true for the $p_T$ distribution for different rapidity intervals, shown in Fig.~\ref{fig:he3_pt_8.8}.
Figure~\ref{fig:8.8_d2nptdptdy} displays the PHQMD $p_T$ spectra at midrapidity for clusters of different size (red squares for protons, green triangles for deuterons and blue upside down triangles for ${}^3$He) as well as an exponential fit
of the form 
\be
\frac{d^2N}{p_Tdp_Tdy} = \frac{dN/dy}{T(m+T)^2} \, m_T \, \exp\left(-\frac{m_T-m}{T}\right)
\label{fexp}
\ee
to the spectra. We see that the 
spectral shapes are very different for the different clusters. In order to better understand this difference we introduce the quantity $\frac{N(A)}{N(p)}$, closely related to \eq{eq:BA}:
\be
\frac{N(A)}{N(p)} = \frac{\frac{d^2N_A}{dP_T^2}}{ \Big(\frac{d^2N_{protons}}{dp_T^2}\Big)^A}
\label{eq:BApt}
\ee
with $P_T= p_T A$. For central Pb+Pb collisions at $\sqrt{s_{NN}}= 8.8$~GeV this quantity is displayed as a function of $P_T/A$ in Fig.~\ref{fig:8.8_pt_ratios} for deuterons (red squares) and for ${}^3$He (green triangles) emitted in central collisions at midrapidity ($|y| < 0.3$).  We see that this ratio is flat up to $p_T = 0.6$~GeV$/c$ and then increases considerably in contradistinction to the reaction at $E_{kin} = 10.6$~$A$GeV, Fig.~\ref{fig:10.6_BA}. This means that high transverse momentum protons have a lower chance to form a cluster. This result of PHQMD follows the experimental findings very well \cite{Anticic:2016ckv}.  

\begin{figure}
    \includegraphics[scale=0.4]{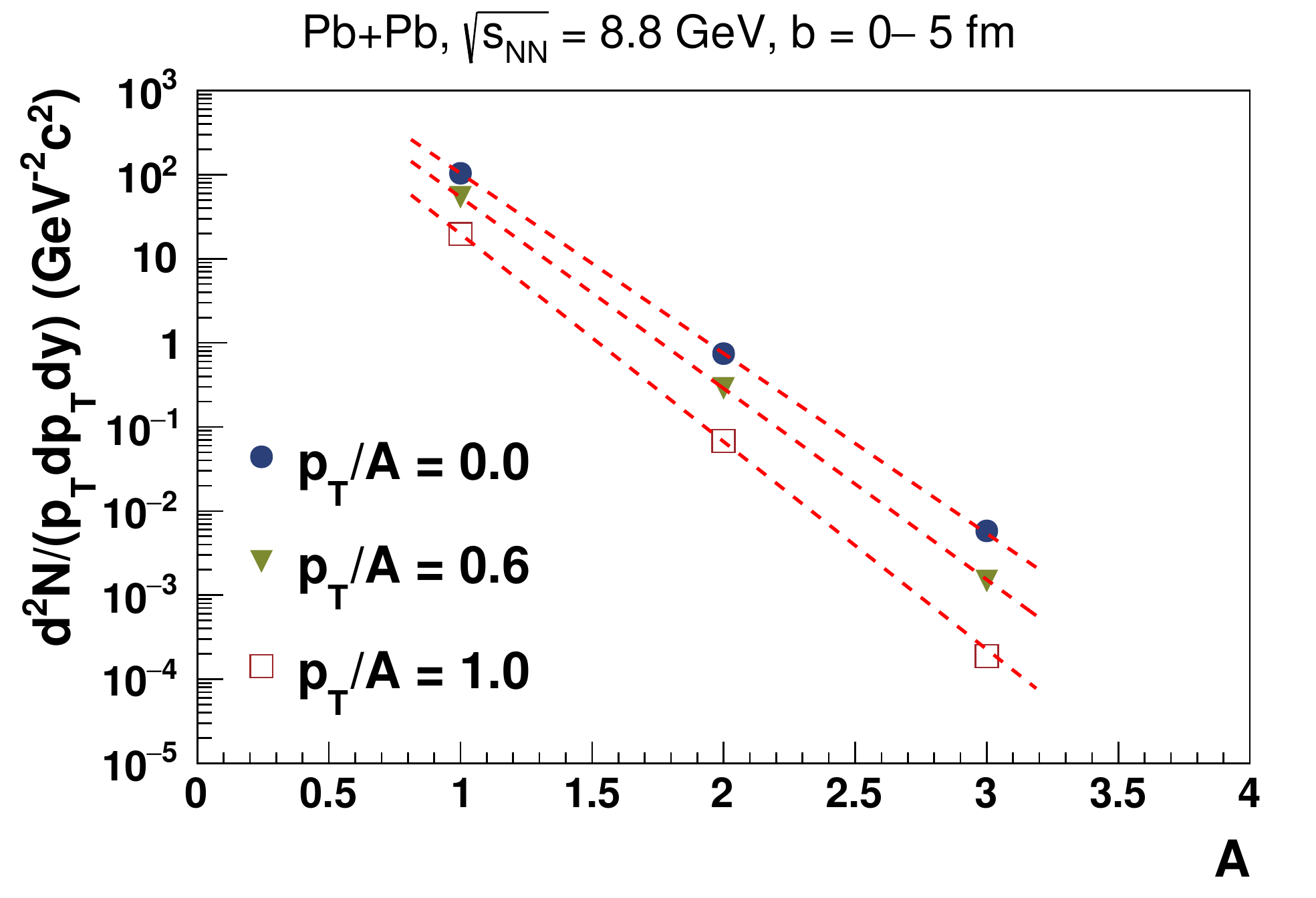}
\caption{\label{fig:8.8_inv_yield} 
Invariant yield of clusters from PHQMD at several values of $p_{T}/A$ in central Pb+Pb collisions at $\sqrt{s_{NN}} = 8.8$~GeV. The dashed lines represent a fit by a function of the form $f(A)=const/P^{A-1}$.}
\end{figure}

\begin{figure}
    \includegraphics[scale=0.4]{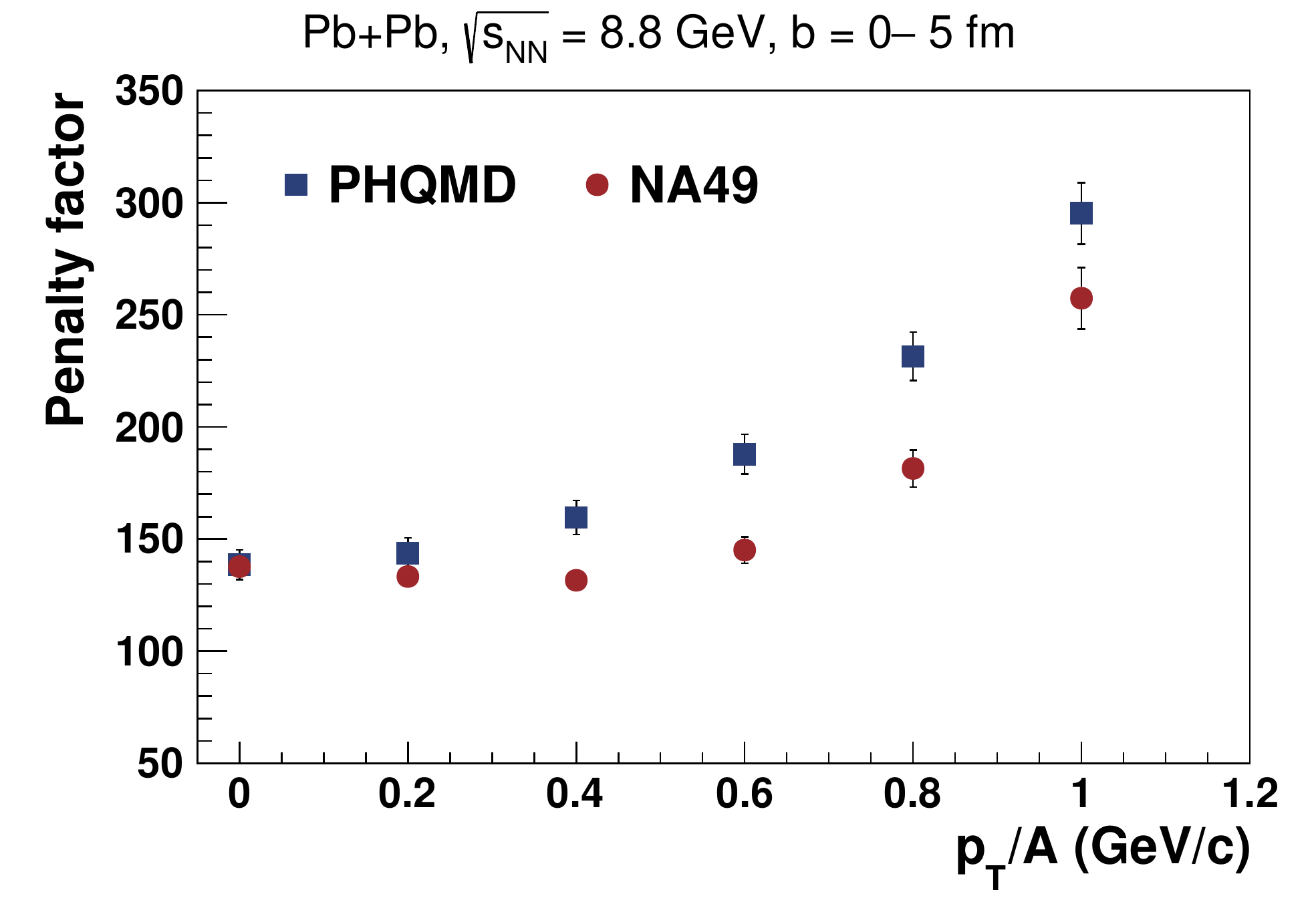}
\caption{\label{fig:8.8_penalty} Penalty factor for the cluster yields at several values of $P_{T}/A$ in central Pb+Pb collisions at $\sqrt{s_{NN}} = 8.8$~GeV. Experimental data are taken from the NA49 Collaboration \cite{Anticic:2016ckv}.}
\end{figure}

This indicates that at $\sqrt{s_{NN}}= 8.8$~GeV the probability that two or three nucleons with $p_T=P_T/A$ form a cluster depends strongly on the transverse momentum, $P_T/A$, and is therefore not proportional to the phase space.
PHQMD predicts that the probability to form a cluster in a given $P_T/A$ bin depends in form of a power law on the size of the cluster, similar as for the Au+Pb collisions at $E_{kin } = 10.6$~$A$GeV. This is shown in Fig.~\ref{fig:8.8_inv_yield}. The blue dots, green upside down triangles and red squares are the PHQMD results for the $P_T/A$ bins around 0.0, 0.6~GeV$/c$ and 1.0~GeV$/c$, respectively. The lines present the results of fit of the form
\be
f(A)=const/P^{A-1}.
\ee
The NA49 Collaboration \cite{Anticic:2016ckv} has presented the $P_T/A$ dependence of the cluster formation probability by introducing a penalty factor, already discussed in the last section. To make a comparison with PHQMD calculations
possible we fit the $d^2N/(p_T dp_T dy)$ spectra of $p$, $d$, ${}^3$He, presented in Fig.~\ref{fig:8.8_d2nptdptdy},
by an exponential function, \eq{fexp}. Then we calculate the ratio as defined in \eq{eq:BApt}. Figure~\ref{fig:8.8_penalty} displays the penalty factor for central Pb+Pb collisions at $\sqrt{s_{NN}} = 8.8$~GeV as open squares. The experimental results from the NA49 Collaboration \cite{Anticic:2016ckv} are given by red points. We obtain a statistical error of the PHQMD calculations of 1\% for protons, of 5\% for deuterons and of 30\% for ${}^3$He. Theory and experiment are in reasonable agreement. Both show a penalty factor for deuterons which strongly increases with $P_T/A$. Neither thermal models nor coalescence models predict such an increase.  


\begin{figure}
    \includegraphics[scale=0.4]{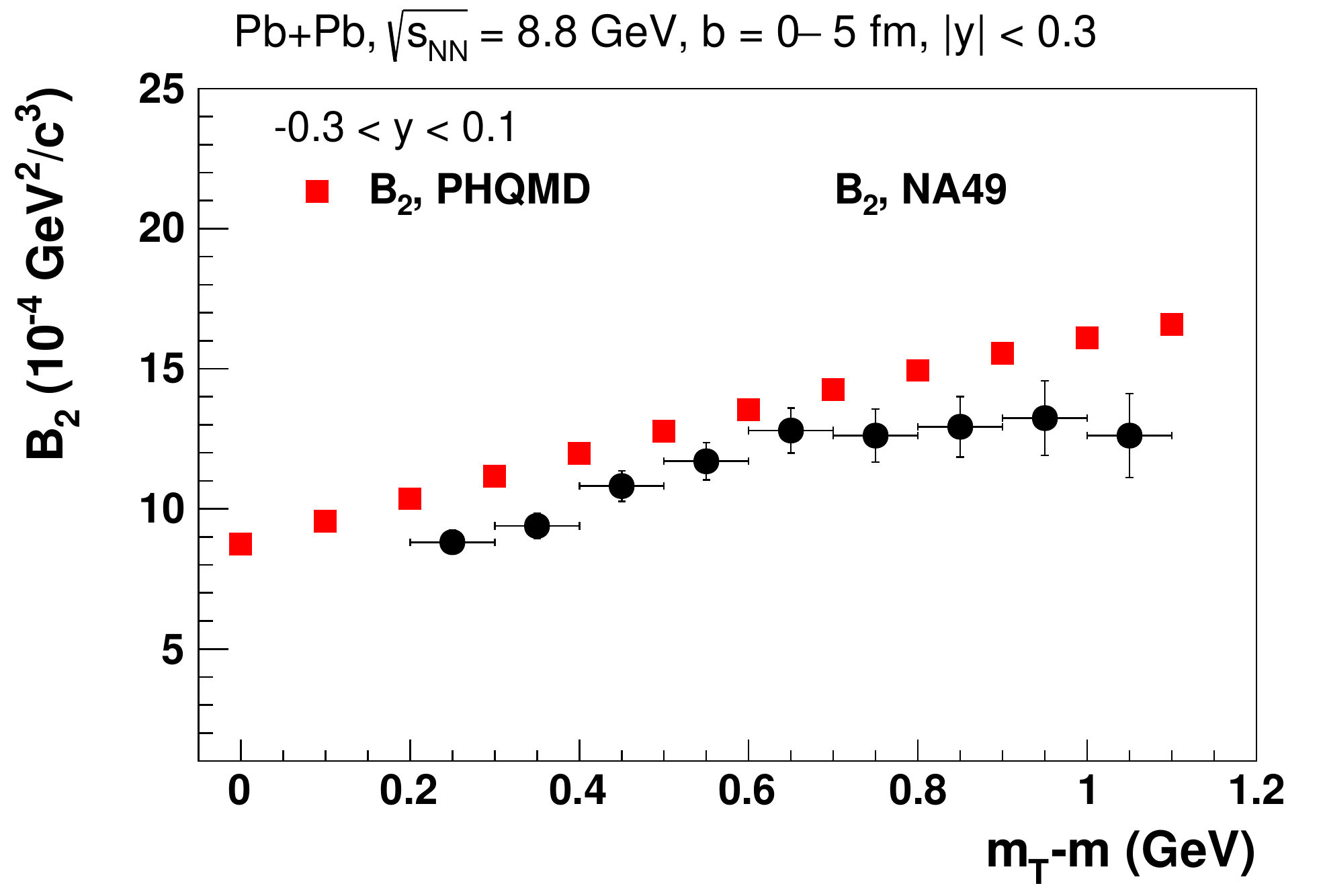}
\caption{\label{fig:8.8_b2} The coalescence factor $B_2$ as a function of $m_{T}-m$ for deuterons in central Pb+Pb collisions at $\sqrt{s_{NN}}=8.8$~GeV. Experimental data are taken from the NA49 Collaboration \cite{Anticic:2016ckv}.}
\end{figure}

The NA49 Collaboration has also published $B_2$, \eq{eq:BA}, as a function of $m_T=\sqrt{p_T^2+m^2}$ for protons and deuterons, observed in the rapidity interval $-0.3<y<0.1$ in Pb+Pb collisions at $\sqrt{s_{NN}} = 8.8$~GeV. We compare these data \cite{Anticic:2016ckv}, shown as red squares, in Fig.\ref{fig:8.8_b2} with the PHQMD calculations (black circles). Also here we see that the increase of $B_2$ with $p_T$ is well reproduced by the simulations.

This section has shown that at two intermediate energies, $E_{kin}=10.6$~$A$GeV and $\sqrt{s_{NN}} = 8.8$~GeV, the PHQMD simulations are in agreement with the available experimental data. This concord includes the cluster rapidity distributions, the $p_T$ distributions at different rapidities, but also ratios between cluster yields and cluster $p_T$ distributions.

\subsection{Energy dependence of cluster production}
\subsubsection{PHQMD results in comparison with thermal model predictions}
\begin{figure}[h!]
\centering
\includegraphics[width=8.6cm]{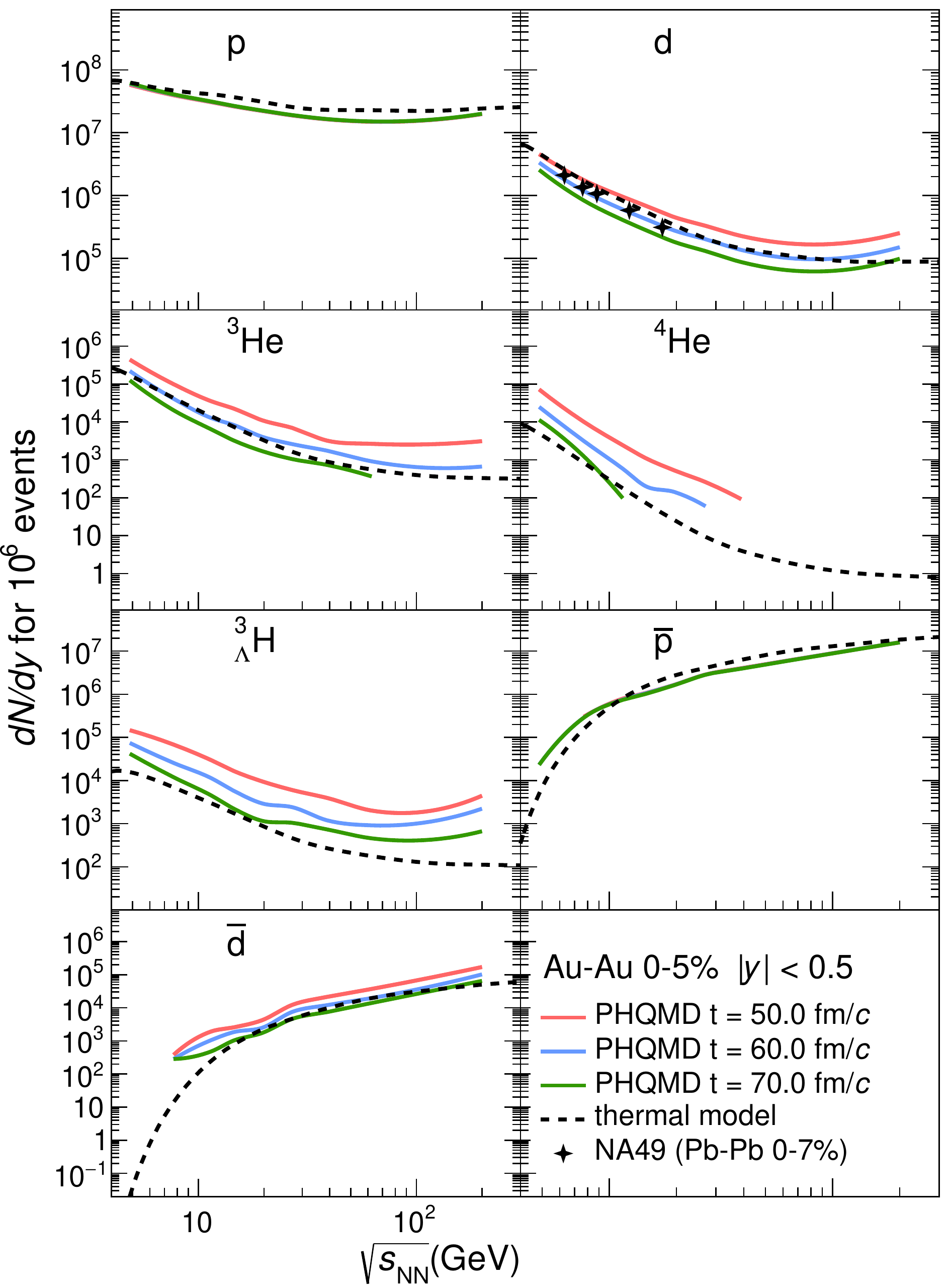}
\caption{\label{fig:th} The comparison of the PHQMD results for $dN/dy$ at midrapidity for (anti-)protons, (anti-)deuterons,  ${}^3$He, ${}^4$He and hypertritons as a function of $\sqrt{s_{NN}}$ with thermal model predictions \cite{Andronic:2010qu}.  All calculations are done  for  central Au+Au collisions (0-5\% most central) in the rapidity interval  $|y|<0.5$. 
The crosses indicate the experimental data for deuterons from the NA49 Collaboration \cite{Anticic:2016ckv}. The PHQMD results are taken at  $t = 50$~fm/$c$ (red lines) and 60~fm/$c$ (blue lines) and 70~fm/$c$ (green lines).}
\end{figure}

It is of interest to see how the excitation function of the cluster yield at midrapidity, as obtained in PHQMD calculations, compares with the prediction of statistical model calculations \cite{Andronic:2010qu}. This is investigated in Fig.~\ref{fig:th} where we display $dN/dy$ at midrapidity as a function of $\sqrt{s_{NN}}$ for different hadrons and clusters. 
All calculations are done for  central Au+Au collisions (0-5\% most central) in the rapidity interval $|y|<0.5$. The solid stars indicate the experimental data for deuterons from the NA49 Collaboration  \cite{Anticic:2016ckv}. 
We refrain from showing the measured (anti)proton data, because the weak decay correction have been done differently by the different collaborations and the results can therefore not be compared directly. The limited computer resources prevent calculations of antideuterons below $\sqrt{s_{NN}} = 8.8$~GeV,
where thermal models (which can be used as a benchmark) predict less than $10^{-5}$ antideuterons per collision.
 We see quite different excitation functions of $dN/dy$ for the different particle species. For all of them the predictions of the statistical model calculations and those provided by PHQMD are rather similar with the exception of $\bar d$ at low center-of-mass energies. This is a remarkable result, because we will see later that
clusters and single baryons come from different regions in coordinate space in contradistinction to the statistical model assumption of a common source of all particles.
In addition, PHQMD is a transport model in which elementary collisions and mutual interactions are responsible for the phase space distributions of clusters and nucleons, not the assumption of a thermal equilibrium.
One has therefore to conclude that thermal spectra or multiplicity distributions are not necessarily a hint for a thermal source.
\subsubsection{PHQMD results in comparison with data}

\begin{figure}
\centering
\includegraphics[width=0.34\textwidth]{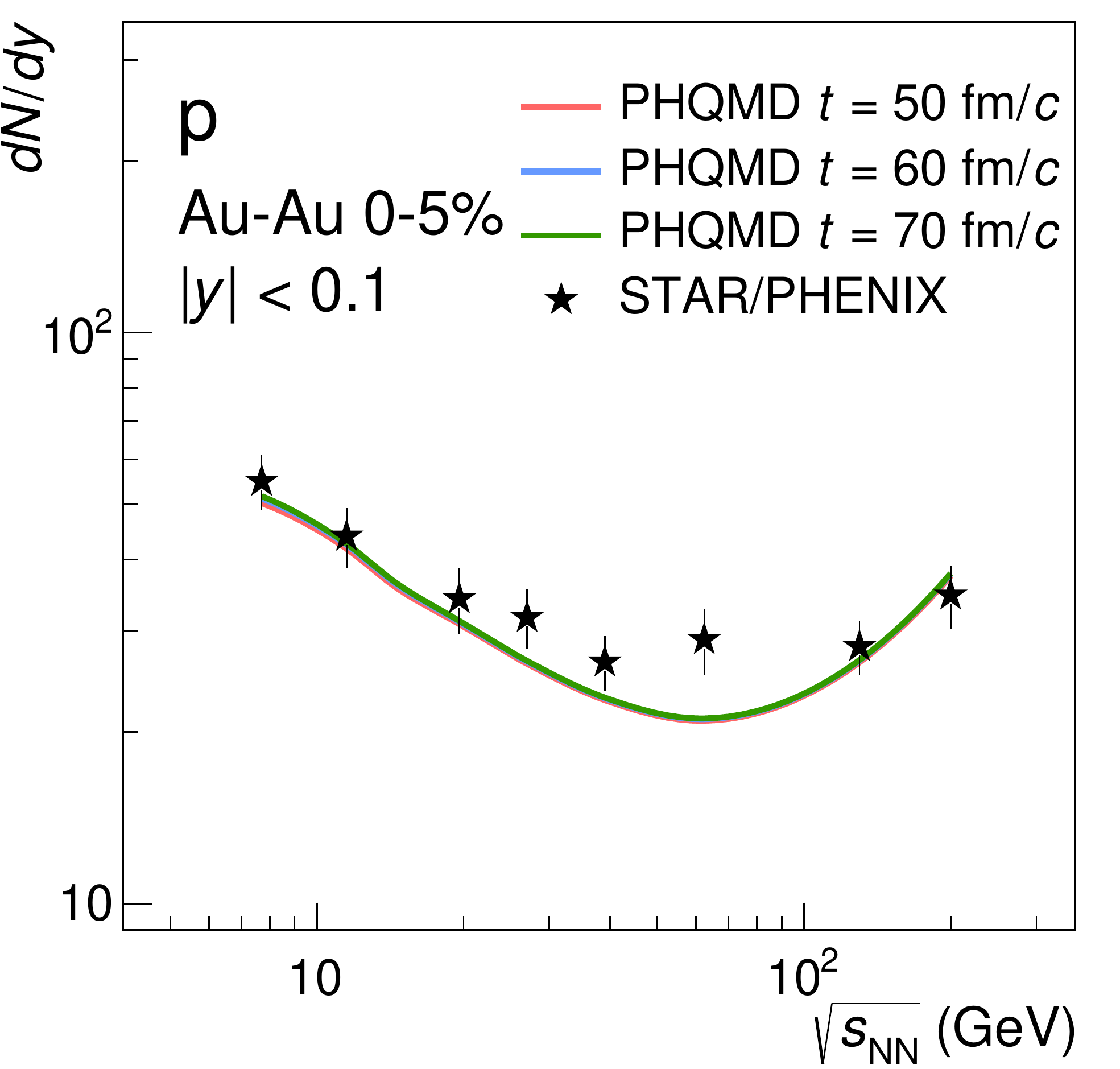}
\includegraphics[width=0.34\textwidth]{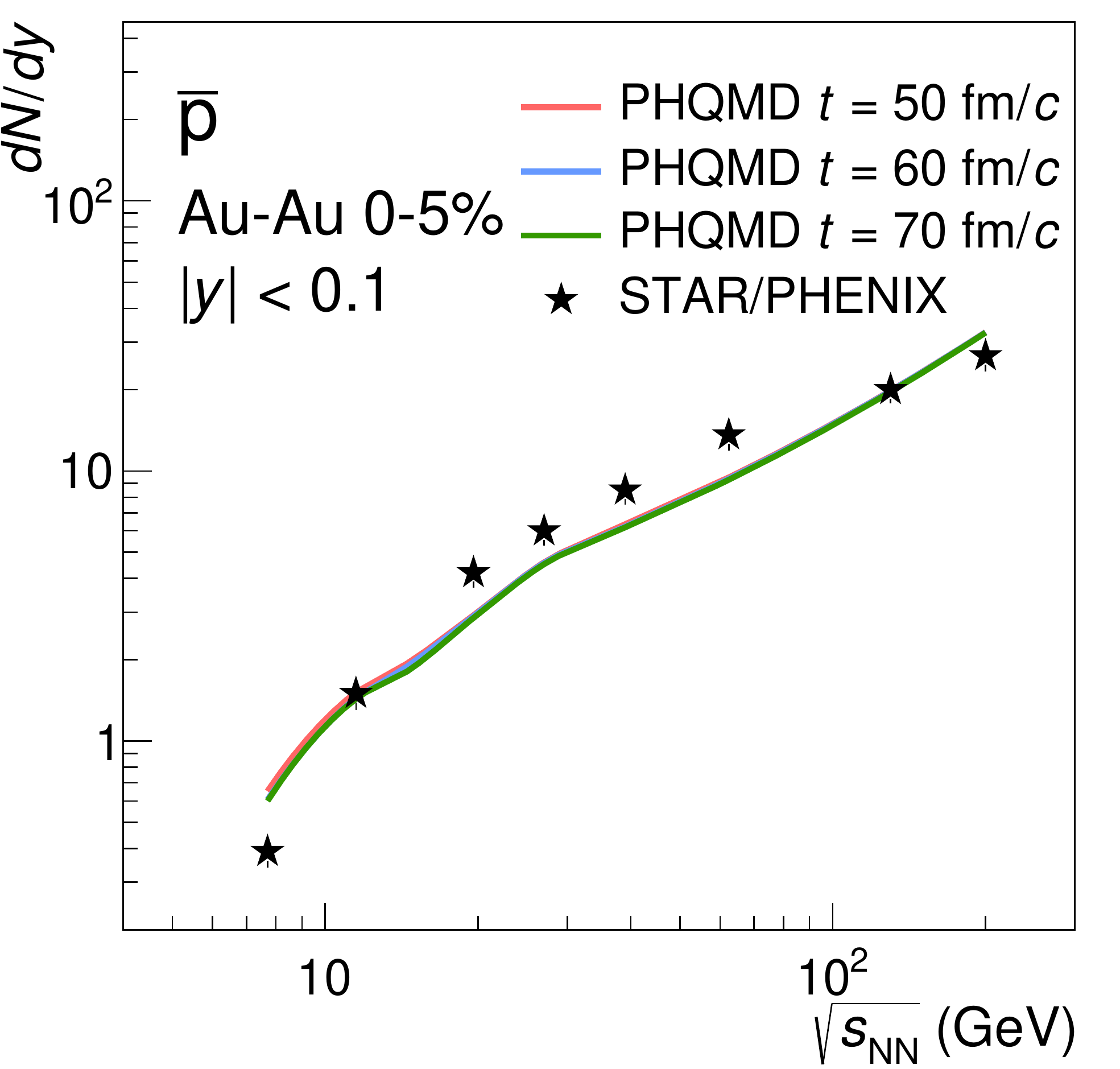}
\includegraphics[width=0.34\textwidth]{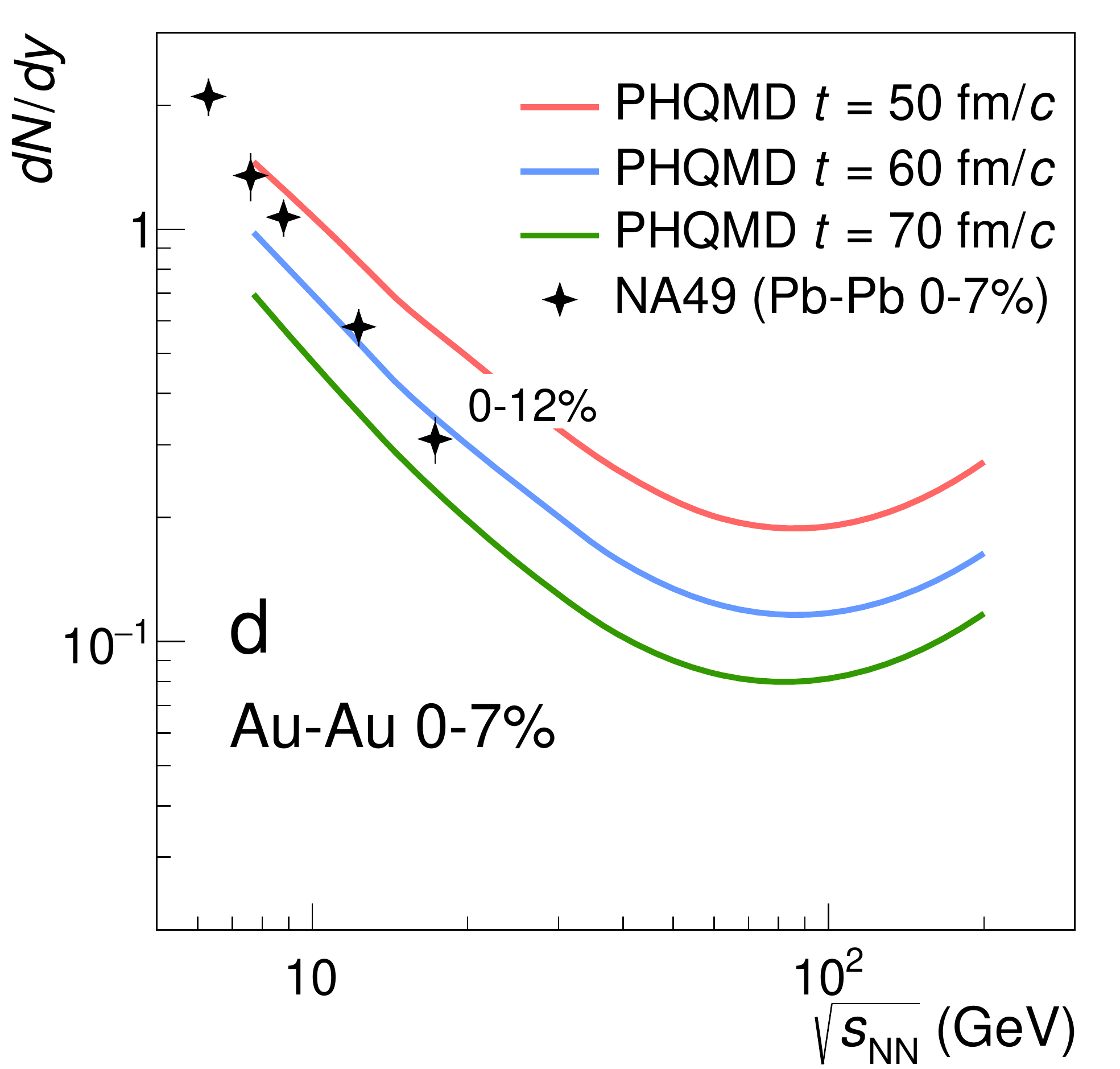}
\caption{\label{fig:WS} The midrapidity excitation function of $dN/dy$ of protons (top), antiprotons (middle) and deuterons (bottom)  as a function of $\sqrt{s_{NN}}$ for central Au+Au collisions (5\% most central (protons and antiprotons) and 7\% most central (deuterons) in comparison with the experimental data from the NA49 Collaboration \cite{Anticic:2016ckv} (crosses) where the midrapidity intervals are taken as $-0.4<y<0.0$ for 
$\sqrt{s_{NN}}= 6.3$~GeV, 7.6~GeV and 8.8~GeV; $-0.6<y<-0.2$ for 12.3~GeV and  $-0.6<y<-0.4$ for 17.3~GeV, 
as well as from the STAR and PHENIX Collaborations \cite{Adamczyk:2017iwn,Adler:2003cb} (stars) for $|y|<0.1$. The rapidity intervals for the PHQMD results correspond to the experimental ones.
The PHQMD results for protons and antiprotons  are scaled to account for the protons from weak decay  feed-down that are included in the experimental data. }
\end{figure}

We proceed now to discuss the energy dependence of several key cluster observables. In Fig.~\ref{fig:WS} we display
the excitation function of $dN/dy$ at midrapidity  of protons (top), antiprotons (middle) and deuterons (bottom) as a function of $\sqrt{s_{NN}}$ for central Au+Au collisions (5\% most central for protons and antiprotons and 7\% most central for deuterons) in comparison to the experimental data from the NA49, STAR and PHENIX  Collaborations. The experimental $p$ and $\bar p$ data from the STAR and PHENIX Collaborations \cite{Adamczyk:2017iwn,Adler:2003cb} are marked as stars. The deuteron results from the NA49 Collaboration  \cite{Anticic:2016ckv} are marked as crosses. The midrapidity intervals are taken as $-0.4<y<0.0$ for the NA49 data at the center-of-mass energies
$\sqrt{s_{NN}}= 6.3$~GeV, 7.6~GeV and 8.8~GeV;  $-0.6<y<-0.2$ for 12.3~GeV and  $-0.6<y<-0.4$ for 17.3~GeV. 
For the STAR and PHENIX data we employ $|y|<0.1$. 
The rapidity intervals for the PHQMD results correspond to the experimental ones.
In this section the PHQMD results are shown as red lines for $t_0=50$~fm$/c$, blue lines for $t_0=60$~fm$/c$ and  green lines  for $t_0=70$~fm$/c$. These data are mostly taken close to midrapidity, therefore the physical time and the
time in the computational frame, $t_0$, do not differ substantially. The PHQMD results for protons and antiprotons are scaled to account for the protons from weak decay feed-down that are included in the experimental data. 

\begin{figure}
\centering
\includegraphics[width=0.4\textwidth]{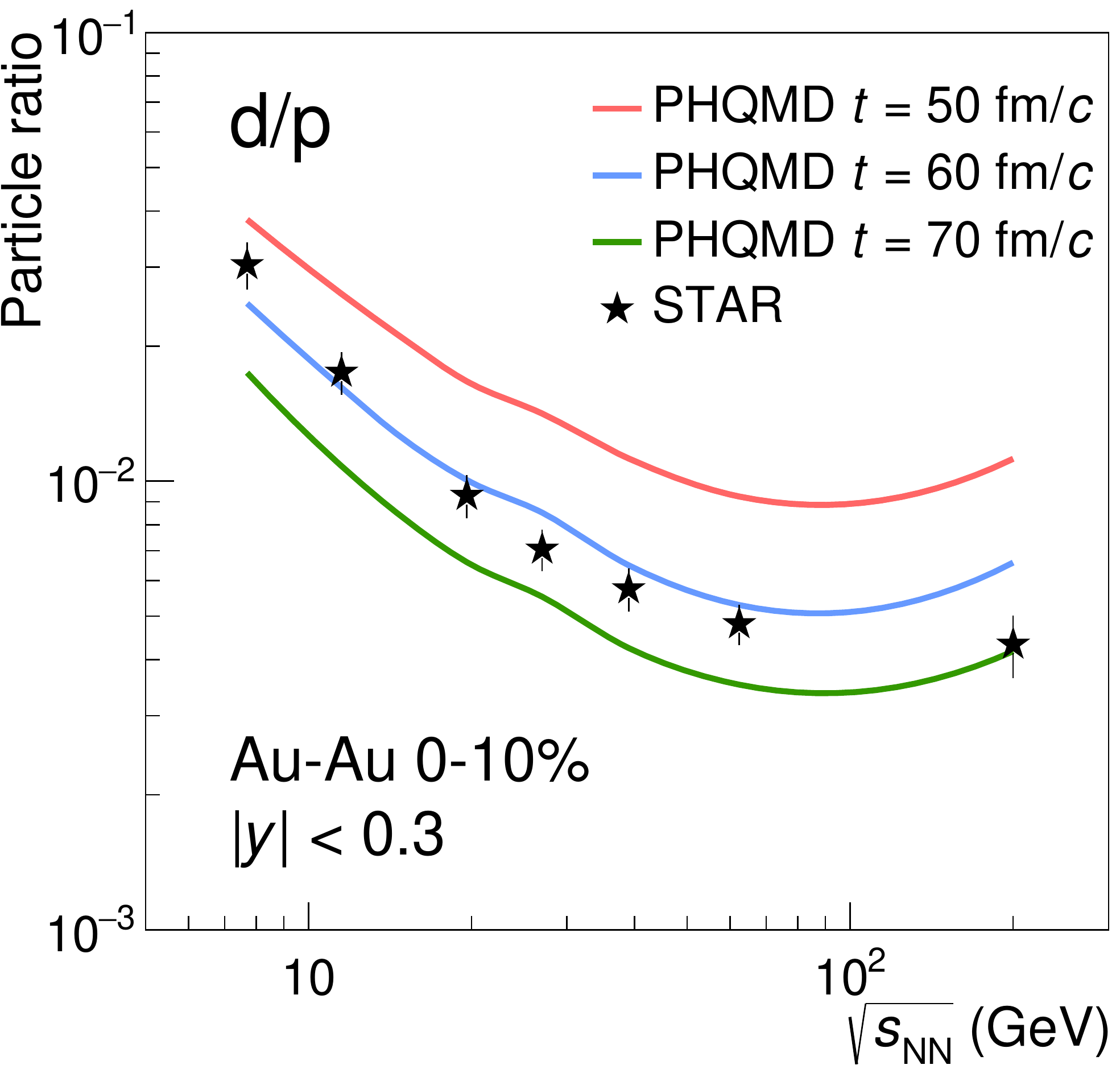}
\includegraphics[width=0.4\textwidth]{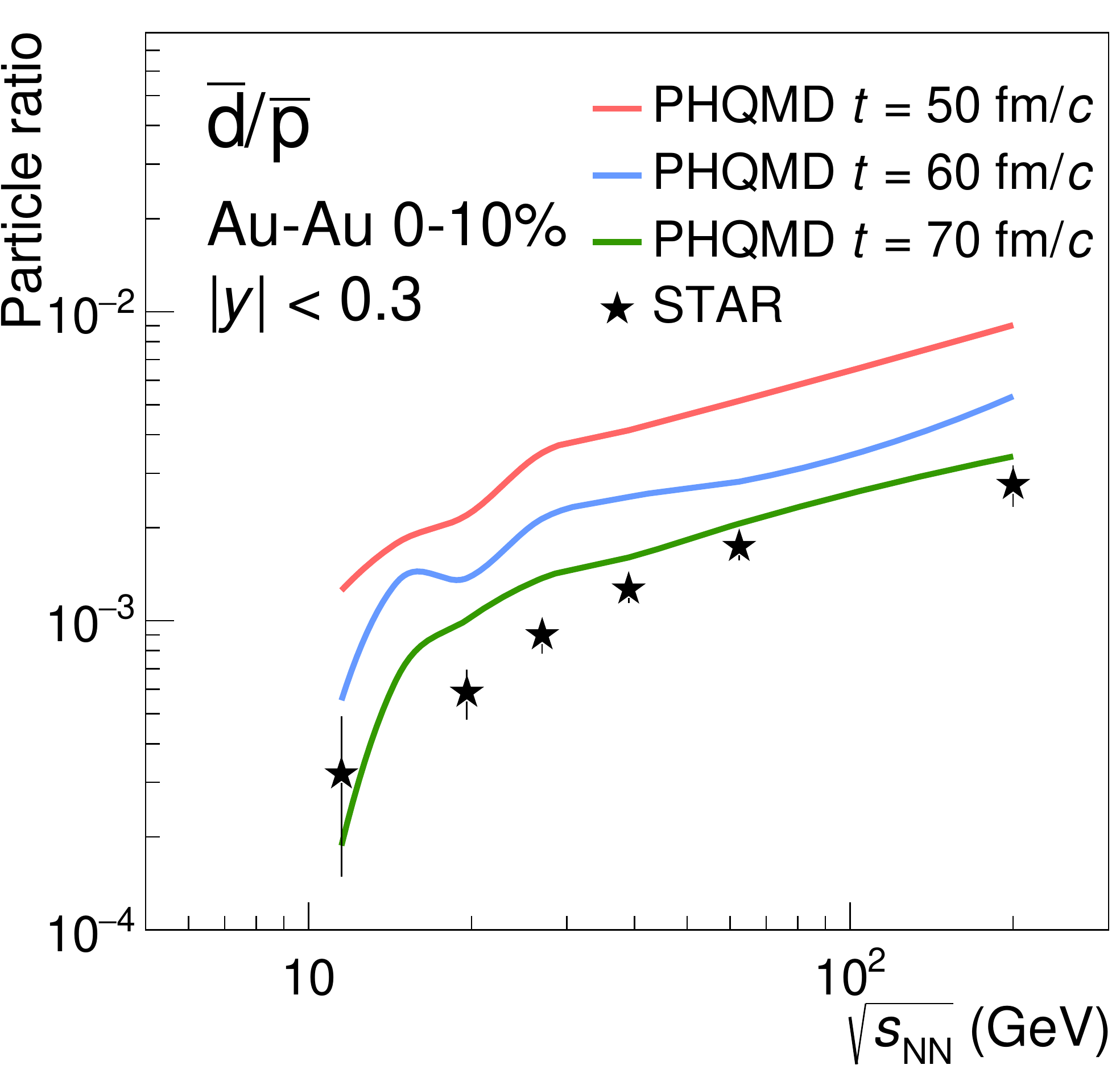}
\caption{\label{fig:rat} The excitation function of the deuteron to proton (top) and antideuteron to antiproton ratios (bottom) for central Au+Au collisions as a function of $\sqrt{s_{NN}}$. 
The experimental data from the STAR Collaboration \cite{Adam:2019wnb} in the rapidity interval $|y|<0.3$ are indicated as stars.
The PHQMD results are shown as red lines for $t=50$~fm$/c$, blue lines for $t=60$~fm$/c$ and green lines for $t=70$~fm$/c$. }
\end{figure}

The proton and antiproton  as well as the deuteron $dN/dy$ are nicely reproduced from the lowest SPS energies up to the highest RHIC energies
(in the present version of PHQMD the calculation of the cluster production at LHC energies is too time consuming and not included in the comparison). As a consequence of the good agreement with the (anti)protons, the experimental $d/p$ ratio as a function of $\sqrt{s_{NN}}$ is also well reproduced. 
In Fig.~\ref{fig:rat} we present the excitation function of the deuteron to proton $d/p$ (top) and antideuteron to antiproton $\bar d/\bar p$ (bottom) ratios from central Au+Au collisions as a function of $\sqrt{s_{NN}}$. 
The experimental data from the STAR Collaboration \cite{Adam:2019wnb} in the rapidity interval $|y|<0.3$ are indicated as stars and compared to the PHQMD results.
The form of the dependence of  $\bar d/ \bar p$ on $\sqrt{s_{NN}}$ is well reproduced although we overpredict this ratio by roughly a factor of two, if we determine the yield of antiparticles at the same time as that of the particles. As Fig.~\ref{fig:WS}, middle, shows that the $\bar p$ distribution does not depend on time, this enhancement is consequently due to a overpredicted $\bar d$ yield.

\begin{figure}
\centering
\includegraphics[width=0.4\textwidth]{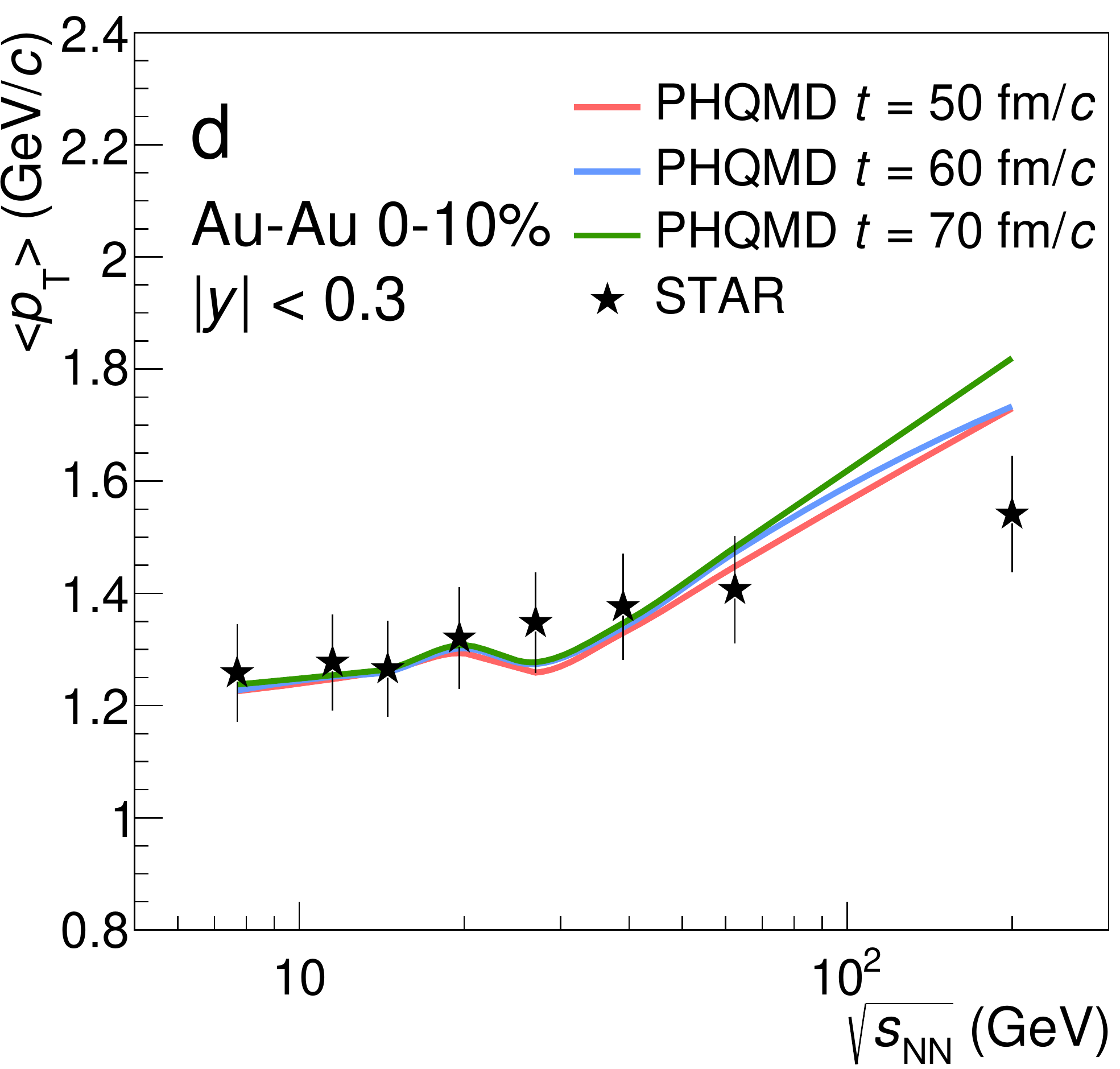}
\includegraphics[width=0.4\textwidth]{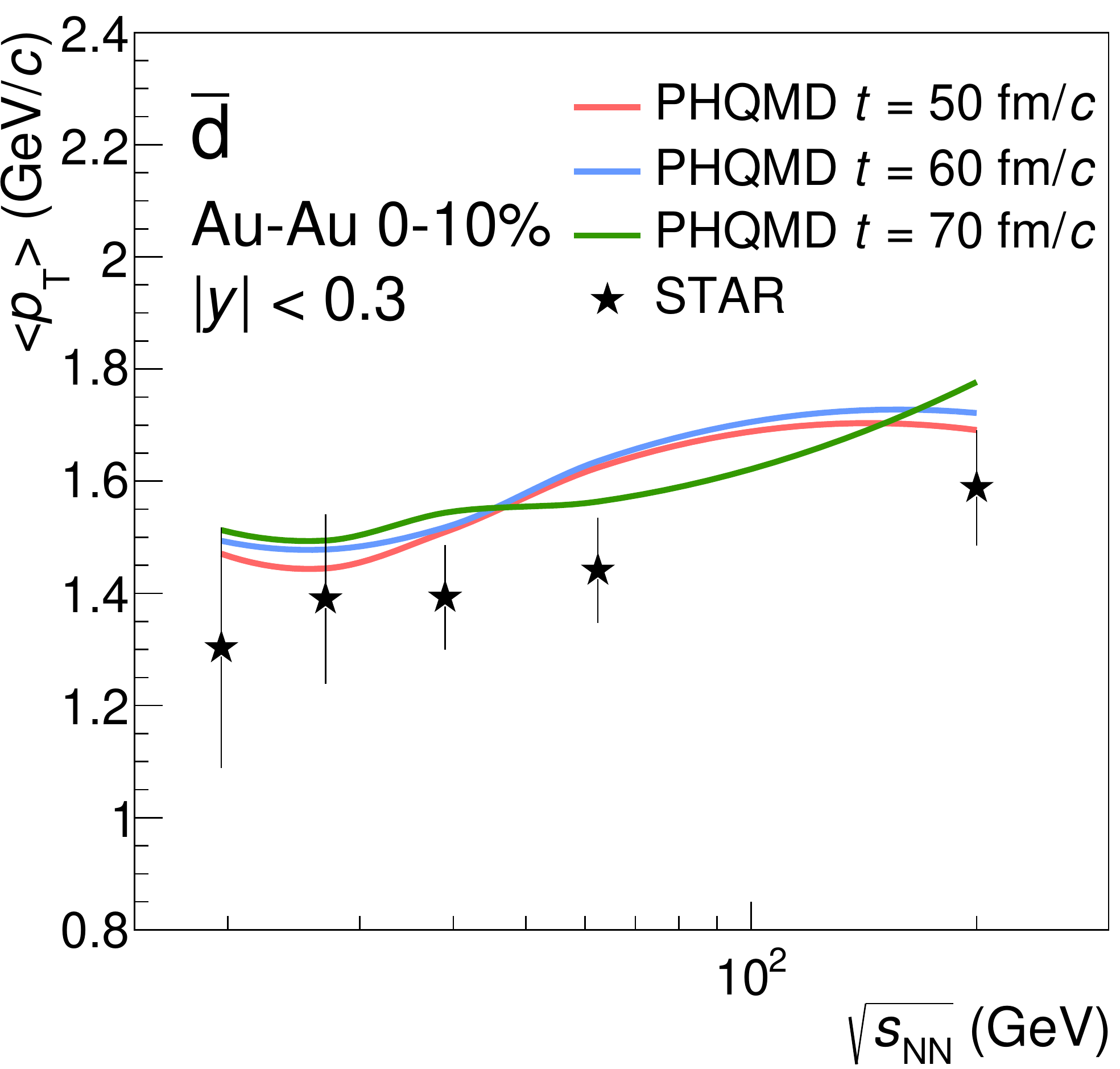}
\caption{\label{fig:pt} The average transverse momentum of deuterons (top) and antideuterons (bottom)  
as a function of $\sqrt{s_{NN}}$ for central Au+Au collisions in the rapidity interval  $|y|<0.3$ as compared with data from the STAR Collaboration. $\langle p_{T} \rangle$ is calculated using on individual Blast-Wave model fits to the $p_T$ spectra.}
\end{figure} 

Besides the excitation function of the rapidity distribution also that of the mean transverse momentum $\langle p_T \rangle$ at midrapidity has been published by the STAR Collaboration for deuterons, as well as for antideuterons.  In Fig.~\ref{fig:pt} we compare the experimental excitation function of $\langle p_T \rangle$ with PHQMD calculations.
We see that the absolute value as well as the form of the excitation function agrees well for deuterons, for antideuterons we overpredict the data slightly.

\begin{figure}[h!]
\centering
\includegraphics[width=8.6cm]{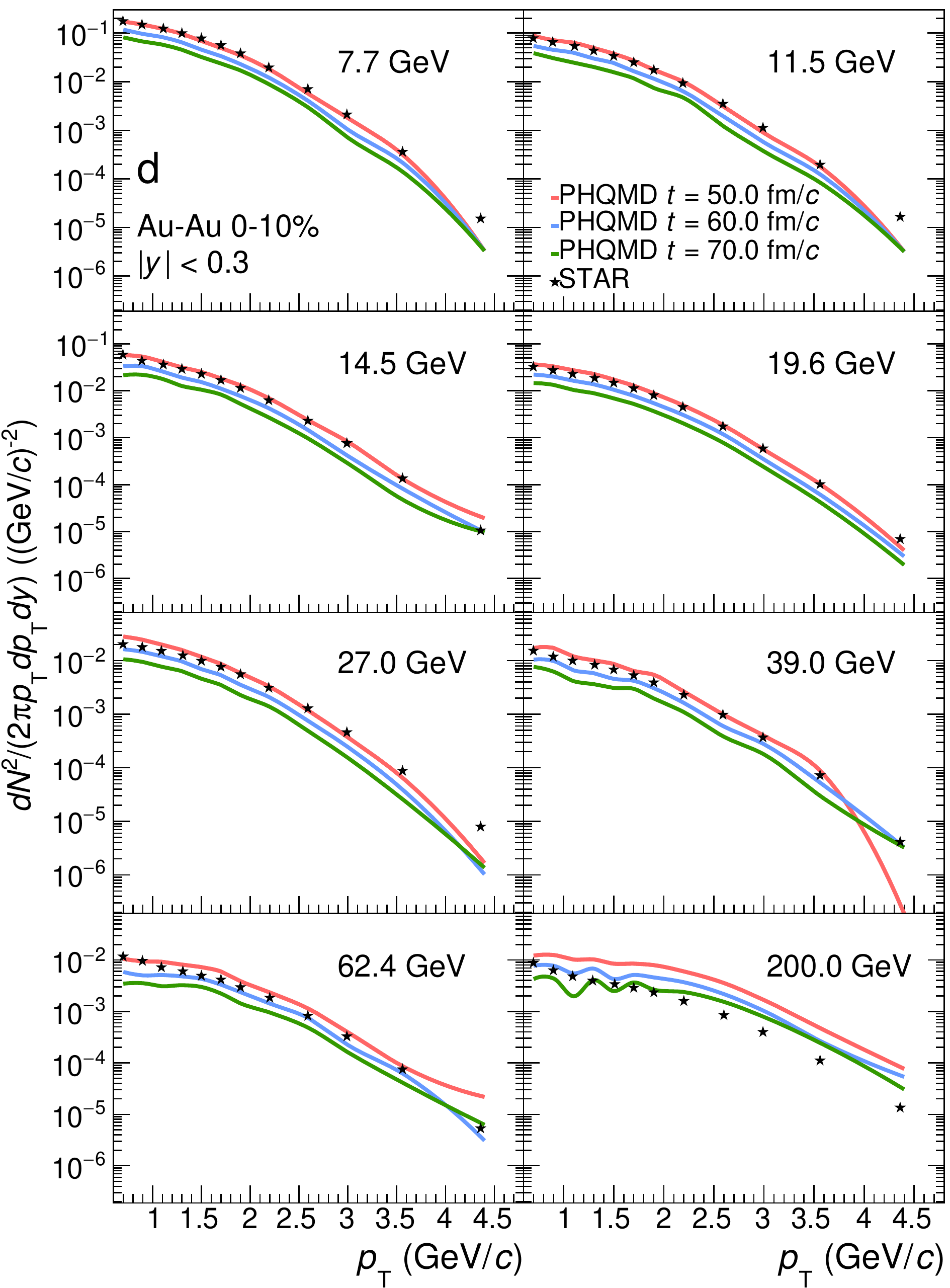}
\caption{\label{fig:ptp}   
The transverse momentum distribution of deuterons in central Au+Au collisions at midrapidity for
$\sqrt{s_{NN}} =7.7$, 11, 14.5, 19, 27, 39, 62 and 200~GeV (as indicated in the legend).
The dots indicate the experimental data from the STAR Collaboration \cite{Adam:2019wnb}. The PHQMD results are
taken at  $t = 50$~fm/$c$ (red lines), 60~fm/$c$ (blue lines)
and 70~fm/$c$ (green lines). }
\end{figure}

The experimental $p_T$ spectra of deuterons from the BES at RHIC, taken by the STAR Collaboration \cite{Adam:2019wnb}, allow for an even more detailed comparison between PHQMD simulations and experimental data. In Fig.~\ref{fig:ptp} we present the transverse momentum distribution of deuterons in central Au+Au collisions at midrapidity for
$\sqrt{s_{NN}} = 7.7$, 11, 14.5, 19, 27, 39, 62 and 200~GeV (as indicated in the legends).
The full dots indicate the experimental data from the STAR Collaboration \cite{Adam:2019wnb}. 
taken at  $t = 50$~fm/$c$ (red lines) and 60~fm/$c$ (blue lines)
and 70~fm/$c$ (green lines). 
We see that from $\sqrt{s_{NN}}= 7.7$~GeV up to $\sqrt{s_{NN}} = 62.4$~GeV the deuteron yield at 50~fm/$c$ agrees within error bars with the experimental results, only at $\sqrt{s_{NN}} = 200$~GeV PHQMD overpredicts the data. This will be subject of future investigations.

\begin{figure}
\centering
\includegraphics[width=0.4\textwidth]{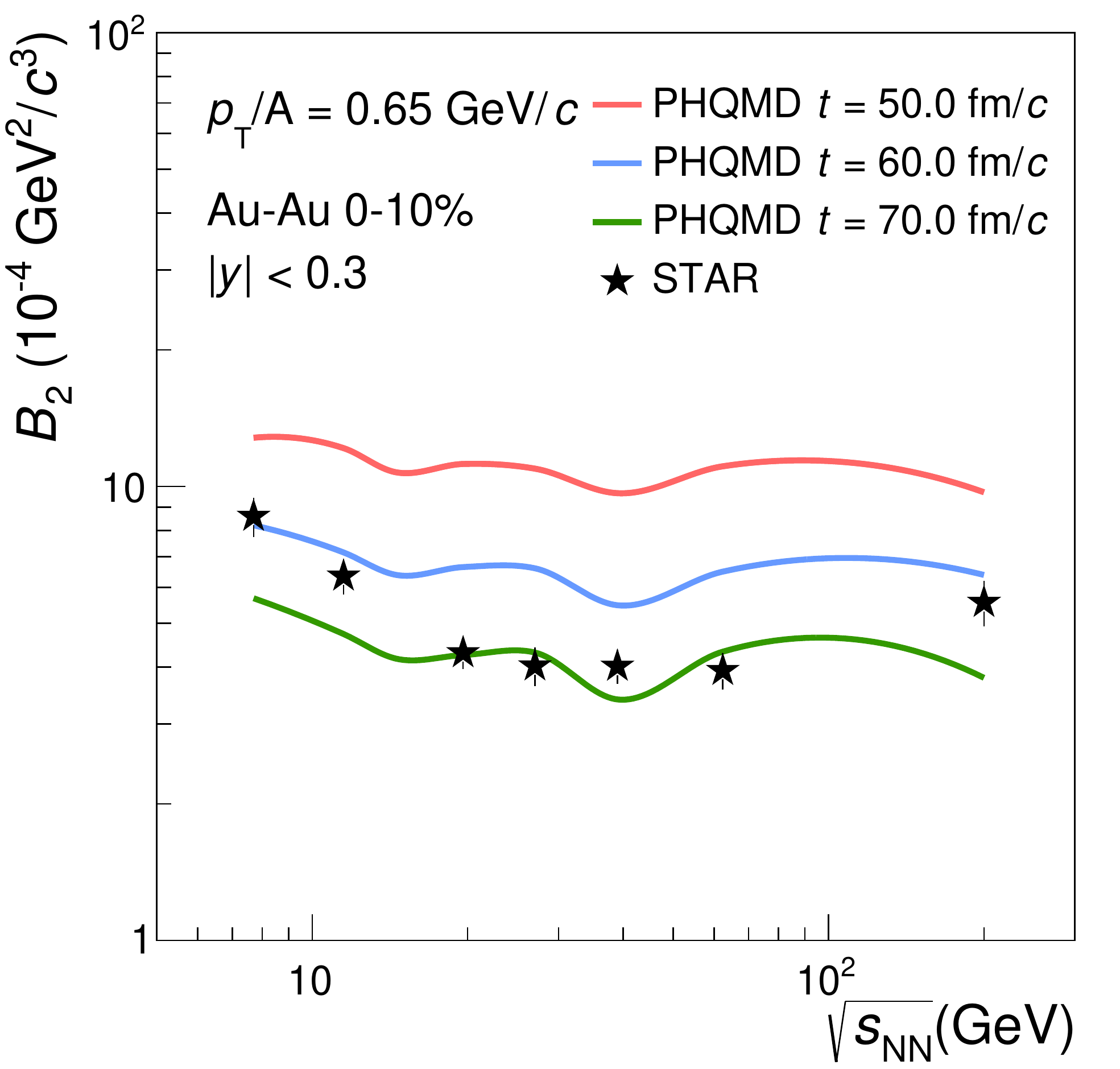}
\includegraphics[width=0.4\textwidth]{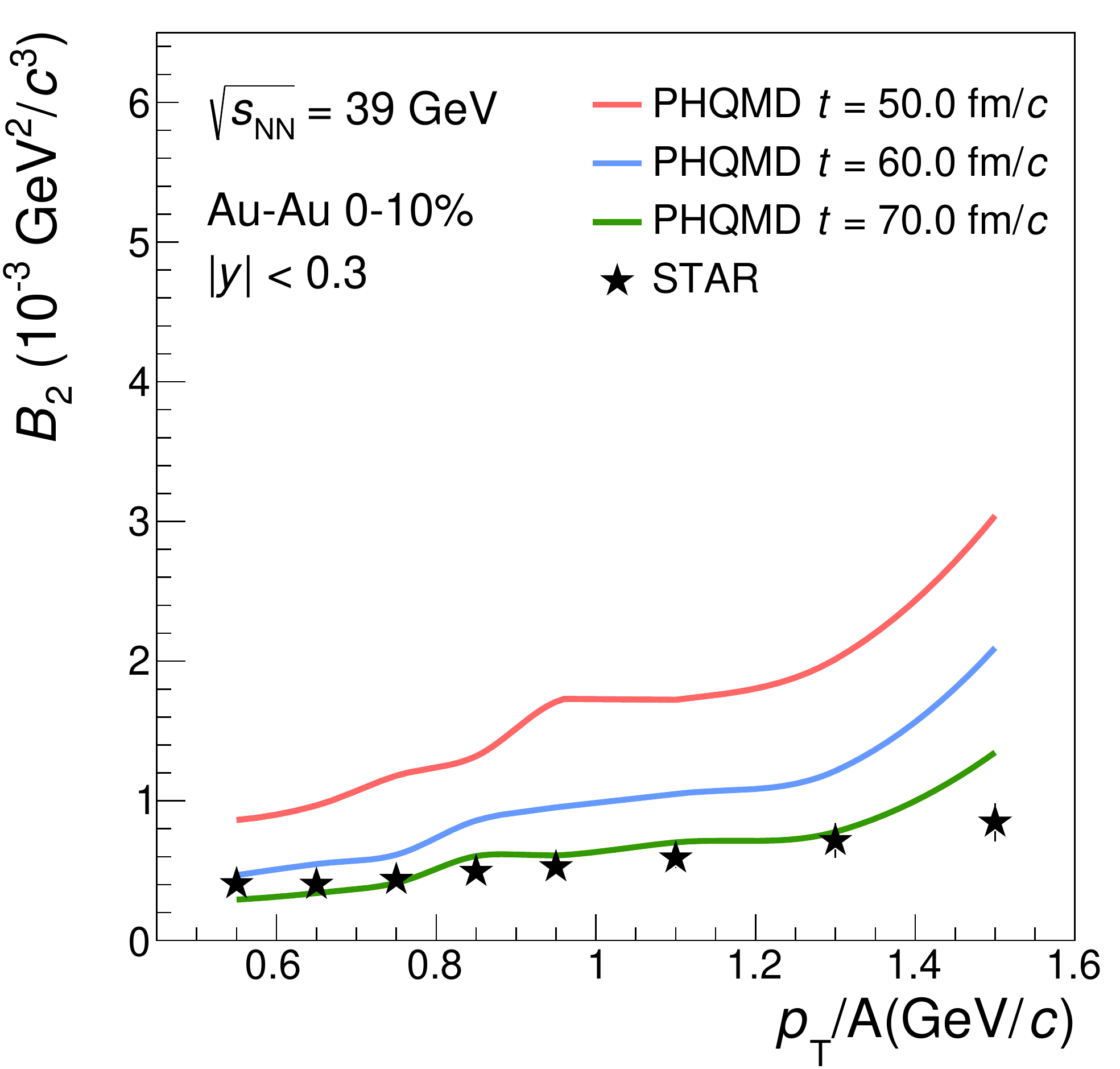}
\caption{\label{fig:B2} Top: The excitation function of $B_2$ for central Au+Au collisions versus the invariant energy $\sqrt{s_{NN}}$
for  $p_T/A = 0.65$~GeV/$c$. The stars indicate the experimental data from the STAR Collaboration \cite{Adam:2019wnb}. 
Bottom: $B_2$ as a function of $p_T$ for central Au+Au collisions at $\sqrt{s_{NN}} = 39$~GeV in the rapidity interval $|y|<0.3$. 
The STAR results \cite{Adam:2019wnb} are shown as star symbols.
The PHQMD results are taken at $t = 50$~fm/$c$ (red lines), 60~fm/$c$ (blue lines) and 70~fm/$c$ (green lines).}
\end{figure}

The knowledge of the $p_T$ spectra of deuterons allows for calculating the $B_2$ factor, \eq{eq:BA}. The STAR Collaboration has published $B_2$ as a function of $\sqrt{s_{NN}}$ for $P_T/A  = 0.65$~GeV/$c$ and $B_2$ as a function of $P_T/A$ for $\sqrt{s_{NN}} = 39$~GeV to which we can compare the PHQMD calculations. Both comparisons are displayed in Fig.~\ref{fig:B2}.


In the top part of Fig.~\ref{fig:B2} we show the excitation function of $B_2$ for central Au+Au collisions as a function of the center-of-mass energy, $\sqrt{s_{NN}}$, for $p_T/A = 0.65$~GeV/$c$. 
The dots indicate the experimental data from the STAR Collaboration \cite{Adam:2019wnb}. 
The lower part of Fig.~\ref{fig:B2} indicates $B_2$ as a function of $p_T$  for central Au+Au collisions at $\sqrt{s_{NN}} = 39$~GeV in the rapidity interval  $|y|<0.3$. 
The STAR results~\cite{Adam:2019wnb} are shown as stars. At 70~fm/$c$ we find a quite good agreement with the experimental results.

\begin{figure}
\centering
\includegraphics[width=0.4\textwidth]{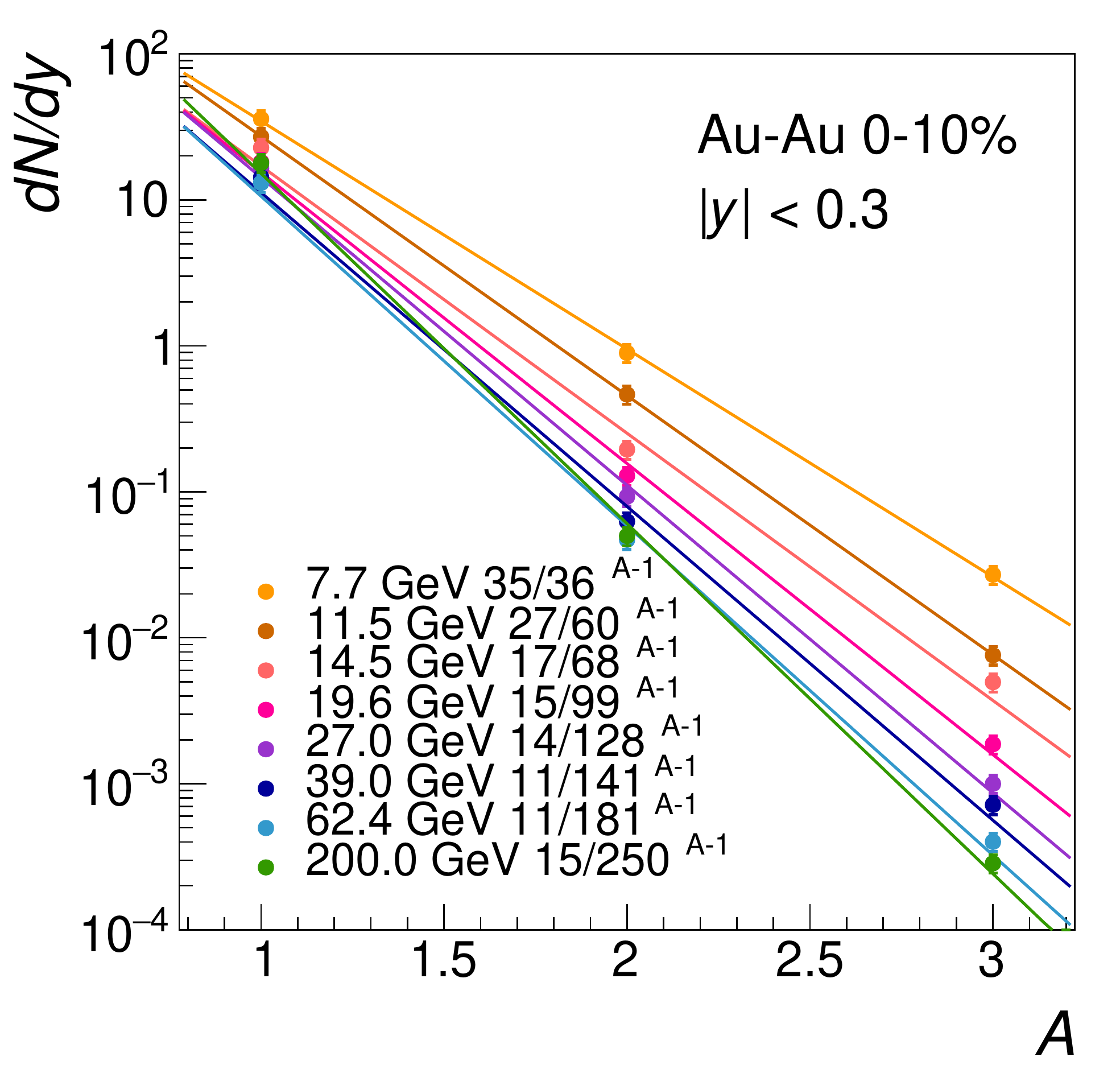}
\includegraphics[width=0.4\textwidth]{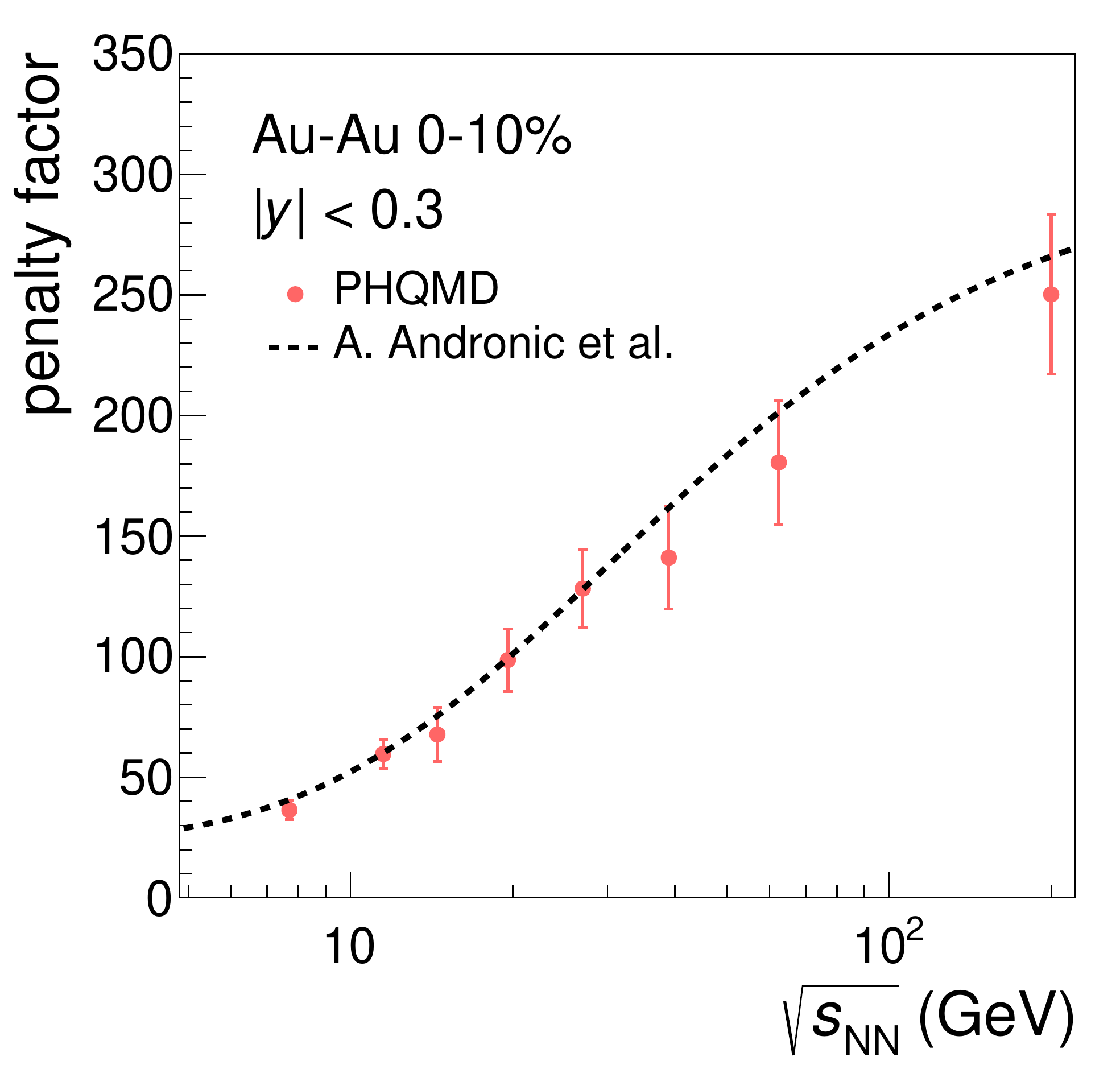}
\caption{\label{fig:STAR_penalty} Top: $dN/dy$ at midrapidity, $|y|<0.3$, as a function of the mass number $A$ ($p$, $d$, ${}^3$He) from central Au+Au collision at invariant energies  7.7~--~200~GeV. The solid lines represent the fit with an exponential function. Values of $dN/dy$ for $d$ were divided by the spin degeneracy factor 3/2. The times at which the PHQMD results were taken are chosen depending on the mass number and the collision energy, according to the results in Figs.~\ref{fig:WS}, \ref{fig:rat} and \ref{fig:th}.
Bottom: Penalty factor from the cluster yields at midrapidity in central Au+Au collisions versus $\sqrt{s_{NN}}$. The dots represent the PHQMD results for the selected times (see above). The dotted line indicates a determination of the penalty factor for cluster yields by the Boltzmann factor $\exp[-(m-\mu_B)/T]$ with $\mu_B$, $T$, and $m$ being the baryo-chemical potential, freeze-out temperature, and nucleon mass, respectively. $T$ and $\mu_B$ were calculated employing the parameterizations for their energy dependence from a thermal statistical model estimate, established in Ref. \cite{Andronic:2008gu}. 
}
\end{figure} 

In Fig.~\ref{fig:STAR_penalty} (top) we display $dN/dy$ at midrapidity ($|y|<0.3$) as a function of mass number $A$ ($p$, $d$, ${}^3$He) for central Au-Au collision from 7.7 to 200~GeV. The midrapdity $dN/dy$ of clusters of size $A$, produced by PHQMD, is an exponential function of the form $\frac{const}{P^{A-1}}$  for all energies covered by the RHIC-BES. The value of the penalty factor $P$ is displayed in the legend, together with the constant. The solid lines represent the fit with the exponential function. The values of $dN/dy$ for $d$ were divided by the spin degeneracy factor 3/2. The times at which the PHQMD results were taken were chosen depending on the mass number and on the collision energy according to the results in Figs.~\ref{fig:WS}, \ref{fig:rat} and \ref{fig:th}. Clearly, with increasing energy the exponential function becomes steeper.

We can compare the penalty factor obtained in the PHQMD calculations to the penalty factor determined from statistical model calculations. This is presented in Fig.~\ref{fig:STAR_penalty} (bottom) where we plot the penalty factor versus the center-of-mass energy $\sqrt{s_{NN}}$ for central Au+Au collisions. The full dots represent the PHQMD results for the selected times (see above). The dotted line indicates a determination of the penalty factor for cluster yields using the Boltzmann factor $\exp[-(m-\mu_B)/T]$ with $\mu_B$, $T$, and $m$ being the baryon chemical potential, freeze-out temperature, and nucleon mass, respectively. $T$ and $\mu_B$ were calculated employing the parameterizations of their energy dependence from a thermal statistical model estimate, established in \cite{Andronic:2008gu}. We observe that both models predict a very similar energy dependence of the penalty factor.
\section{Hypernuclei}

At the beam energies considered here hyperons are produced copiously in elementary hadron-hadron collisions or during the QGP hadronization. They have the possibility to become part of hypernuclei if there are other nucleons around with a momentum not too much different from the hyperon momentum. So hyper nuclei production studies the local phase space density of non-strange baryons, as well as the phase space density of the hyperons itself. This makes the study of hypernucleus production very interesting. Unfortunately, the presently available data is very scarce at these energies and therefore the possibility to compare our results to data is very limited.

\begin{figure}
\centering
  \includegraphics[width=0.4\textwidth]{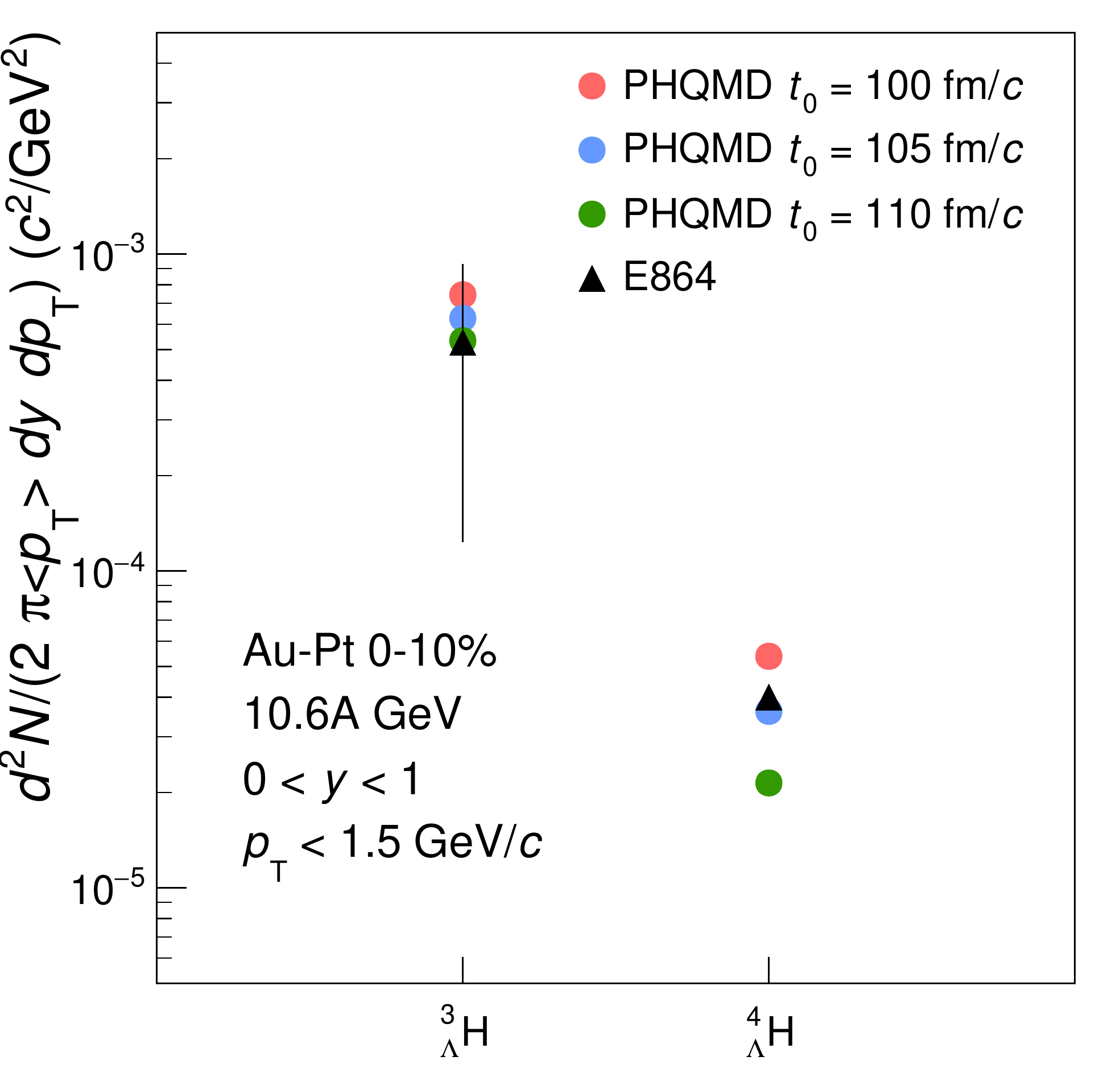}
\caption{\label{fig:hyper_10.6} 
Invariant yields of light hypernuclei  at $0 < y < 1$  and $p_T < 1.5$~GeV/$c$ in central Au+Pt collisions at the beam energy $E_{kin} = 10.6$~$A$GeV. 
The filled triangles indicate the experimental data from the E864 Collaboration \cite{Armstrong:2002xh}. The PHQMD results are taken at the physical time $t= t_0 \cosh(y)$ with $t_0 = 100$~fm/$c$ (red circles), 105~fm/$c$ (blue circles) and 110~fm/$c$ (green circles).}
\end{figure}

Until recently the only data set which exists has been measured by the E864 Collaboration \cite{Armstrong:2002xh} and is displayed in Fig.~\ref {fig:hyper_10.6} in comparison with PHQMD calculations. There the invariant yield of light hypernuclei at $0 < y < 1$  and $p_T < 1.5$~GeV/$c$ in central Au+Pt collisions at a beam energy of $E_{kin} = 10.6$~$A$GeV is shown. The filled triangles indicate the experimental data. The PHQMD results are taken at the physical time $t= t_0 \cosh(y)$ with $t_0 = 100$~fm/$c$ (red circles), 105~fm/$c$ (blue circles) and 110~fm/$c$ (green circles). The large rapidity interval does not allow
to determine exactly the physical time and introduces uncertainties
into the comparison of the PHQMD predictions with data.
This result shows, however, that PHQMD can also be used to study hypernuclei, whose production is one of the major goals of the upcoming experiments at NICA and FAIR.

Recently, first preliminary data of hypernuclei production at a center-of-mass energy of $\sqrt{s_{NN}}$ = 3~GeV were made public. In their fixed target program the STAR collaboration measured for two hypernuclei, ${}^3_\Lambda$H and ${}^4_\Lambda$H, the transverse momentum distribution $d^2N/p_T dp_T$ in different rapidity bins \cite{Abdallah:2021ab}. 

\begin{figure}
\centering
    \includegraphics[width=0.4\textwidth]{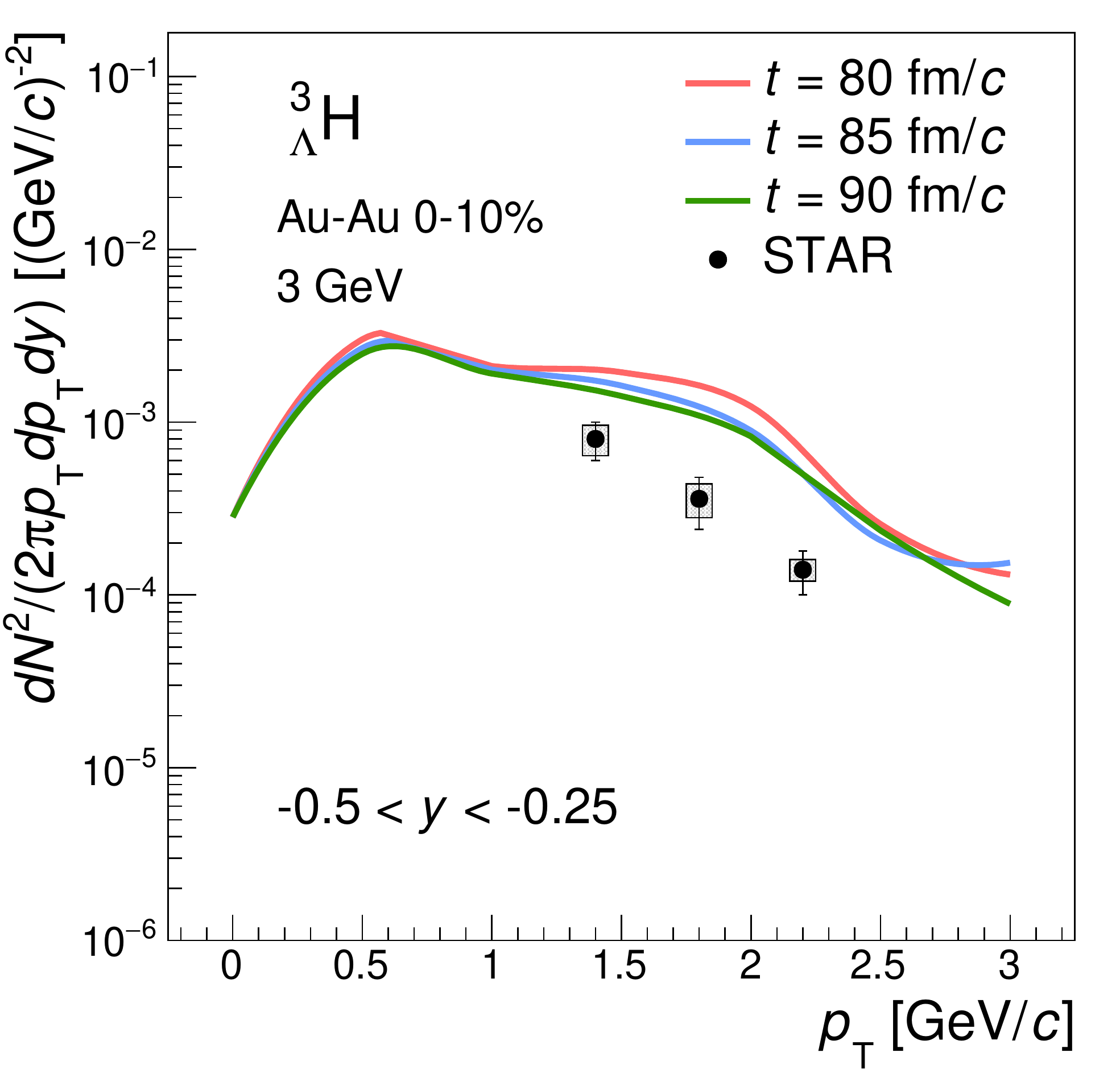}
    \includegraphics[width=0.4\textwidth]{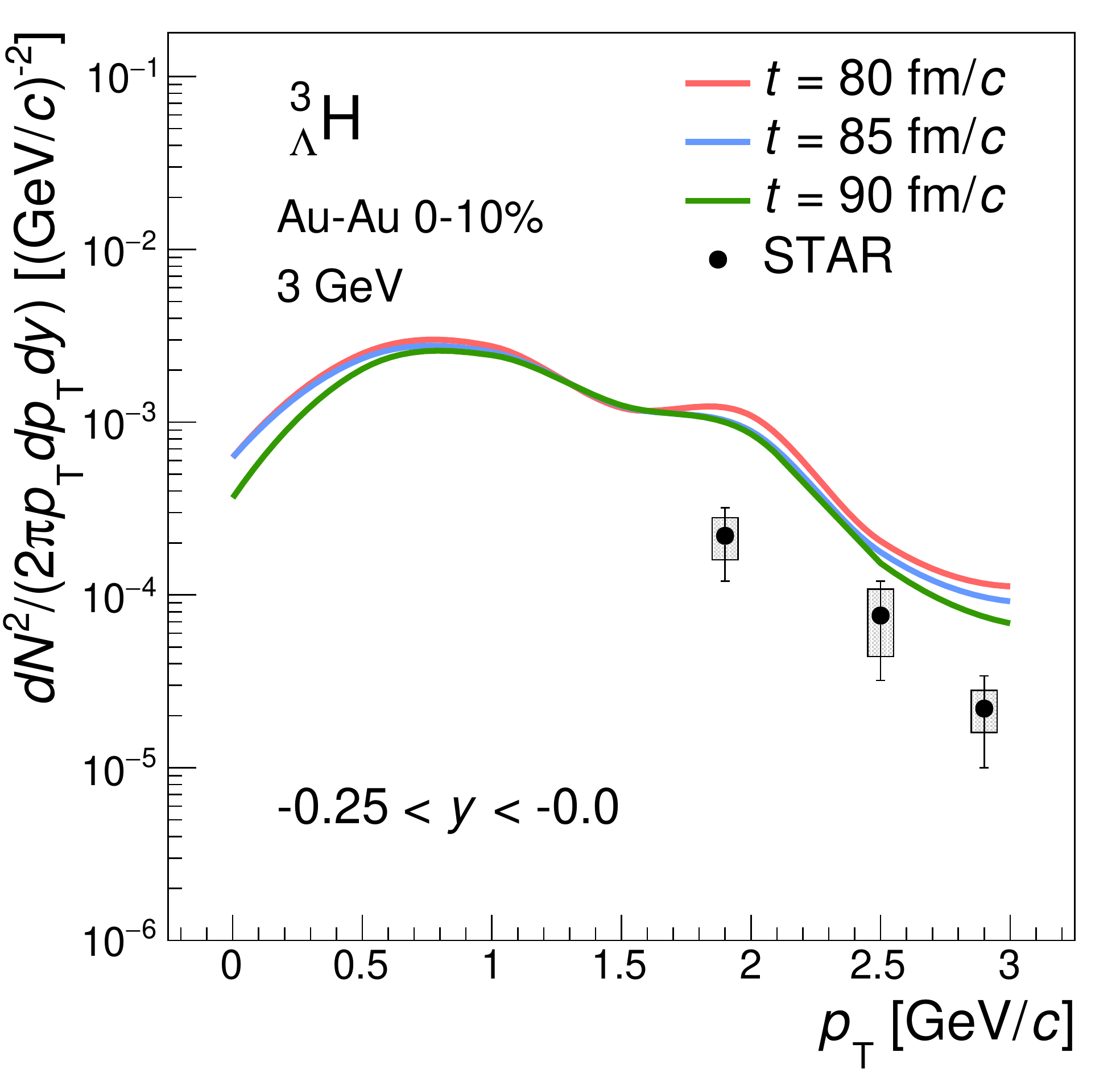}
\caption{\label{fig:H3L_pT_3} 
Transverse momentum distribution of $^{3}_{\Lambda}$H for different rapidity intervals as indicated in the legends in central Au+Au collisions at $\sqrt{s_{NN}}= 3 $~GeV. 
The filled circles indicate the preliminary experimental data from the STAR Collaboration \cite{Abdallah:2021ab}. The PHQMD results are taken at the times $t = 80$~fm/$c$ (red lines), 85~fm/$c$ (blue lines) and 90~fm/$c$ (green lines).}
\end{figure}

\begin{figure}[h!]
    \includegraphics[width=0.4\textwidth]{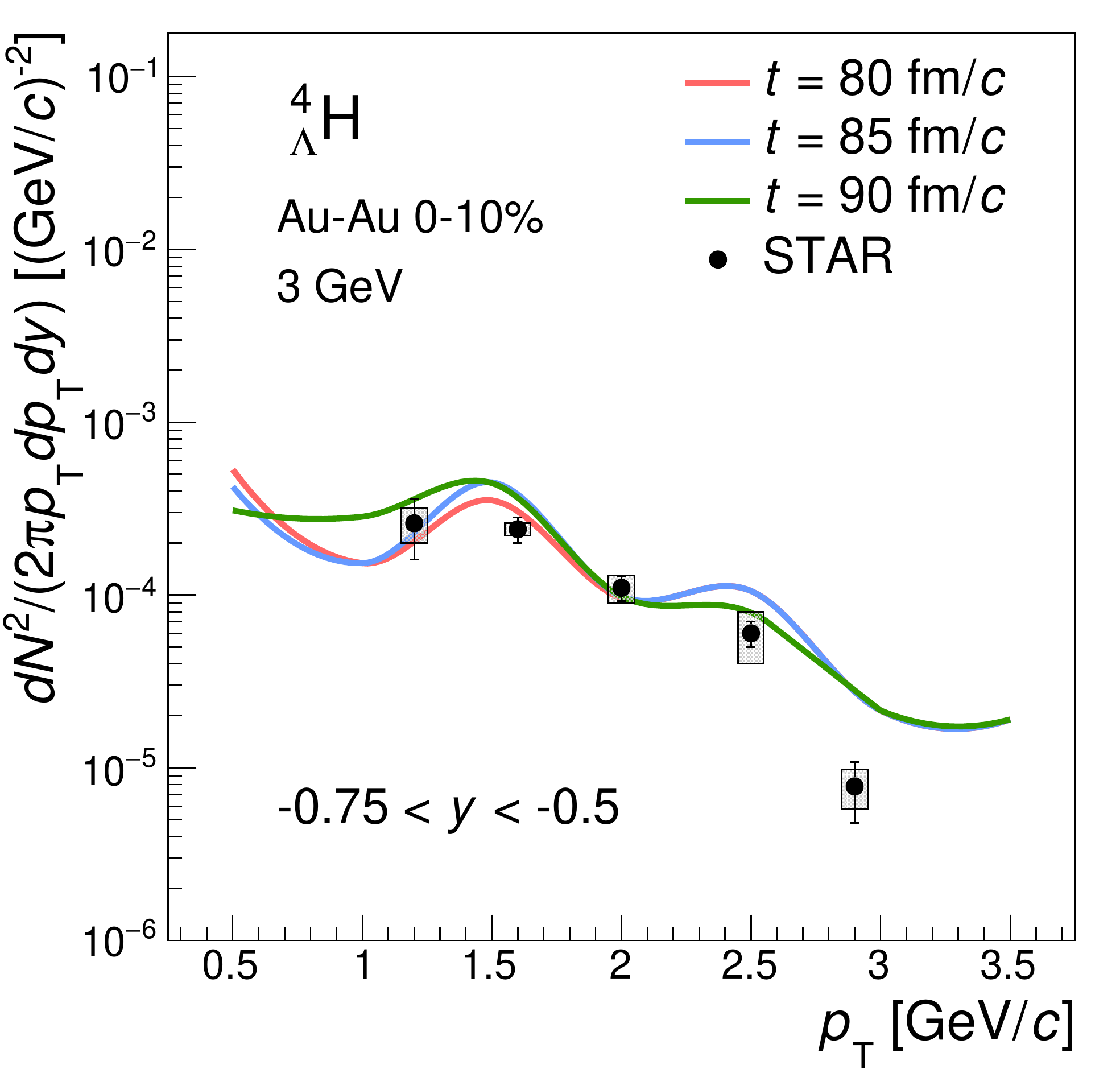}
    \includegraphics[width=0.4\textwidth]{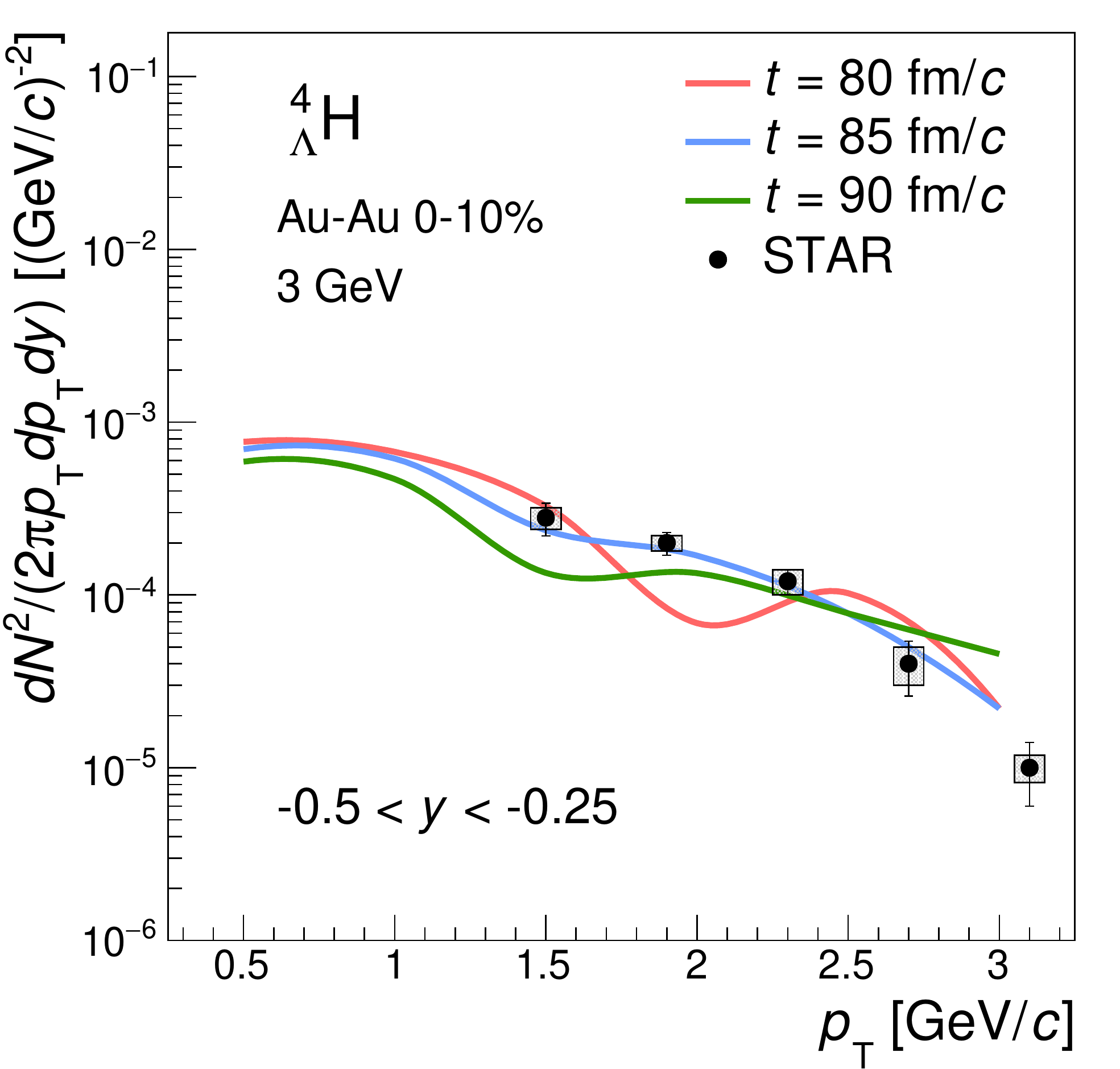}
   \includegraphics[width=0.4\textwidth]{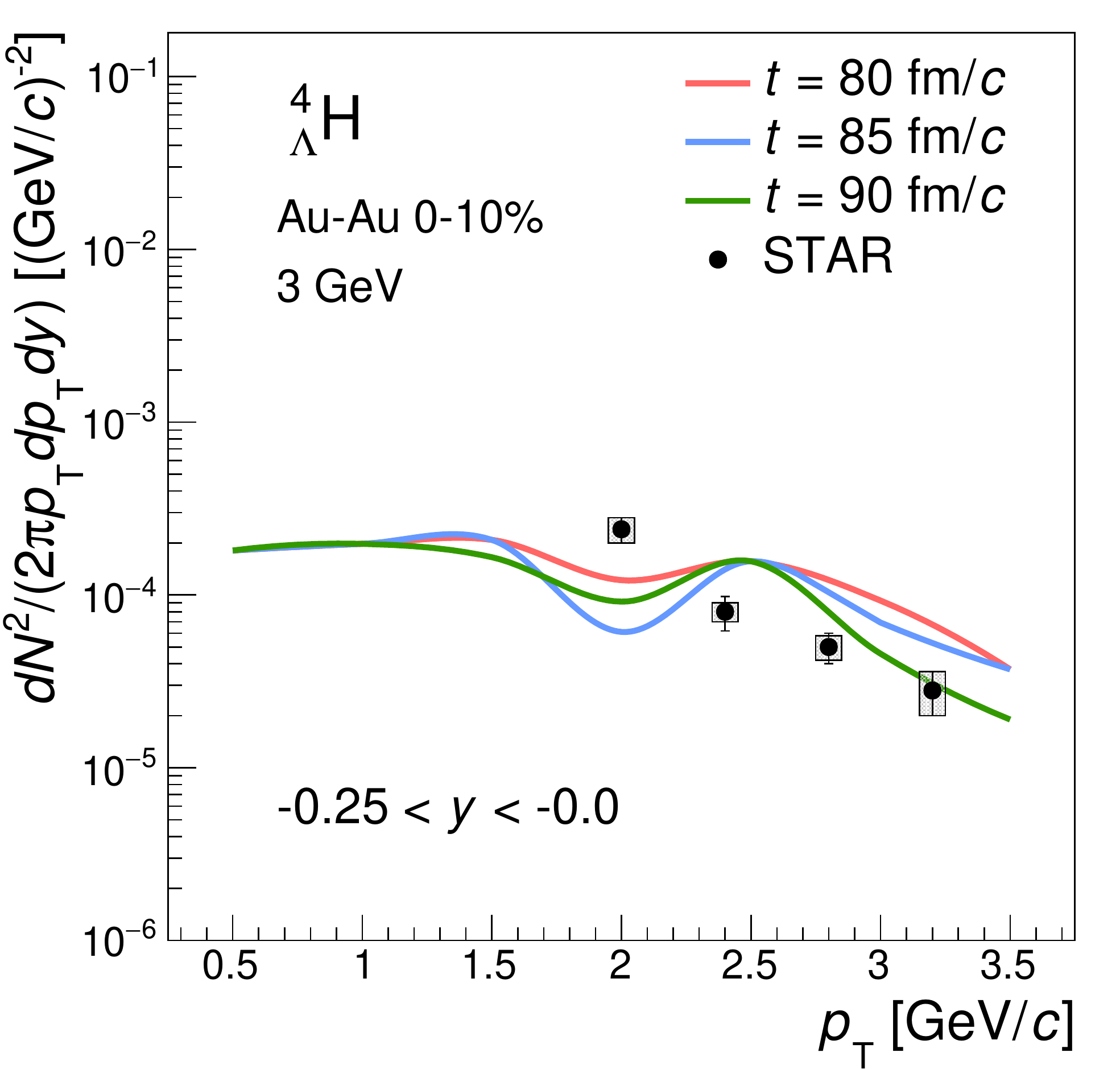}
\caption{\label{fig:H4L_pT_3} 
Transverse momentum distribution of $^{4}_{\Lambda}$H for different rapidity intervals as indicated in the legends in central Au+Au collisions at $\sqrt{s_{NN}}= 3 $~GeV. 
The filled circles indicate the preliminary experimental data from the STAR Collaboration \cite{Abdallah:2021ab}. The PHQMD results are taken at the times $t = 80$~fm/$c$ (red lines), 85~fm/$c$ (blue lines) and 90~fm/$c$ (green lines).}
\end{figure}

Figures~\ref{fig:H3L_pT_3} and \ref{fig:H4L_pT_3} display the comparison of the PHQMD transverse momentum distribution with the preliminary experimental data from the STAR Collaboration. In Fig.~\ref{fig:H3L_pT_3} we show the results for ${}^3_\Lambda$H, while in Fig.~\ref{fig:H4L_pT_3} those for ${}^4_\Lambda$H for different rapidity bins are presented. The filled circles indicate the experimental data from the STAR Collaboration \cite{Abdallah:2021ab}. The PHQMD results are taken at the times $t = 80$~fm/$c$ (red lines), 85~fm/$c$ (blue lines) and 90~fm/$c$ (green lines), but at this low energy the time dependence of the cluster yield is weak. The calculations show that the trend of the experimental $p_T$ spectra is well reproduced. We stress that the $\Lambda N$ potential implemented in PHQMD is presently quite simple ($=2/3 \, V_{NN}$), so we do not expect an exact quantitative agreement.  We overpredict the differential yield of ${}^3_\Lambda H$, but reproduce the yield of ${}^4_\Lambda H$. This demonstrates that PHQMD in its present version is a good starting point for more sophisticated studies of the hypernucleus production in this energy regime. 
We will address the origin of the different quality of agreement for ${}^{3}_{\Lambda}$H and for ${}^{4}_{\Lambda}$H in a future study.

\begin{figure}[h!]
    \includegraphics[scale=0.4]{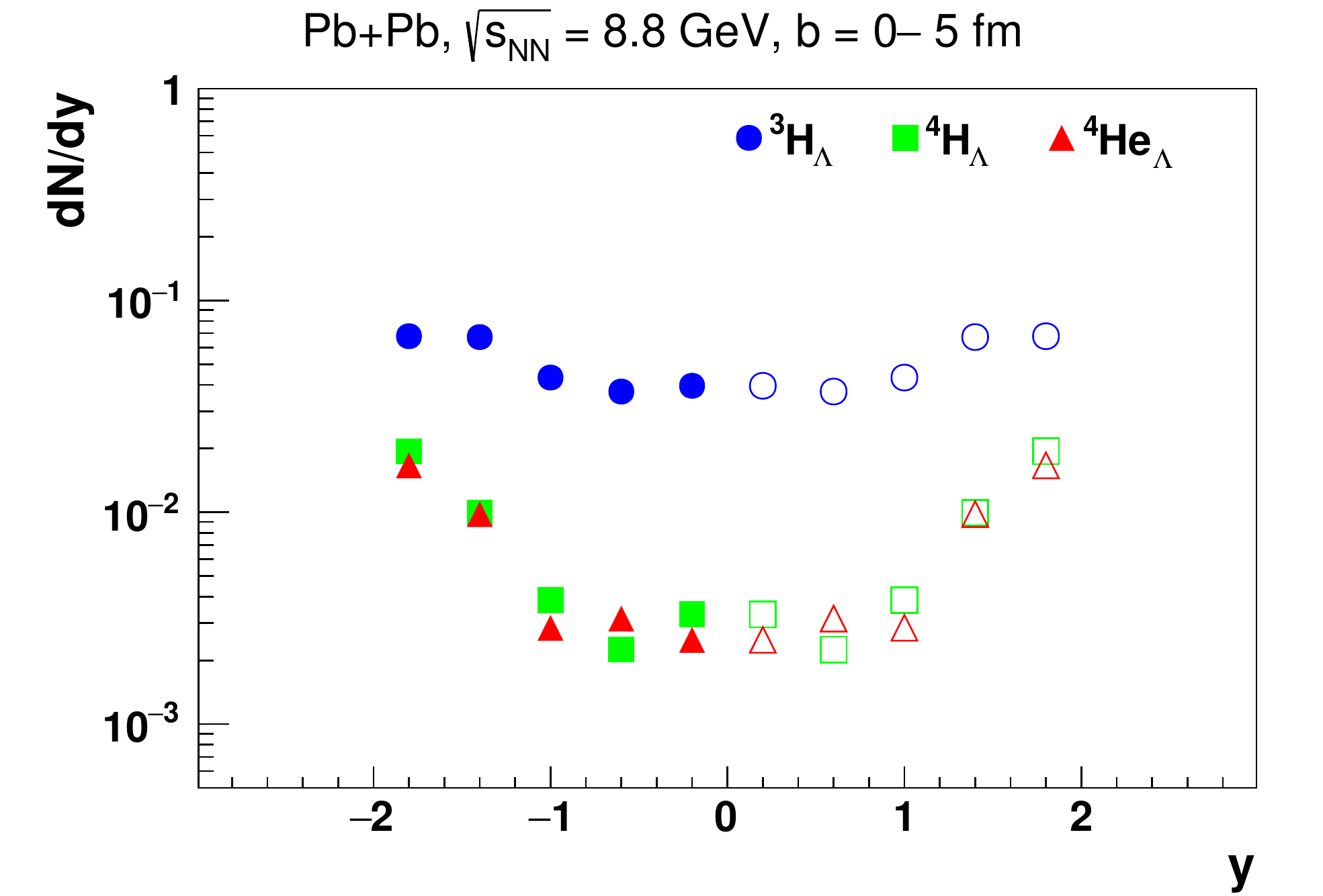}
\caption{\label{fig:hyper_dndy_8.8_cosh} 
The rapidity distribution of ${}^3_\Lambda$H, ${}^4_\Lambda$H and ${}^4_\Lambda$He from central Pb+Pb collisions at $\sqrt{s_{NN}} = 8.8$~GeV calculated at the physical time $t= t_0 \cosh(y)$ for $t_0 = 53$~fm/$c$.}
\end{figure}

In Fig.~\ref{fig:hyper_dndy_8.8_cosh} we display the rapidity distribution of our calculations for ${}^3_\Lambda$H (blue dots), ${}^4_\Lambda$H (green squares) and ${}^4_\Lambda$He (red triangles)
for central collisions at $\sqrt{s_{NN}}= 8.8$~GeV, taken at 53~fm/$c$. We see a distinct mass dependence. The heavier the particle, the more difficult it is to form it at midrapidity as is expected from the coalescence ansatz (even if the coalescence model misses important features of the spectra). When moving towards projectile and target rapidity clusters from an additional source will become more important. These include hypernuclei originating from the combination of projectile and target fragments with produced hyperons. As a consequence the mass dependence of the multiplicities is being reduced when moving away from midrapidity.

\begin{figure}[h!]
    \includegraphics[scale=0.3]{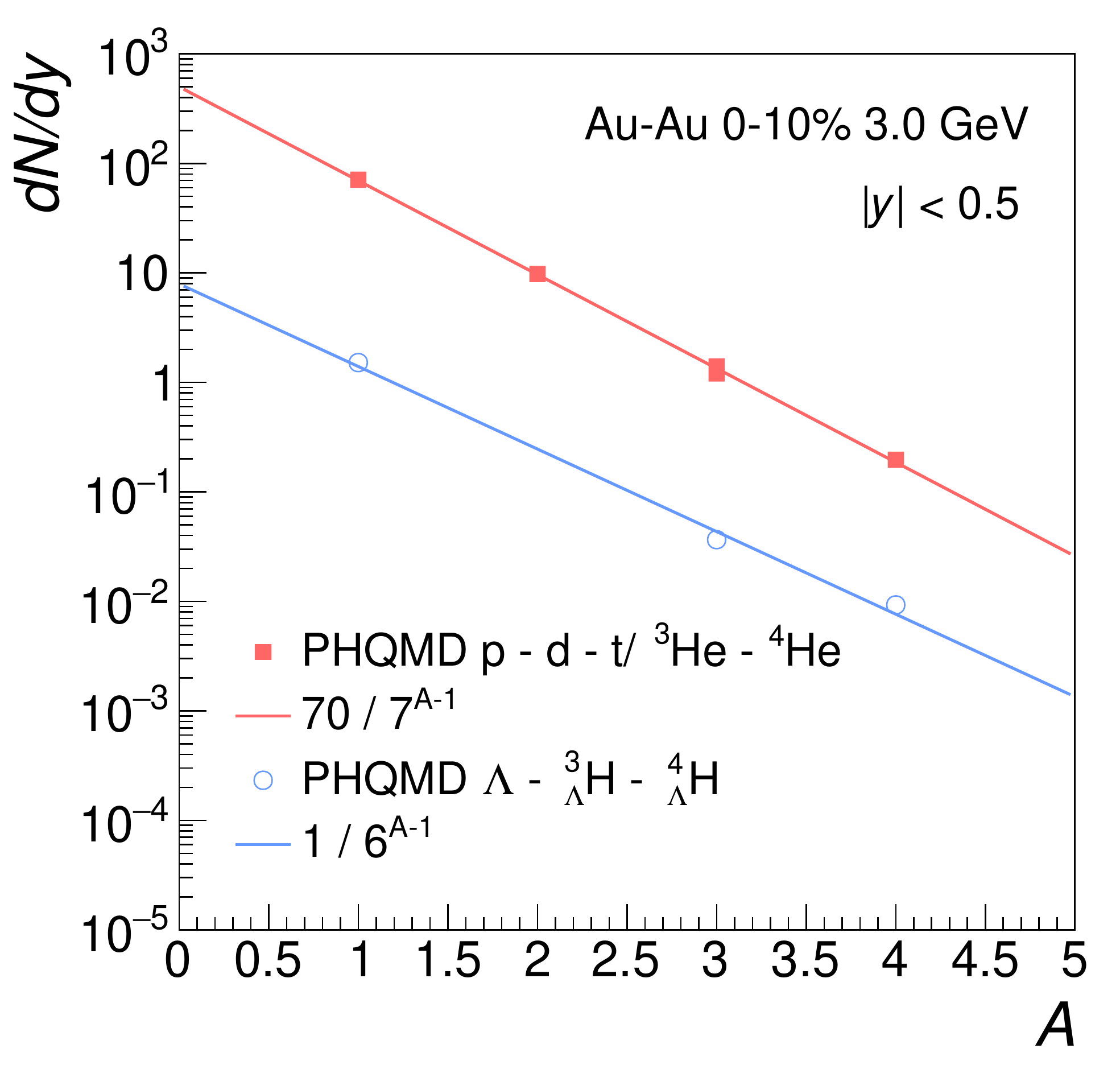}
    \includegraphics[scale=0.3]{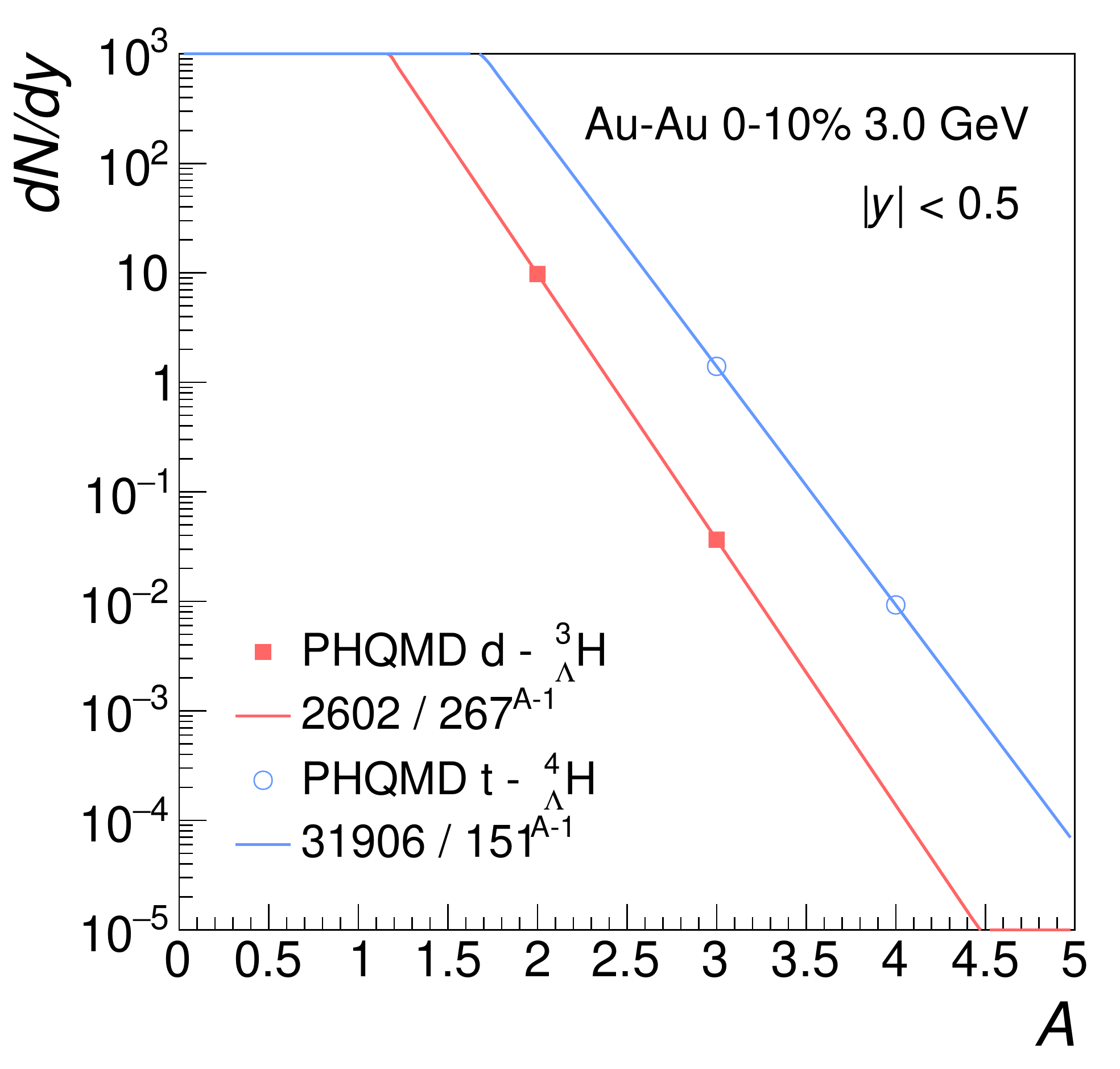}
\caption{\label{fig:penaltyfactor_hype_3GeV} 
$dN/dy$ at $|y|<0.5$ versus the cluster size $A$ for central Au+Au collisions at $\sqrt{s_{NN}} = 3$~GeV. The penalty factors $P$ for the production of hypernuclei,  
extracted from an exponential fit $dN/dy = const/P^{A-1}$, are indicated  
i) (top)  for adding additional nucleons to a proton (red) and to a hyperon (blue), respectively; ii) (bottom)  for adding an additional $\Lambda$ to a deuteron (red) and
to a triton (blue), respectively. Nuclei with $A = 2$ are taken at $t = 50$~fm/$c$, nuclei with $A = 3$ at $t = 60$~fm/$c$, nuclei with $A = 4$ and hypernuclei at $t = 70$~fm/$c$.}
\end{figure}
\begin{figure}[h!]
    \includegraphics[scale=0.3]{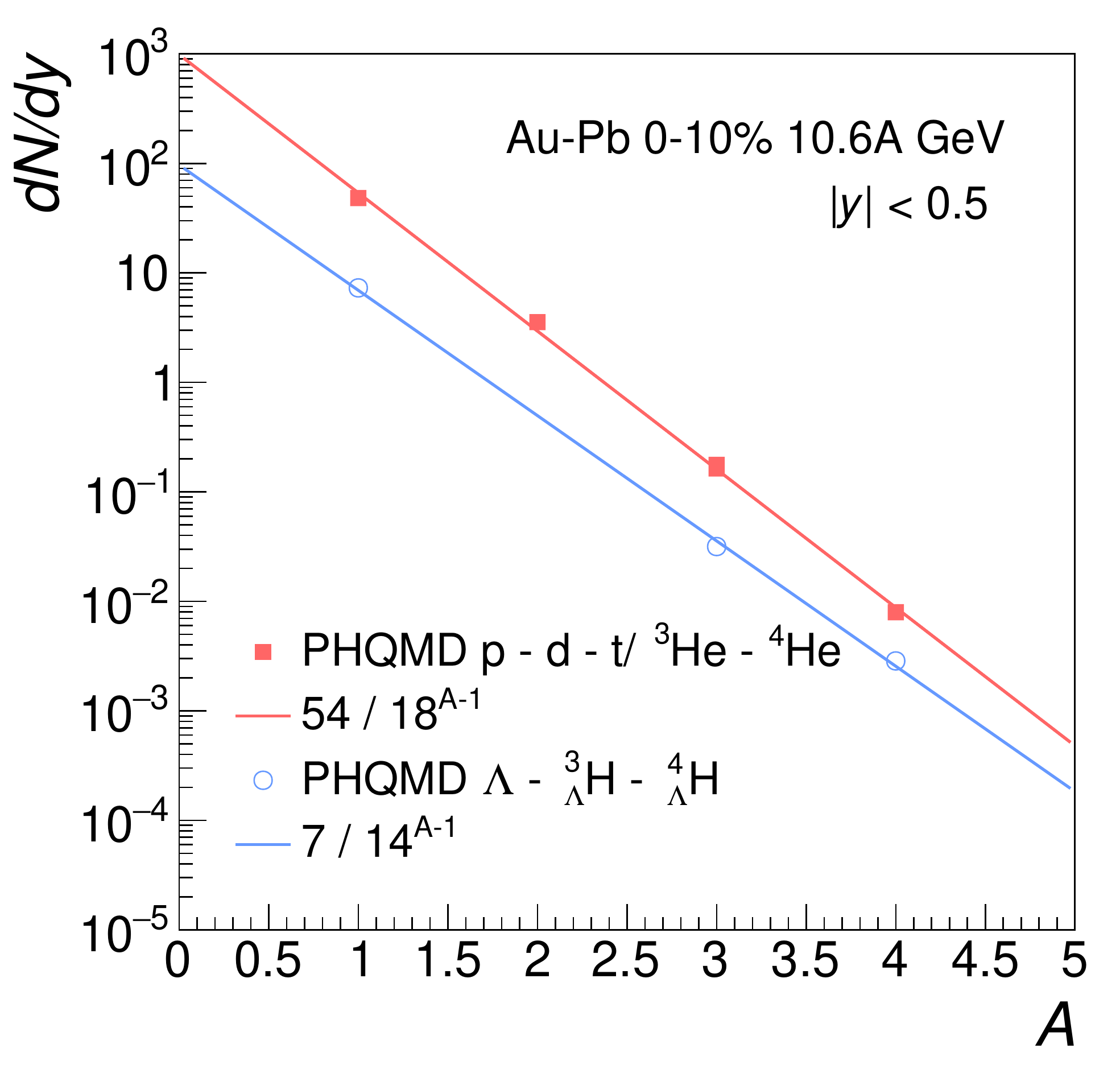}
    \includegraphics[scale=0.3]{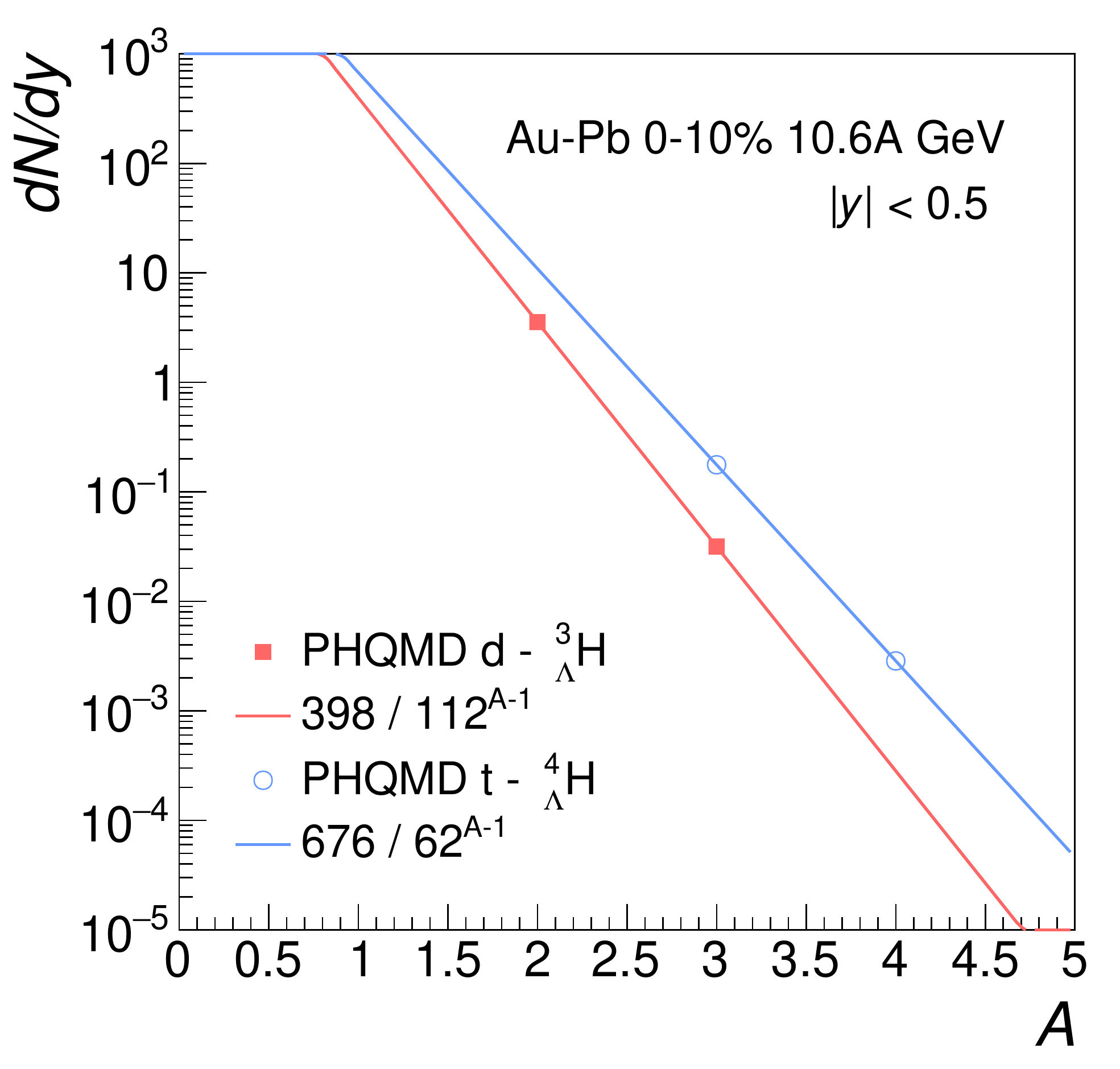}
\caption{\label{fig:penaltyfactor_hype_10.6AGeV} 
$dN/dy$ at $|y|<0.5$ versus the cluster size $A$ as well as the penalty factors for the production of (hyper)nuclei for central Au+Pb collisions at  $E_{kin} = 10.6$~$A$GeV.
For details see the caption of Fig.~\ref{fig:penaltyfactor_hype_3GeV}.
}
\end{figure}

Similar to the non-strange clusters we can also calculate for the hypernuclei a
penalty factor $P$ by $dN/dy = const /P^{A-1}$ and compare the both. This is done
in Figs.~\ref{fig:penaltyfactor_hype_3GeV} and \ref{fig:penaltyfactor_hype_10.6AGeV} for $\sqrt{s_{NN}} = 3$~GeV and $E_{kin} = 10.6$~$A$GeV, the energy of the fixed target STAR experiment and the AGS experiment. 
On the top we see the penalty factor for adding a nucleon, on the bottom that for adding a $\Lambda$. 
If one starts with a proton, the penalty factor for adding a nucleon is larger in comparison to the chain which starts with a $\Lambda$. The penalty factor for adding a $\Lambda$ depends of course on the $\Lambda$ multiplicity and is therefore very large at  $\sqrt{s_{NN}}= 3$~GeV where the probability that 2 $\Lambda$'s are produced in 
a heavy-ion reaction is small. It becomes considerably lower at $E_{kin} = 10.6$~$A$GeV. A more realistic implementation of the $\Lambda N$ potential, planned for the future, will modify the detail but we expect that the 
general trend will be conserved.

\section{When are the clusters formed?}

As demonstrated above,  the experimental observables - such as multiplicities, rapidity and transverse momentum distributions of clusters - are reasonably well reproduced within the PHQMD approach. However, based only on the final observables it is difficult to answer the question: when are the observed clusters formed? In order to shed light on this issue we explore the advantages of a microscopic transport description of heavy-ion dynamics, which allows to follow the time evolution of each particle and thus to study the origin of cluster formation and their stability over time.
Using the recorded history of all baryons in the system, we can trace back the time history of the baryons (nucleons and hyperons) which are finally embedded in the final clusters.
In this Section we use this information to study when and how  clusters,  observed finally at midrapidity, are formed during the fireball expansion. In contradistinction to clusters, which are observed at projectile and target rapidity and present surviving initial correlations of nucleons in the spectator matter, the clusters at midrapidity are newly formed by participants during the reaction, i.e. they contain a mixture of baryons from projectile and target. 

\begin{figure}
\centering
\includegraphics[scale=0.4]{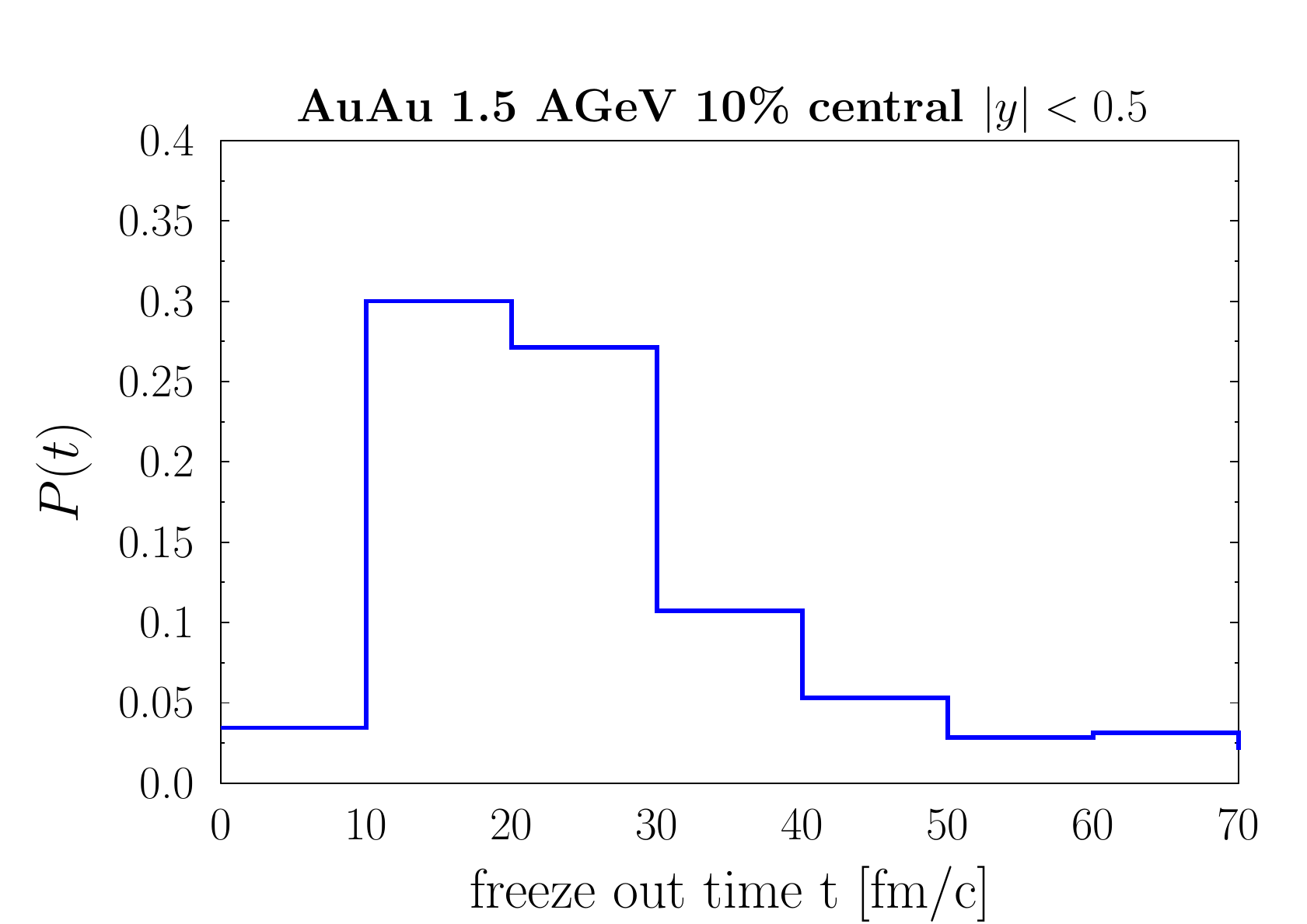}
\includegraphics[scale=0.4]{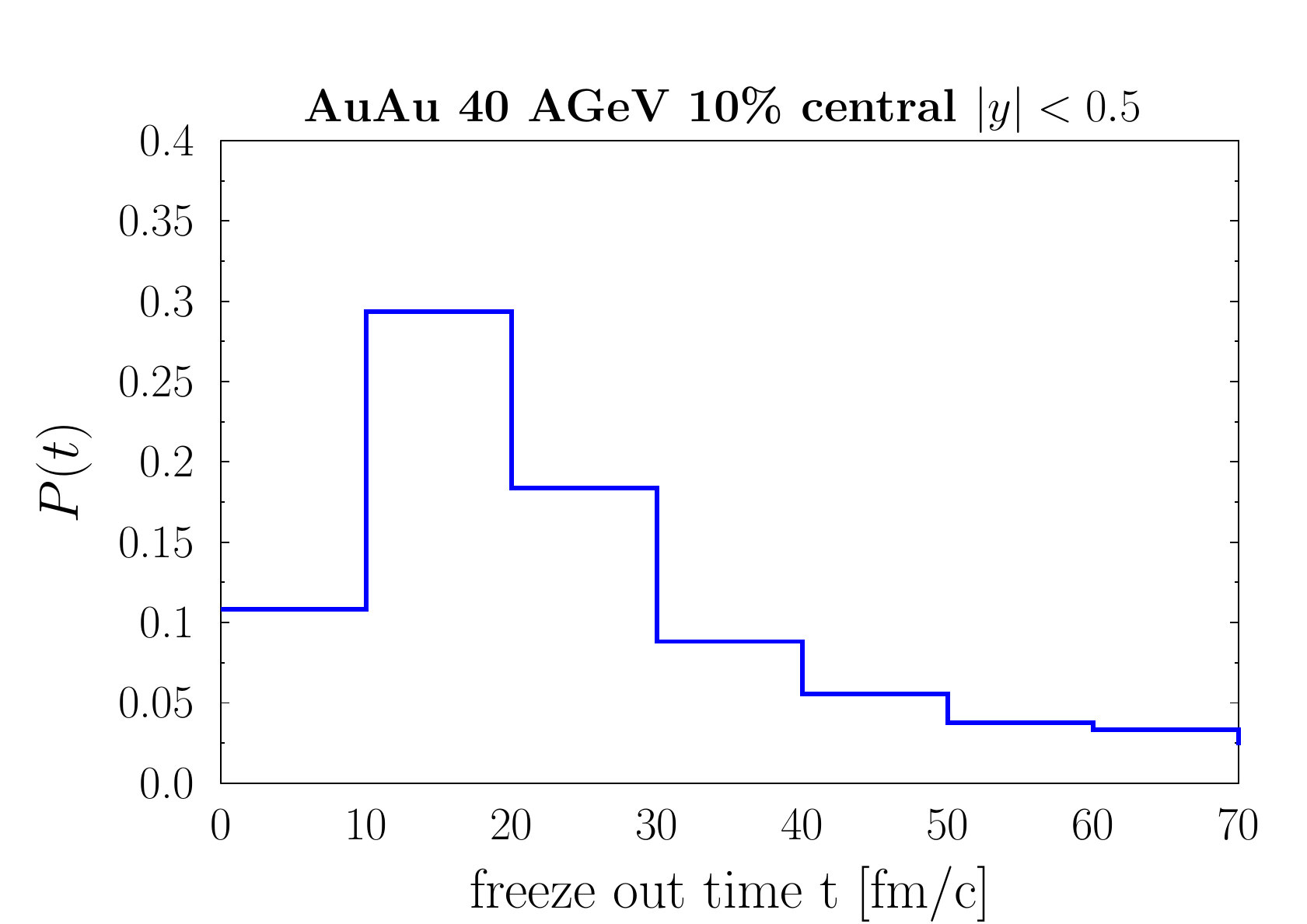}
\caption{\label{fig:frz} The normalized distribution of freeze-out times of baryons (nucleons and hyperons), i.e. the time of the last hadronic collision of the baryons, observed at midrapidity, $|y|<0.5$, for 10\% central Au+Au collisions at beam energies of $E_{kin} = 1.5$~$A$GeV (top) and of $E_{kin} = 40$~$A$GeV (bottom).}
\end{figure}

We start out with the investigation of the freeze-out time distribution of final baryons, i.e. the time at which the baryons made their last collision with the surrounding hadrons. This distribution for baryons with a rapidity $|y|<0.5$, normalized to one, is shown in Fig.~\ref{fig:frz} for 10\% central Au+Au collisions at beam energies of $E_{kin}=1.5$~$A$GeV (upper plot) and  $E_{kin}= 40$~$A$GeV (lower plot). On can see a striking similarity between the two plots which evidences that the expansion velocity of the midrapidity fireball is not very different at the different beam energies. Without showing explicitly, we mention that even for the highest energy studied here,
$\sqrt{s_{NN}} = 200$~GeV, the distribution is very similar to other energies. This result is compatible with the observation that above $E_{kin}=1.5$~$A$GeV the average transverse momentum of the midrapidity baryons does not change substantially with increasing beam energy. It can as well be seen in Fig.~\ref{fig:frz} that after 40~fm/$c$ the collisions among the expanding hadrons are essentially over and only potential interactions continue to be present. 

\begin{figure}
\centering
\includegraphics[scale=0.45]{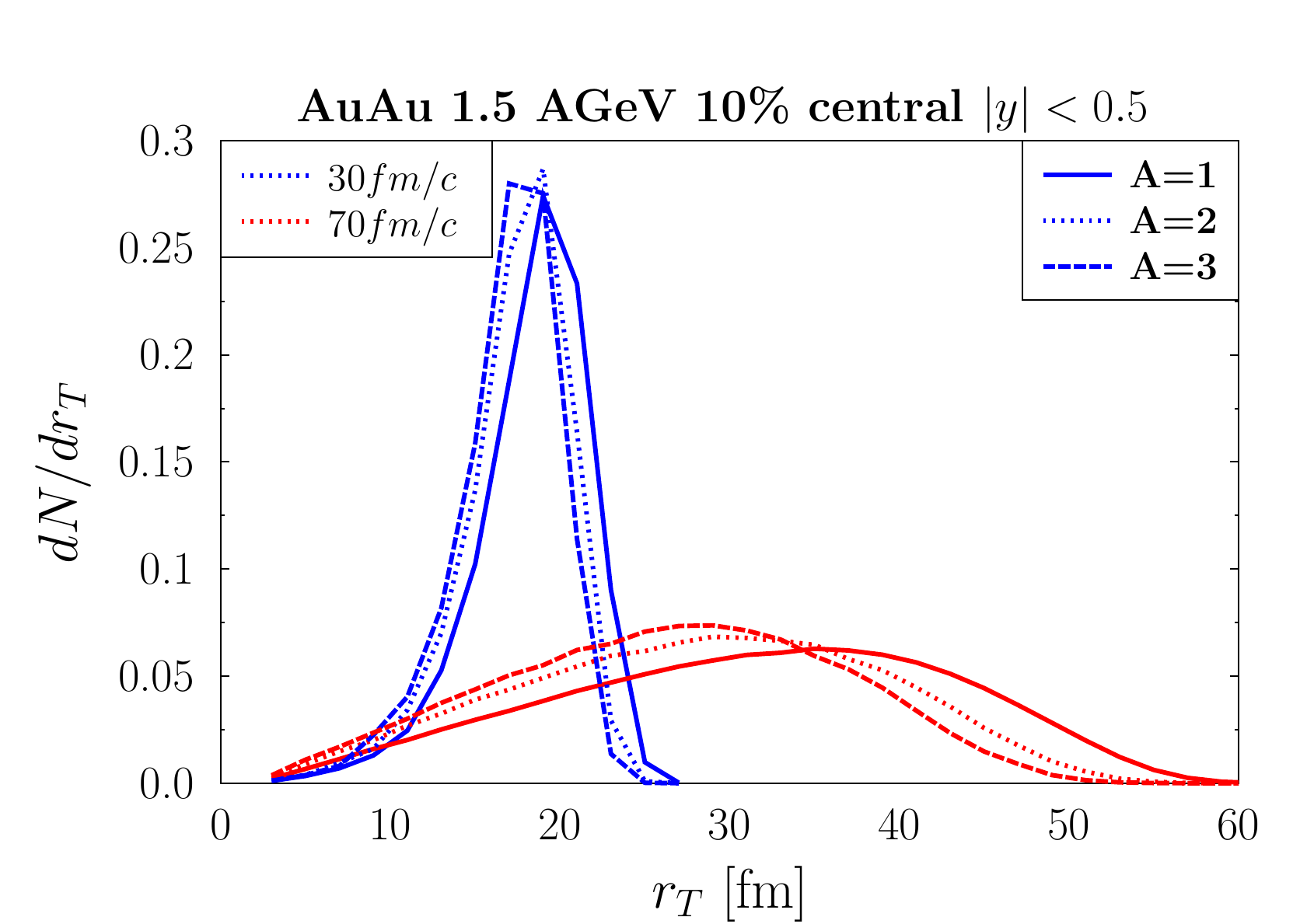}
\includegraphics[scale=0.45]{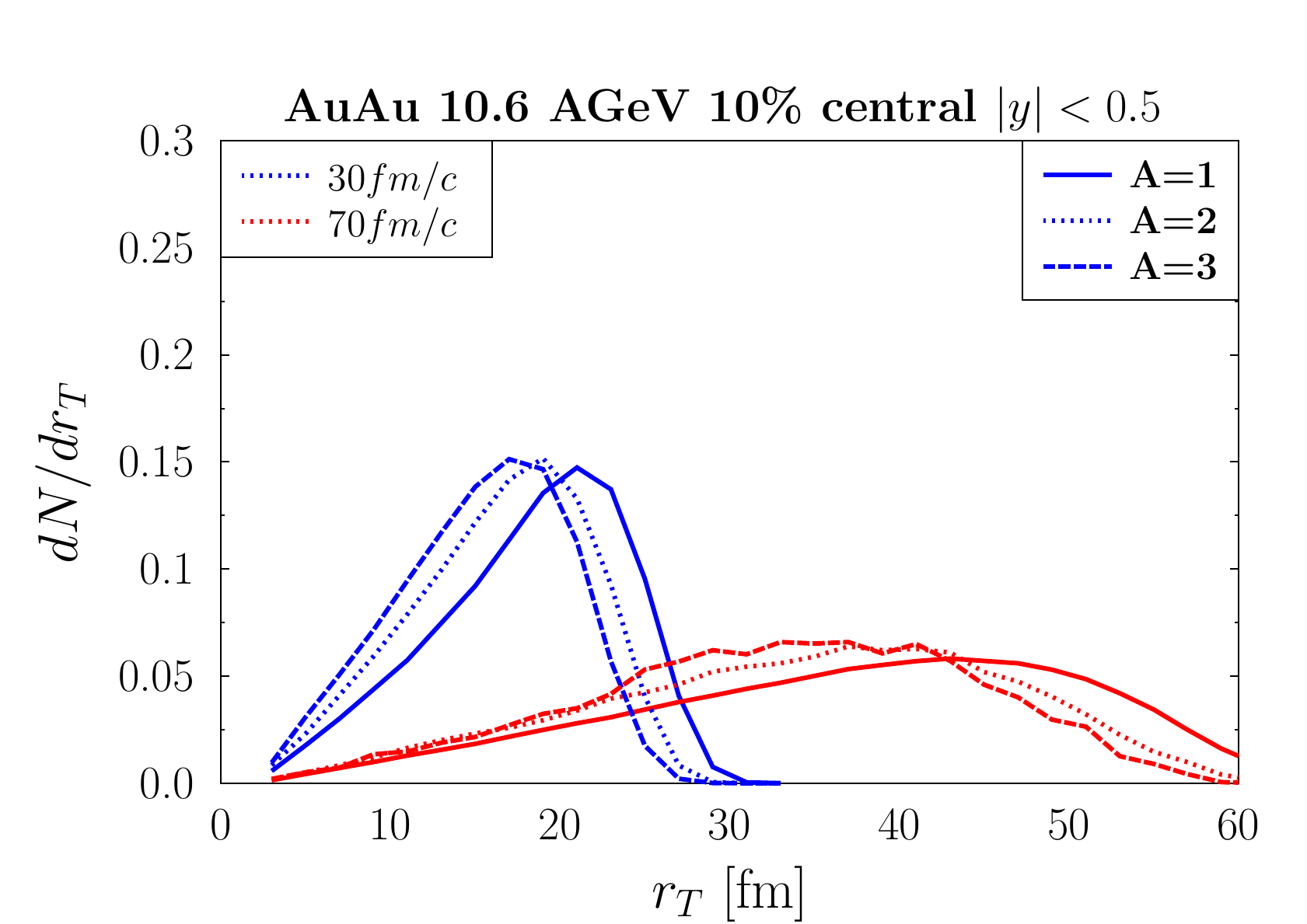}
\includegraphics[scale=0.45]{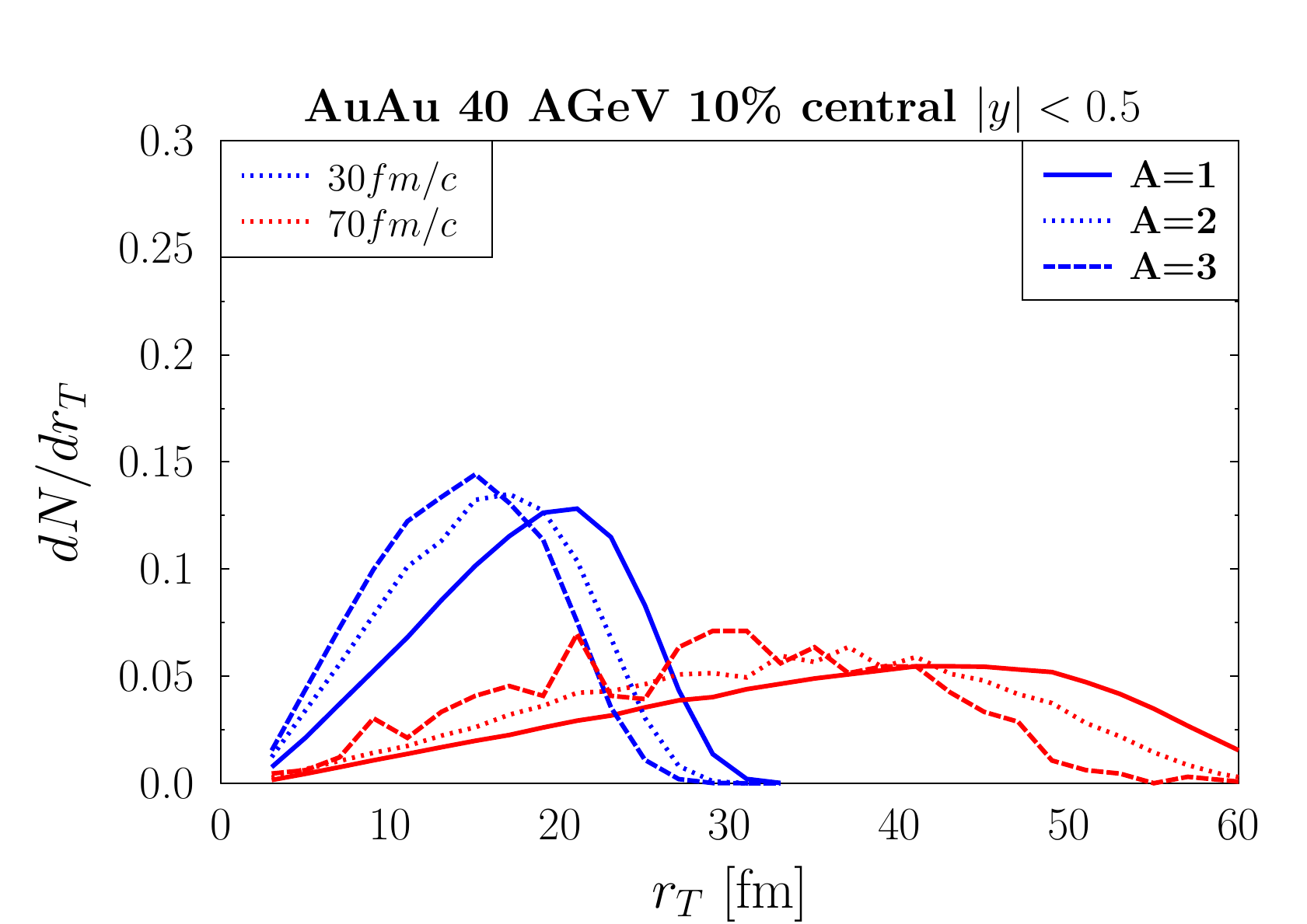}
\caption{\label{fig:RT} The normalized distribution of the transverse distance of the nucleons, observed at midrapidity ($|y|<0.5$), to the reaction center at time 30~fm/$c$ (blue lines) and 70~fm/$c$ (red lines) for for 10\% central Au+Au collisions at beam energy $E_{kin} = 1.5$~$A$GeV (top), $E_{kin} = 10$~$A$GeV (middle) and $E_{kin} = 40$~$A$GeV (bottom).  The dotted lines display the distributions of the nucleons which belong to the clusters of size $A=2$, the dashed lines to $A=3$ clusters and the solid lines indicate the distributions of the free nucleons $A=1$.}
\end{figure}

Next, we investigate the transverse distance of nucleons, observed at midrapidity ($|y|<0.5$), to the center of the heavy-ion reaction  (defined in the calculational frame which is the initial NN center-of-mass frame), i.e relative to the center of the fireball. 
In Fig.~\ref{fig:RT} we show the time evolution of the distributions (normalized to unity) of the transverse distance 
of the baryons which belong to clusters of size $A=2$ (dotted lines) and $A=3$ (dashed lines),
as well as for free nucleons $A=1$ (solid lines) for 10\% central Au+Au collisions 
at beam energies of $E_{kin} = 1.5$~$A$GeV (top), $E_{kin} = 10$~$A$GeV (middle) and $E_{kin} = 40$~$A$GeV (bottom).
The blue lines display the distributions calculated at 30~fm/$c$ and the red lines that at 70~fm/$c$.
The similarity of these distributions is again striking and points to a very similar mechanism for cluster production for all the energies investigated here. Clusters are formed from nucleons which have a smaller distance to the reaction center than the average and, even more, the larger the cluster, the smaller is their distance to the center of the reaction.
This shows clearly that after the violent phase of the heavy-ion reaction the distribution of clusters of size $A$ is not proportional to the $A$-th power of the nuclear density $(\rho_N(x))^A$ or of the nuclear phase space density $(f_N(p,x))^A$. Since we use in this analysis MST to identify the clusters we cannot say whether at an earlier time such a proportionality could be observed.  

\begin{figure}
\centering
\includegraphics[scale=0.45]{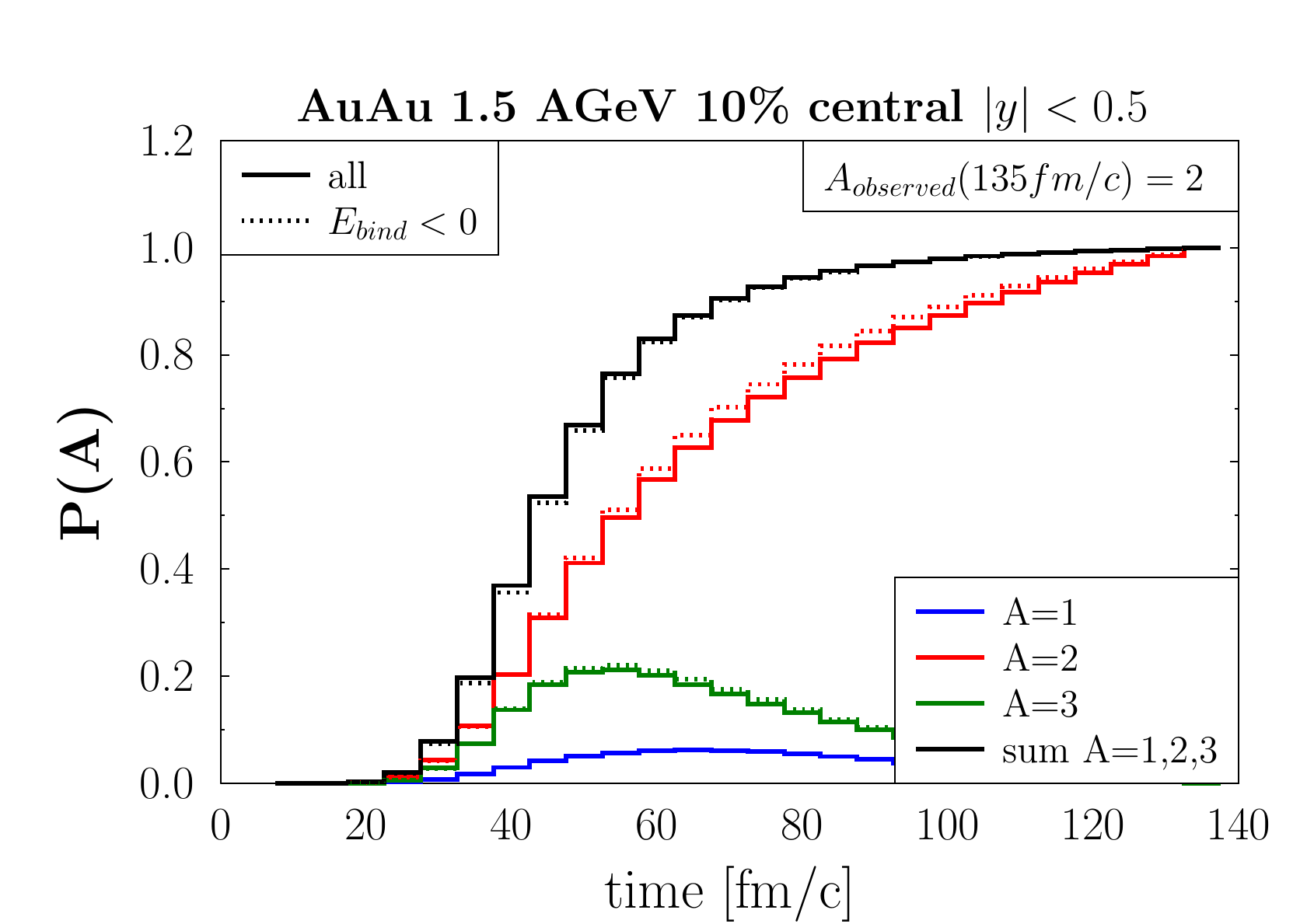}
\includegraphics[scale=0.45]{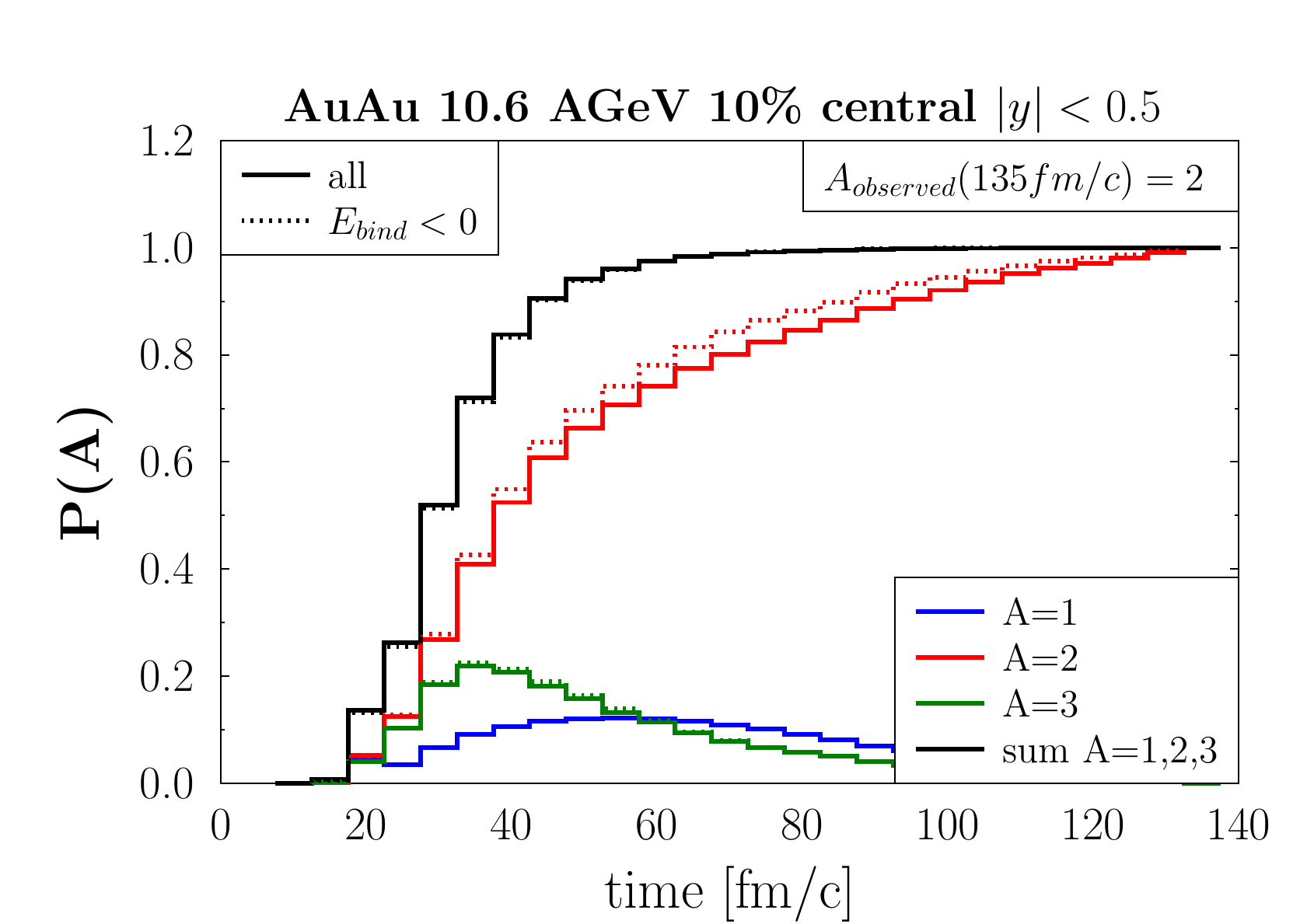}
\includegraphics[scale=0.45]{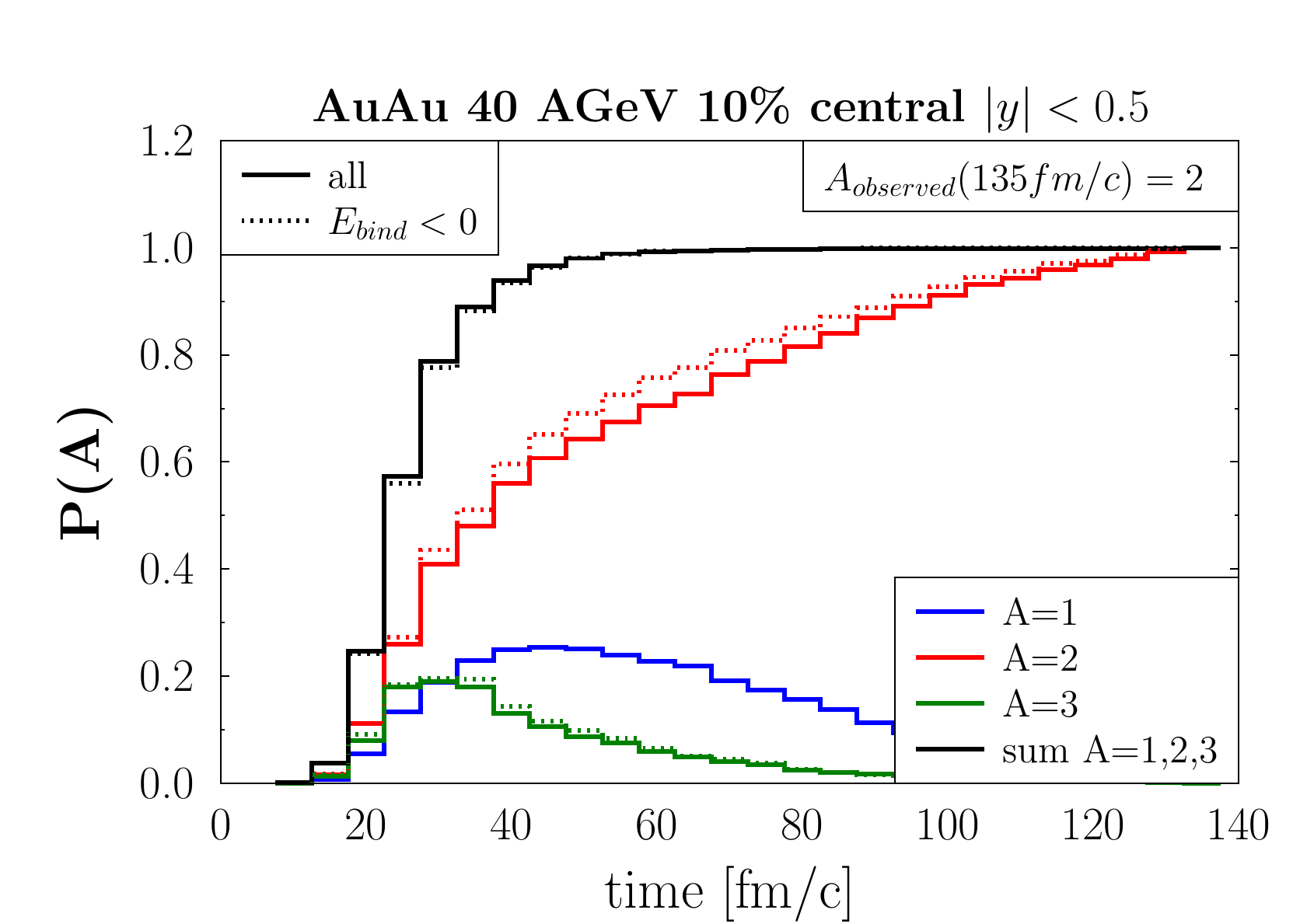}
\caption{\label{fig:formtime2} https://edt.univ-nantes.fr/sciences/s85965.html
The probability distribution $P(A)$ of the formation time of clusters at midrapidity, $|y|<0.5$, for the 10\% most central Au+Au collisions at beam energies of $E_{kin} = 1.5$~$A$GeV (top), of $E_{kin} = 10$~$A$GeV (middle) and of $E_{kin} = 40$~$A$GeV (bottom).  The probabilities that a finally (i.e. at $t=135$~fm/$c$) observed $A=2$ cluster has been identified already at time $t$ are shown as red lines. The green lines show the probabilities that a nucleon of the finally observed $A=2$ cluster has been at time $t$ a part of the $A=3$ cluster;  the blue lines show the probability that a $A=2$ cluster nucleon has been a single nucleon $A=1$ at time $t$. Black lines display the sum $P(1)+P(2)+P(3)$. This analysis is done for clusters which are bound at $t=135$~fm/$c$ (dotted lines) and all clusters (full lines).}
\end{figure}

\begin{figure}
\centering
\includegraphics[scale=0.45]{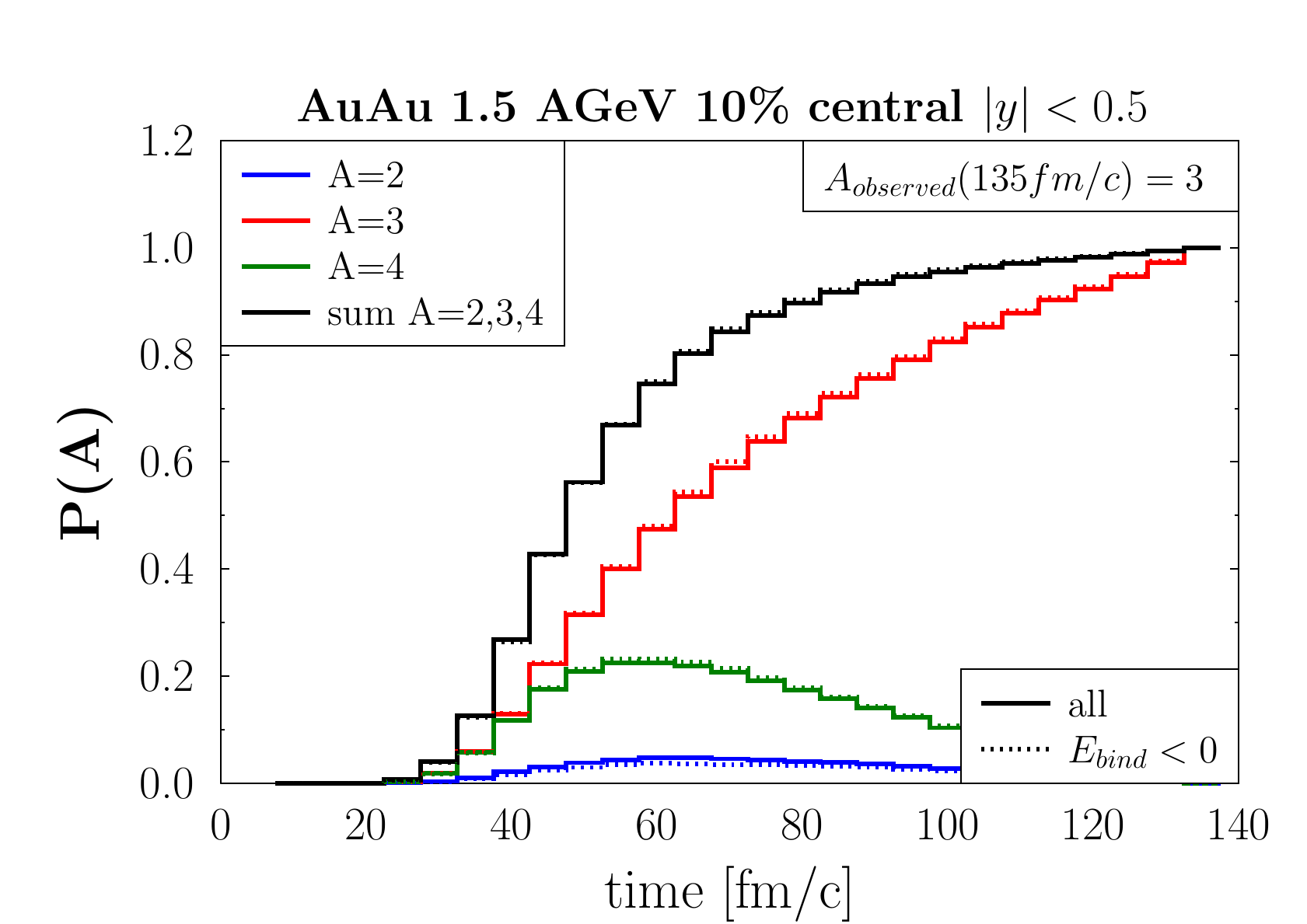}
\includegraphics[scale=0.45]{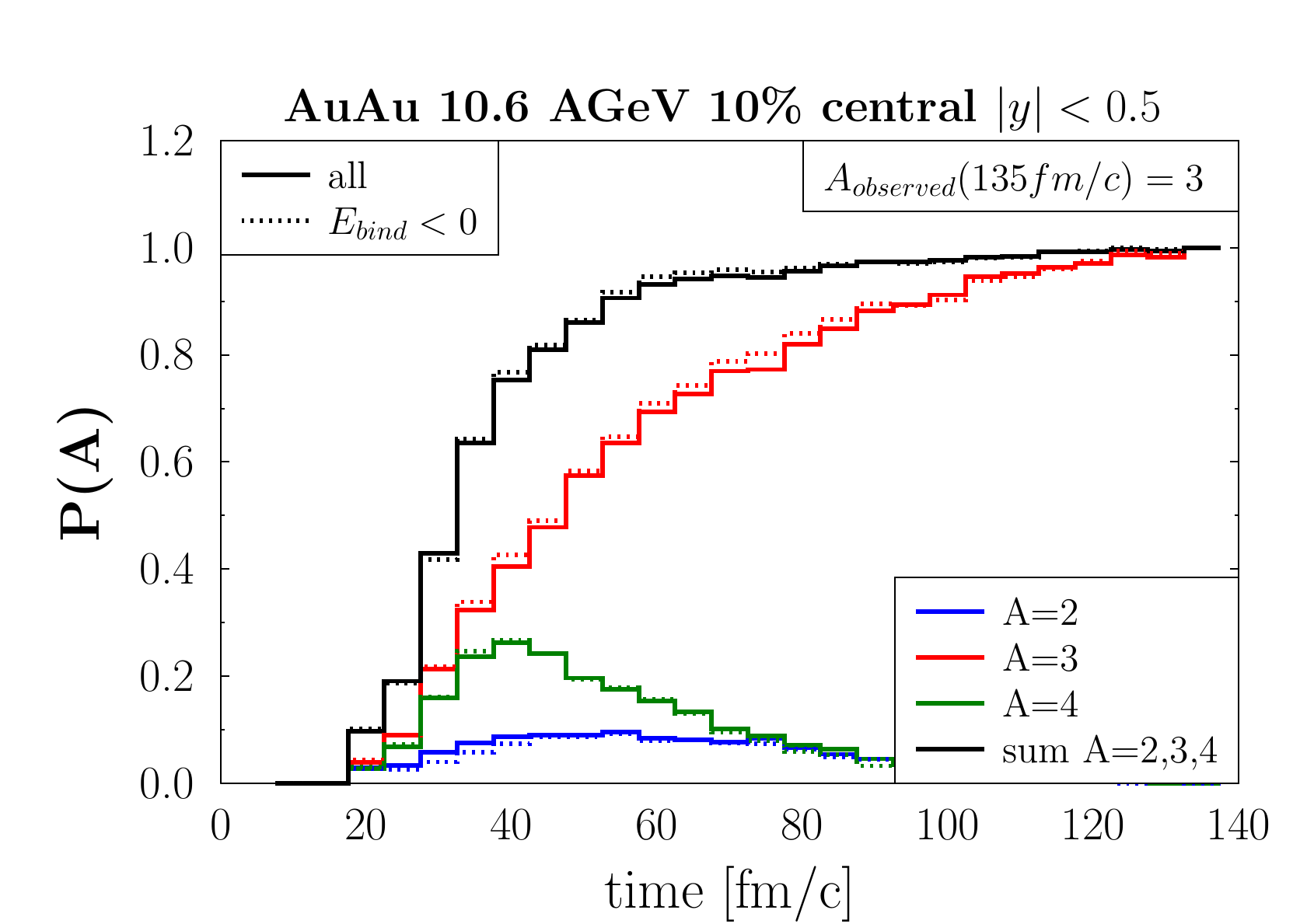}
\includegraphics[scale=0.45]{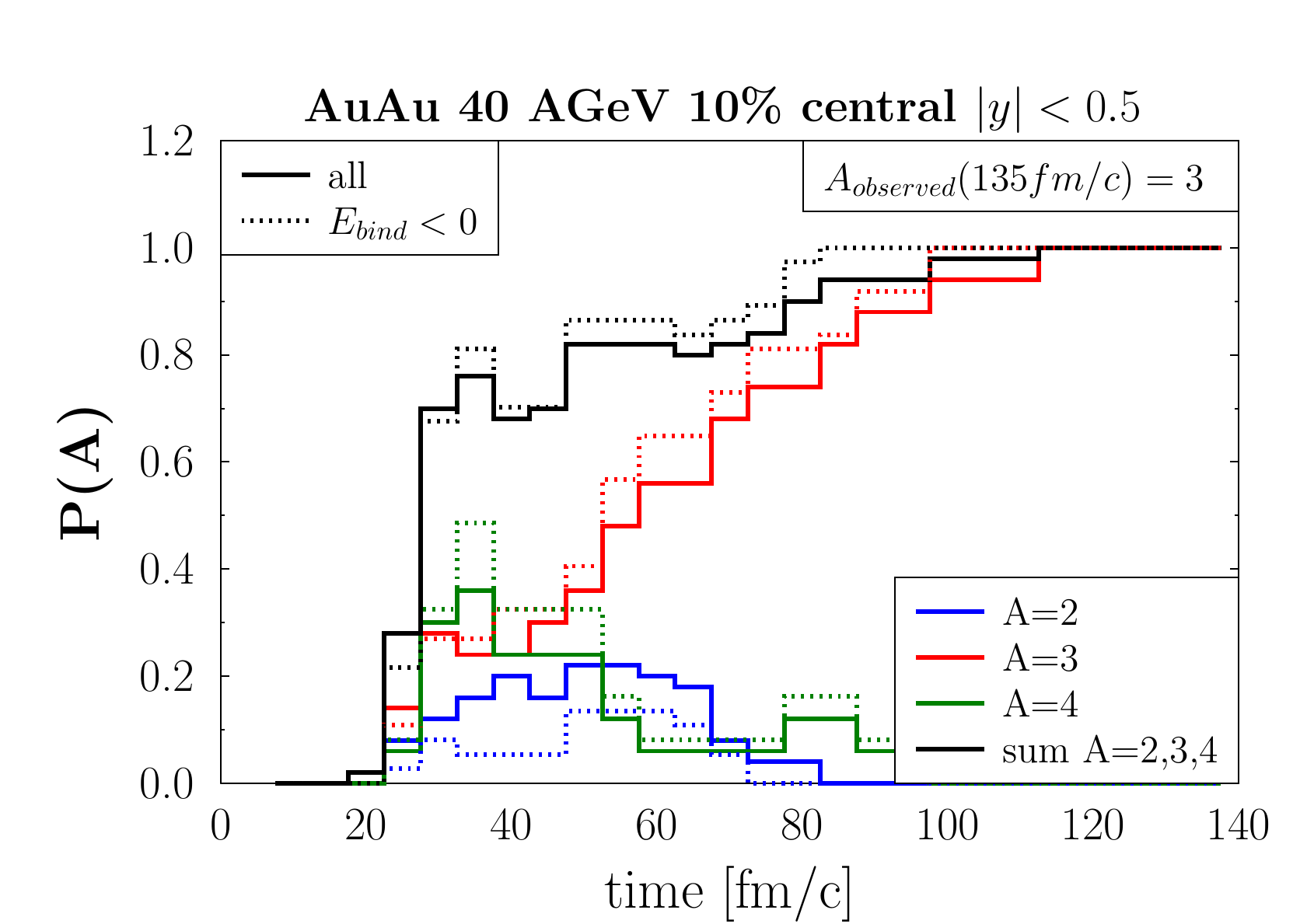}
\caption{\label{fig:formtime3} 
The probability distribution $P(A)$ of the formation time of clusters at midrapidity, $|y|<0.5$, for the 10\% most central Au+Au collisions at beam energies of $E_{kin} = 1.5$~$A$GeV (top), of $E_{kin} = 10$~$A$GeV (middle) and of $E_{kin} = 40$~$A$GeV (bottom).  The probabilities that a finally (i.e. at $t=135$~fm/$c$) observed $A=3$ cluster has been identified already at time $t$ are shown as red lines. The green lines show the probabilities that a nucleon of the finally observed $A=3$ cluster has been at time $t$ a part of the $A=4$ cluster;  the blue lines show the probability that a $A=3$ cluster nucleon has been part of a $A=2$ at time $t$. Black lines display the sum $P(2)+P(3)+P(4)$. This analysis is done for clusters which are bound at $t=135$~fm/$c$ (dotted lines) and all clusters (full lines).}
\end{figure}

Using the recorded  history of baryons in the PHQMD,  we address now the question: When have the clusters been formed? To study this we trace back the nucleons which form finally (i.e. at $t=135$~fm/$c$) clusters  
of size $A=2$ and $A=3$.
We identify the time  when the nucleons fulfil for the first time the condition to be considered as a members of a
$A=2$ or $A=3$ cluster, respectively. The probability that a cluster has already its final size (defined at $t=135$~fm/$c$) is displayed as a function of time in Fig.~\ref{fig:formtime2} for final $A=2$ clusters and in Fig.~\ref{fig:formtime3} for final $A=3$ clusters as red lines. We show as well the probability that the final cluster of size $A$ has been at that time part of a cluster of size $A+1$ (green line) or of a $A-1$ cluster (blue lines), which in the case of $A=2$ is a single nucleon. 
The black line is the sum of all three probabilities. The figure shows that the clusters are produced shortly after the collisional interactions of baryons (cf. Fig.~\ref{fig:frz}) have practically ceased, i.e. near the thermal freeze-out. From then on clusters may still lose or gain one nucleon but the probability for larger changes is very small. 
Again we observe the similarity of the formation time of clusters produced at different beam energies
$E_{kin} = 1.5, 10$ and 40~$A$GeV. The cluster formation time decreases a bit with increasing beam 
energy because at lower energies the baryon density is higher and, therefore, the formation of larger, unstable clusters is more  probable.
We can conclude that cluster formation at midrapidity is a process which occurs  after the collisional interactions of baryons have ceased. They are dominantly formed from nucleons which are closer to the reaction center than free nucleons. 
Therefore, our findings are different from the models which assume that the probability of the formation of a cluster of size $A$ is proportional to the $A$-th power of the single particle density. 

\section{Conclusions}

In this paper we have employed the recently advanced PHQMD transport approach to study the cluster formation at midrapidity in the energy range from AGS to top RHIC with the special focus on the energies between $E_{kin} = 10.6$~$A$GeV and $E_{kin} = 40$~$A$GeV, relevant for the upcoming experiments at the FAIR and NICA facilities. Clusters are identified with a minimum spanning tree (MST) in coordinate space (additional cuts in momentum space do, however, not change much the cluster distributions found by the MST \cite{Kireyeu:2021igi}). Small semi-classical clusters are not very stable and therefore we have to chose
a time at which we identify the clusters. This time has exclusively an influence on the multiplicity, but not on dynamical variables and is almost the same for the energy range investigated here.

The major findings of our investigation are the following:
\begin{itemize}
    \item The mutual potential interactions between baryons lead naturally to the formation of midrapidity clusters. No additional coalescence procedure is necessary to identify the clusters.
    
    \item At energies of $E_{kin}=10.6$~$A$GeV and $\sqrt{s_{NN}} = 8.8$~GeV our calculations agree quite well with the experimental data. We reproduce the rapidity distribution of light clusters as well as their $p_T$ distribution in different rapidity intervals. The agreement of our results with the experimental data show that cluster production is not a simple phase space coalescence: we find that the probability to form clusters depends on the transverse momentum of the cluster which  is not predicted by simple coalescence models. 
    
    \item We obtain also good agreement between theory and experimental data for the excitation function of the cluster yield and for the average $p_T$ of clusters from  $E_{kin}=10.6$~$A$GeV up to the top RHIC energies.
    
    \item The few existing experimental data on hypernuclei are reasonable well described, even if we use in this study the simplifying assumption that $V_{N\Lambda}=2/3 \, V_{NN}$. This means that not only the creation of $\Lambda$'s is well reproduced \cite{Aichelin:2019tnk} but also the formation of the hypernuclei themselves, i.e. the absorption of the $\Lambda$'s by nucleons to form a hypernucleus. The error bars of the experiments and that of the theory (it is very costly to produce hypernuclei at AGS energies and below) are still too large for drawing detailed conclusions about the production process.
    
    \item We used the registered cluster history in PHQMD to investigate how and when clusters are formed. We find an unexpected similarity of the cluster production at midrapidity from $E_{kin}=1.5$~$A$GeV up to $\sqrt{s_{NN}} = 8.8$~GeV. The clusters are produced in between 15 fm/c after
    collisions between the hadrons have ceased (around 30-40 fm/c). At 50 fm/c 70\%, 90\% and 95\% of the pre-clusters have been formed at 1.5, 10.6 and 40 AGeV, respectively. Pre-cluster we name those cluster which differ by only one
    baryon from the finally (at 135 fm/c) observed clusters.
    
    and, most important, in coordinate space they are formed behind the front of the fast expanding hadrons. This is the reason why clusters can survive. 
    
    \item We compared the $dN/dy$ at midrapidty with thermal model predictions and found an unexpected similarity. Unexpected because, as said in the introduction, due to their low binding energy clusters do not survive in a thermal source of a temperature larger than 100~MeV. Therefore, the experimental observation that their multiplicity corresponds to the thermal expectation value
    is puzzling. In PHQMD the final rapidity distribution is a consequence of the potential and collisional interactions among the baryons and collisions of cluster nucleons with other hadrons would destroy the clusters.   
    
    \item Hence the 'ice in the fire' puzzle is only a puzzle due to the unobservable assumption of thermal models that baryons and clusters are distributed homogeneously in space. The PHQMD calculations show that clusters come from different space regions than free nucleons which does not exclude that they have $p_T$ spectra compatible with that expected from a homogeneous thermal source.
\end{itemize}

\section*{Acknowledgements}
The authors acknowledge inspiring discussions with  W. Cassing, C. Hartnack, Yv. Leifels,  A. Le F\'evre,
O. Soloveva, T. Song and  Io. Vassiliev. 
Furthermore, we acknowledge support by the Deutsche Forschungsgemeinschaft 
(DFG, German Research Foundation): grants BR~4000/7-1 and BL~982/3-1,  by the Russian Science Foundation grant 19-42-04101
and   by the GSI-IN2P3 agreement under contract number 13-70.
Also we thank the COST Action THOR, CA15213 and the European Union’s Horizon 2020 research and innovation program under grant agreement
No 824093 (STRONG-2020).
The computational resources have been provided by the LOEWE-Center for Scientific Computing
and the "Green Cube" at GSI, Darmstadt.

\bibliography{PHQMDII_1}

\end{document}